%% file: main.tex
\documentclass[%
 reprint,
 amsmath,amssymb,
 prd,
]{revtex4-2}

\usepackage{graphicx}
\usepackage{makecell}
\usepackage{dcolumn}
\usepackage{bm}
\usepackage[export]{adjustbox} 
\usepackage{array}
\usepackage{amsmath}
\usepackage{amsfonts} 
\usepackage{float} 
\usepackage{comment}
\usepackage[caption=false]{subfig}

\usepackage[usenames,dvipsnames,svgnames,table]{xcolor}
\usepackage{hyperref}

\definecolor{verdeoscuro}{rgb}{0, 0.5, 0}

\begin{document}

\preprint{APS/123-QED}

\title{Resonant heterodyne conversion applied to a low-frequency haloscope \\ for dark matter axion searches in the 1-35 MHz range}

\author{Jose R. Navarro-Madrid$^{a}$}
\thanks{Corresponding author: joser.navarro@upct.es} 
\author{Jos\'e Reina-Valero$^{b,c}$} 
\author{Alejandro D\'iaz-Morcillo$^{a}$} 
\author{Benito Gimeno$^{b}$}

\affiliation{$^a$Departamento de Tecnolog\'ias de la Informaci\'on y las Comunicaciones,
Universidad Polit\'ecnica de Cartagena,
Plaza del Hospital 1, 30202 Cartagena, Spain. \\ 
$^b$Instituto de Física Corpuscular (IFIC), CSIC-University of Valencia, Calle Catedr\'atico Jos\'e Beltr\'an Martínez, 2, 46980 Paterna (Valencia), Spain. \\
$^c$Laboratorio Subterráneo de Canfranc (LSC), 22880 Canfranc-Estación, Spain.}

\date{\today}

\begin{abstract}

We study resonant heterodyne up-conversion in the RADES-BabyIAXO haloscope as a method to search for low-mass dark matter axions using microwave cavities. Starting from axion electrodynamics, we derive the axion-induced source term and the power extracted through a readout mode, explicitly accounting for the finite axion linewidth. This leads to effective quality factors that determine the pump–axion mixing, detection bandwidth, and detected signal power. We extend the BI-RME 3D full-wave formulation to heterodyne axion detection in a realistic two-port cavity, including pump leakage into the readout channel. Applying the formalism to the largest RADES-BabyIAXO cavity identifies the $\mathrm{quasi\textrm{-}TE}_{011}-\mathrm{quasi\textrm{-}TM}_{010}$ mode pair as a favorable configuration, enabling sensitivity to axion frequencies between 0.9 and 34.6 MHz. Analytical and full-wave predictions show excellent agreement at resonance, while the full-wave model provides a more accurate description off resonance and allows a precise characterization of the pump leakage. We also derive the optimal port couplings that maximize the scanning rate. Upper-limit sensitivity for cryogenic copper and superconducting niobium cavities indicate that, under thermal-noise-limited conditions and assuming sufficient pump-leakage rejection, the experiment could probe axion-photon couplings down to $10^{-15}~\mathrm{GeV}^{-1}$ at 90\% confidence level, representing a significant improvement over previous heterodyne-based searches.

\end{abstract}

\maketitle


\section{Introduction} \label{sec:intro}

Strong interest in axion searches has arisen in recent years as a possible constituent of dark matter (DM). Among the possible candidates for explaining the nature of this component of the Universe, the axion is probably the most theoretically motivated one: it was proposed to solve the puzzling strong CP problem in the context of Quantum Chromodynamics (QCD). R. Peccei and H. Quinn proposed an additional $U(1)$ symmetry and promoted the $\bar{\theta}$ parameter to a field, giving a dynamical explanation for the cancellation of $\bar{\theta}$ \cite{Peccei_Quinn, Kim:2008hd}. Weinberg \cite{Weinberg} and Wilczek \cite{Wilczek} noticed that, when this new symmetry is spontaneously broken, a pseudo-Nambu-Goldstone boson arises: the axion. It was later realized that it could explain the observed DM abundance of the Universe today \cite{Willy_1, Willy_2, Willy_3}.

The most straightforward method for detecting this particle is through its coupling to photons. To this end, one of the most widely used setups is the haloscope, in which a resonant cavity is employed for amplifying the resulting radiation. The main way of implementing this method is through the so-called homodyne detection, in which the cavity is placed inside a uniform, high-intensity magnetostatic field \cite{ADMX, CAPP_2020, CAPP_2024, HAYSTAC, QUAX:2024fut, ORGAN, RADES_1, RADES_2, CADEx}. The resultant radiation excites an electromagnetic mode of the cavity, which is then detected. One of the difficulties of this method relies on its own nature: one has to apply a very intense magnetic field for the conversion to be significant enough. Increasing this intensity can be very challenging \cite{IRASTORZA201889, RevModPhys.75.777}, even more when trying to increase the magnet volume: the amount of detected signal will increase linearly with the cavity volume, implying that large magnets would also be needed. In addition, if dielectric materials are avoided inside the cavity, the explored mass range determines the volume of the haloscope and, therefore, of the magnet. This has currently set the lower limit for DM axion detection with resonant cavities around 100 MHz. In this context, an alternative detection strategy at lower frequencies can be developed: the heterodyne detection.

Heterodyne detection experiments rely on the presence of two microwave cavity modes with a finite electromagnetic field overlap, such that the electric field of one mode and the magnetic field of the other mode are non-orthogonal. In this configuration, one of the modes is driven with a strong input power (called the pump mode) at angular frequency $\omega_p$, while the second mode serves as the readout channel (readout mode) at angular frequency $\omega_r$. This second mode is used to measure the signal that can be generated through axion-induced mixing with the pump mode magnetic field. Therefore, the up/down-conversion approach provides a heterodyne strategy for axion DM detection, in which the axion is not searched through direct conversion into a single resonant mode, but rather through the mode mixing of two operating modes. This method provides the ability to search for axions in the angular frequency range of $\omega_a<<\omega_p, \omega_r$ by using cavity haloscopes, since the explored axion angular frequency range is $\omega_a=|\:\omega_p-\omega_r\:|$, thereby avoiding the use of LC resonators \cite{LC_resonator1, LC_resonator2} and shifting the problem into the microwave range. Two techniques can be identified for detecting axions by employing the heterodyne detection: the ``power technique'', in which the axion perturbs the power amplitude of the readout mode \cite{Lasenby_2020,Lasenby_2021,Heterodyne_2021}, and the ``frequency technique'', in which the phase or frequency undergoes a variation \cite{Upconversion_original, Upconversion_frequency}. In the present article, we focus on the first one. The up/down-conversion method was first proposed in \cite{Sikivie_SRF} and experimentally demonstrated for axions in \cite{Upconversion_australianos}, although this method can also be applied to other types of phenomena, such as the detection of gravitational waves (GWs) \cite{MAGO_2.0,GW_upconversion, Gue:2026kga}.

This article is organized as follows: in Sec. \ref{sec:birme}, a review of axion electrodynamics in the presence of an external radio frequency (RF) magnetic field is presented, followed by the derivation of the detected power for the heterodyne detection of a monochromatic axion. Then, an analysis of non-monochromatic sources is carried out, which is particularly important in the context of Superconducting RF (SRF) cavities. At the end of this section, the scanning rate expression is derived. In Sec. \ref{sec:birmecircuit}, the application of the BI-RME 3D method to heterodyne detection is investigated, which takes on special relevance in the study of the signal leakage from the pump signal to the readout mode. Section \ref{sec:Results} applies this detection scheme to a realistic resonant haloscope, which includes tuning mechanical elements; here, numerical simulations are performed to obtain both the overlap factor as well as the detected power. Furthermore, we conduct a study on the optimal coupling to maximize the scanning rate, and finally, upper-limit sensitivity estimates are provided for the considered frequency range.

\section{Theory} \label{sec:birme}

\subsection{Axion electrodynamics}

The effective Lagrangian density $\mathcal{L}$ describing the axion electromagnetic interaction can be derived, in SI units, as
\begin{eqnarray}
\mathcal{L} \, = \, \mathcal{L}_0 \, + \, \mathcal{L}_a \, + \,  \mathcal{L}_U \nonumber
\end{eqnarray}
where $\mathcal{L}_0$ is the classical electromagnetic Lagrangian density, $\mathcal{L}_U$ is the axion Lagrangian density with a potential $U(\mathtt{a}) \, = \, (1/2) \, \omega_a^2 \, \mathtt{a}^2$, and $\mathcal{L}_a$ is the axion interaction term,
%
\begin{align}
\mathcal{L}_0 \, & = \, \frac{-1}{4 \, \mu_0} \, F^{\mu \nu} \, F_{\mu \nu} \, - \, A_{\mu} \, J_e^{\mu}  \nonumber \\
\mathcal{L}_U \, & = \, \frac{1}{2} \, (\partial^{\mu} \, \mathtt{a}) \, (\partial_{\mu} \, \mathtt{a}) \, - \, \frac{1}{2} \, \omega_a^2 \, \mathtt{a}^2 \nonumber \\
\mathcal{L}_a \, & = \, \frac{g_{a \gamma \gamma}}{4 \mu_0} \, \mathtt{a} \, F^{\mu \nu} \, \overset{\thicksim}{F}_{\mu \nu} \nonumber
\end{align}
%
where $g_{a \gamma \gamma}$ is the two-photon coupling to an axion field $\mathtt{a}$; $\mu_0$ is the magnetic permeability of vacuum; $\omega_a$ is the axion angular frequency; $F_{\mu \nu}$ and $\overset{\thicksim}{F}_{\mu \nu}$ are the electromagnetic field tensor and their dual, respectively; $A_{\mu}$ is the four-potential; and $J_e^{\mu}$ is the four-current of the external source. Time-domain Maxwell's equations in vacuum, expressed in SI units and considering the axion-photon interaction, can be directly derived from the Lagrangian density $\mathcal{L}$, obtaining

\begin{eqnarray}
	\nabla \cdot (\vec{\mathcal{E}} - c \, g_{a \gamma \gamma} \, \mathtt{a} \, \vec{\mathcal{B}}) \, & = & \, \frac{\mathcal{\varrho}_e}{\varepsilon_0}   \nonumber   \\
	\nabla \cdot \vec{\mathcal{B}} \, & = & \, 0  \nonumber  \\
	\nabla \times \vec{\mathcal{E}} \, & = & \, - \frac{\partial \, \vec{\mathcal{B}}}{\partial \, t}   \nonumber \\
	\nabla \times (c \, \vec{\mathcal{B}} + g_{a \gamma \gamma} \, \mathtt{a} \, \vec{\mathcal{E}}) \, & = & \, \frac{1}{c} \, \frac{\partial}{\partial \, t} (\vec{\mathcal{E}} - c \, g_{a \gamma \gamma} \, \mathtt{a} \, \vec{\mathcal{B}}) + c \, \mu_0 \, \vec{\mathcal{J}_e}   \nonumber
\end{eqnarray}
where $\varepsilon_0$ is the electric permittivity of vacuum; $c = 1/\sqrt{\varepsilon_0 \, \mu_0}$ is the speed of light in vacuum; $t$ is the time measured in the laboratory reference system; and $\vec{\mathcal{E}}$ and $\vec{\mathcal{B}}$ are the electric and magnetic fields, respectively. Assuming that the axion-photon interaction slightly modifies the electromagnetic field, these equations can be decoupled into two parts \cite{kim_CAPP_2019}: one part for the external electromagnetic field $\vec{\mathcal{E}_e}$, $\vec{\mathcal{B}_e}$ generated by the classical charge $\mathcal{\varrho}_e$ and current $\vec{\mathcal{J}_e}$ densities, given by
\begin{eqnarray}  \label{maxwell_eq_td_classical}
	\nabla \cdot \vec{\mathcal{E}_e} \, & = & \, \frac{\mathcal{\varrho}_e}{\varepsilon_0}   \nonumber   \\
	\nabla \cdot \vec{\mathcal{B}_e} \, & = & \, 0            \nonumber  \\
	\nabla \times \vec{\mathcal{E}_e} \, & = & \, - \frac{\partial \, \vec{\mathcal{B}_e}}{\partial \, t}   \nonumber  \\
	\nabla \times \, \vec{\mathcal{B}_e} \, & = & \, \frac{1}{c^2} \, \frac{\partial \, \vec{\mathcal{E}_e}}{\partial \, t}  \,  +  \, \mu_0 \, \vec{\mathcal{J}_e}   
\end{eqnarray}
and another set of Maxwell's equations for the reacted fields $\vec{\mathcal{E}_a}$, $\vec{\mathcal{B}_a}$:  
\begin{eqnarray}  \label{maxwell_eq_td_axion}
	\nabla \cdot (\vec{\mathcal{E}_a} - c \, g_{a \gamma \gamma} \, \mathtt{a} \, \vec{\mathcal{B}_e}) \, & = & \,  0     \nonumber   \\
	\nabla \cdot \vec{\mathcal{B}_a} \, & = & \, 0   \nonumber  \\
	\nabla \times \vec{\mathcal{E}_a} \, & = & \, - \frac{\partial \, \vec{\mathcal{B}_a}}{\partial \, t}   \nonumber  \\
	\nabla \times (\vec{\mathcal{B}_a} + \frac{1}{c} \, g_{a \gamma \gamma} \, \mathtt{a} \, \vec{\mathcal{E}_e}) \, & = & \,  
    \frac{1}{c^2} \, \frac{\partial}{\partial \, t} (\vec{\mathcal{E}_a} - c \, g_{a \gamma \gamma} \, \mathtt{a} \, \vec{\mathcal{B}_e}) .\nonumber\\
\end{eqnarray}
The definition of the equivalent axion charge $\mathcal{\varrho}_a$ and current $\vec{\mathcal{J}_a}$ densities as 
\begin{eqnarray}  \label{axion_rho_J_densities}
	\mathcal{\varrho}_a  \, & \equiv & \,  g_{a \gamma \gamma} \, \sqrt{\frac{\varepsilon_0}{\mu_0}} \, \nabla \cdot (\mathtt{a} \, \vec{\mathcal{B}_e})   \\
	\vec{\mathcal{J}_a} \,  & \equiv & \, -  g_{a \gamma \gamma} \, \sqrt{\frac{\varepsilon_0}{\mu_0}} \,   
	\left( \frac{\partial \, (\mathtt{a} \, \vec{\mathcal{B}_e})}{\partial \, t} + \nabla \times (\mathtt{a} \, \vec{\mathcal{E}_e}) \right)   
\end{eqnarray}
allows to rewrite the axion Maxwell's equations in free space (\ref{maxwell_eq_td_axion}) in the conventional form,
\begin{eqnarray}
	\nabla \cdot \vec{\mathcal{E}_a} \, & = & \, \frac{\mathcal{\varrho}_a}{\varepsilon_0}    \nonumber     \\
	\nabla \cdot \vec{\mathcal{B}_a} \, & = & \, 0   \nonumber   \\
	\nabla \times \vec{\mathcal{E}_a} \, & = & \, - \frac{\partial \, \vec{\mathcal{B}_a}}{\partial \, t}   \nonumber   \\
	\nabla \times \, \vec{\mathcal{B}_a} \, & = & \, \frac{1}{c^2} \, \frac{\partial \, \vec{\mathcal{E}_a}}{\partial \, t}  \,  +  \, \mu_0 \, \vec{\mathcal{J}_a}   \, .
\end{eqnarray}
Note that both sets of charge and current densities satisfy the time-domain continuity equations:
\begin{eqnarray}
\nabla \cdot  \, \vec{\mathcal{J}_e} \, + \, \frac{\partial \, \mathcal{\varrho}_e}{\partial \, t}\,  =  \, 0 \, \, \,  \,  ; \, \,	\,  \,  \nabla \cdot  \, \vec{\mathcal{J}_a} \, + \, \frac{\partial \, \mathcal{\varrho}_a}{\partial \, t}\,  =  \, 0   \,  .  
\end{eqnarray}

Axion haloscopes aim to detect the electromagnetic response induced by the axion field inside a microwave cavity in the presence of external electric, $\vec{\mathcal{E}_e}$, and/or magnetic, $\vec{\mathcal{B}_e}$, fields. In the conventional experimental implementations, only an external strong magnetostatic field is considered, while the external electric field is zero (the signal arises through the inverse Primakoff effect \cite{Primakoff_1,Sikivie_1983}). By contrast, the heterodyne technique relies on the excitation of a resonant pump mode of the cavity, whose associated electric and magnetic fields, $\vec{\mathcal{E}_e}$ and $\vec{\mathcal{B}_e}$, satisfy the classical Maxwell's equations within a microwave resonator (\cite{jackson,collin_FMI,collin_FTGW}).

Throughout this work, the axion-induced response is described within the framework of axion electrodynamics (cf. Eq.~(22) of Ref. \cite{Sikivie_2021}), 
\begin{eqnarray}    \label{equation_sikivie_1}
\frac{1}{c^2}\frac{\partial^2 \mathtt{a}}{\partial t^2} \, - \, \nabla^2 \mathtt{a} \, + \, \left(\frac{m_a c}{\hbar}\right)^2 \, \mathtt{a} \, = \, - g_{a \gamma \gamma} \, \hbar c^2\, \varepsilon_0 \, \vec{\mathcal{E}} \cdot \vec{\mathcal{B}},
\end{eqnarray}
where $m_a$ is the axion mass, and $\hbar$ is the reduced Planck constant. By inserting the first-order expansion proposed in subsection 2.2 of Ref. \cite{kim_CAPP_2019} for the electric $\vec{\mathcal{E}}	 \,  \approx \, \vec{\mathcal{E}}_e \, + \, g_{a \gamma \gamma} \, \vec{\mathcal{E}}_1 \, = \,  \vec{\mathcal{E}}_e \, + \, \vec{\mathcal{E}}_a $ and the magnetic $\vec{\mathcal{B}}	 \,  \approx \, \vec{\mathcal{B}}_e \, + \, g_{a \gamma \gamma} \, \vec{\mathcal{B}}_1 \, = \, \vec{\mathcal{B}}_e \, + \, \vec{\mathcal{B}}_a$ fields in Eq. (\ref{equation_sikivie_1}), and neglecting contributions of order $g_{a \gamma \gamma}^2$ and higher, the field equation reduces to
\begin{eqnarray}    \label{equation_sikivie_2}
\frac{1}{c^2}\frac{\partial^2 \mathtt{a}}{\partial t^2} \, - \, \nabla^2 \mathtt{a} \, + \, \left(\frac{m_a c}{\hbar}\right)^2 \, \mathtt{a} \, \approx \, - g_{a \gamma \gamma} \, \hbar \,  c^2\varepsilon_0 \,\left( \vec{\mathcal{E}_e} \cdot \vec{\mathcal{B}_e}\right).  
\end{eqnarray}
To simplify the formulation, we impose that the electric and magnetic fields of each cavity mode be orthogonal pointwise throughout the resonator: $\vec{\mathcal{E}}_e \, \cdot \vec{\mathcal{B}}_e \, = \, 0$. As it is shown in Appendix~\ref{Appendix_modes_cavity}, this condition is satisfied by the TE and TM modes of a microwave cavity. As a consequence, by assuming that the pump mode is a TE or TM mode, it is found that
\begin{eqnarray}
	\frac{1}{c^2}\frac{\partial^2 \mathtt{a}}{\partial t^2} \, - \, \nabla^2 \mathtt{a} \, + \, \left(\frac{m_a c}{\hbar}\right)^2 \, \mathtt{a} \, = 0
\end{eqnarray}
which can be easily expressed in frequency-domain as
\begin{eqnarray}
\left[\nabla^2 + \left(\frac{\omega}{c}\right)^2 - \left(\frac{m_a c}{\hbar}\right)^2\right] \, a \, = \, 0.  
\end{eqnarray}
where $\omega$ is the angular frequency, and the Fourier Transform (FT) of the function $\mathtt{f}(t)$ is defined as,
\begin{eqnarray}  \label{Fourier_transform_definition}
	f(\omega)  \,  \equiv  \int_{- \infty}^{+ \infty}  \, \mathtt{f}(t) \, e^{- \mathrm{i} \omega t} \, dt  \, \, \Rightarrow  \, \, f = FT[\mathtt{f}]
\end{eqnarray} 
$\mathrm{i}$ being the imaginary unit ($\mathrm{i} \equiv \sqrt{-1}$), and $a = FT[\mathtt{a}]$. 
The solution of this differential equation in the frequency-domain can be written, in complex phasor notation, as a plane wave \cite{Sikivie_2021}. Assuming a pure monochromatic axion field, it can be written as
\begin{eqnarray}    \label{axion_field_phasor}
	a(\vec{r}) \, = \, a_0 \, e^{-\mathrm{i} (\vec{k}_a \cdot \vec{r} - \varphi)}   
\end{eqnarray}
where $a_0$ is the amplitude of the axion field, $\varphi$ is the axion field intrinsic phase, and  $\vec{k}_a = (m_a \vec{v})/\hbar$ is the axion De Broglie wavenumber, with $\vec{v}$ the axion velocity in the Galactic halo ($||\vec{v}|| \approx 10^{-3} c$). Throughout this manuscript, the time-harmonic factor $e^{\mathrm{i} \omega t}$ is omitted in the complex-phasor representation. The corresponding real axion field is then obtained as
\begin{eqnarray}    \label{axion_field_timedomain}   
	\mathtt{a}(\vec{r},t) \, = \, \mathfrak{Re} (a \, e^{\mathrm{i} \omega t}) \, = \,   \mathfrak{Re} (a_0  \, e^{\mathrm{i} (\omega t -\vec{k}_a \cdot \vec{r} + \varphi)}) =\nonumber \\ = a_0  \, \cos(\omega t -\vec{k}_a \cdot \vec{r} + \varphi)  
\end{eqnarray}

The next step is to calculate the axion charge and current densities defined in (\ref{axion_rho_J_densities}) as
\begin{eqnarray}     
	\mathcal{\rho}_a \,  \equiv  \, g_{a \gamma \gamma} \, \sqrt{\frac{\varepsilon_0}{\mu_0}} \,   
	\, \nabla \cdot (\mathtt{a} \, \vec{\mathcal{B}_e})  \, = \,  g_{a \gamma \gamma} \, \sqrt{\frac{\varepsilon_0}{\mu_0}} \,  \vec{\mathcal{B}_e} \cdot \nabla \mathtt{a}   
\end{eqnarray}
\begin{eqnarray}    \label{axion_current_timedomain}	
	\vec{\mathcal{J}_a} \,  & \equiv & \, -  g_{a \gamma \gamma} \, \sqrt{\frac{\varepsilon_0}{\mu_0}} \,   
	\left( \frac{\partial \, (\mathtt{a} \, \vec{\mathcal{B}_e})}{\partial \, t} \, + \nabla \times (\mathtt{a} \, \vec{\mathcal{E}_e}) \right) =\nonumber \\
    & = &  \, -  g_{a \gamma \gamma} \, \sqrt{\frac{\varepsilon_0}{\mu_0}} \, \frac{\partial \, \mathtt{a}}{\partial \, t} \, \vec{\mathcal{B}_e}
\end{eqnarray}
where we have assumed that the axion de Broglie wavelength is much larger than the characteristic size of the haloscope, so that $\nabla \mathtt{a} \approx 0$. Under this approximation, the effective axion charge and current densities still satisfy the time-domain continuity equation: $\nabla \cdot \vec{\mathcal{J}_a} + \frac{\partial \mathcal{\rho}_a}{\partial t} = 0$.

We now evaluate the effective axion-induced electric current density using the axion-field expression in Eq.~\eqref{axion_field_timedomain}. In the present treatment, the axion de Broglie wavenumber is neglected $\vec{k}_a \approx \vec{0}$. Assuming that the external time-harmonic magnetic field can be written as $\vec{\mathcal{B}_e}(\vec{r},t) = \vec{\mathtt{B}}_{e}(\vec{r}) \cos(\omega_e t + \xi(\vec{r}))$, where $\vec{\mathtt{B}}_{e}(\vec{r})$ is a real 3D vector which accounts the geometrical field distribution of the magnetic field, $\xi(\vec{r})$ is the spatial phase and $\omega_e$ is the angular frequency of the external field, and considering that it is lossless, it yields
\begin{eqnarray}    \label{axion_current_timedomain_2}	
	\vec{\mathcal{J}_a}(\vec{r},t)  \,  =  \, g_{a \gamma \gamma} \, \sqrt{\frac{\varepsilon_0}{\mu_0}} \, a_0 \, \omega_a  \,  \vec{\mathtt{B}}_{e}(\vec{r})  \, \cos(\omega_e t + \xi(\vec{r})) \, \nonumber  \\
    \sin(\omega_a t + \varphi)   \nonumber   \\
\end{eqnarray}
with $m_a c^2 = \hbar\,\omega_a$. In the frequency-domain, Eq.~\eqref{axion_current_timedomain_2}
can be written as
\[
\vec{J}_a(\vec r,\omega)=FT\!\left[\vec{\mathcal{J}}_a(\vec r,t)\right]
=\vec{J}_{a_u}(\vec r,\omega)+\vec{J}_{a_d}(\vec r,\omega),
\]
with the convention defined in Eq.~\eqref{Fourier_transform_definition}. Thus, the up-conversion and down-conversion terms are given by
\begin{align}
\vec{J}_{a_u}(\vec r,\omega)
&=
-\mathrm{i}\,\frac{1}{2}\,g_{a\gamma\gamma}
\sqrt{\frac{\varepsilon_0}{\mu_0}}\,
a_0\,\omega_a \,\vec{B}_{e}(\vec r)\,e^{i\varphi}
\notag\\
&\quad\ \
\delta\!\bigl(\omega-(\omega_e+\omega_a)\bigr),
\label{eq:current_density_up}
\\[4pt]
\vec{J}_{a_d}(\vec r,\omega)
&=
\operatorname{sgn}(\omega_e-\omega_a)\,
\mathrm{i}\,\frac{1}{2}\,g_{a\gamma\gamma}
\sqrt{\frac{\varepsilon_0}{\mu_0}}\,
a_0\,\omega_a \, \vec{\mathtt{B}}_{e}(\vec{r})
\notag\\
&\quad\ \
e^{\,\mathrm{i}(\xi(\vec r)-\varphi)\operatorname{sgn}(\omega_e-\omega_a)}
\delta\!\bigl(\omega-|\omega_e-\omega_a|\bigr).
\label{eq:current_density_down}
\end{align}
where $\vec{B}_{e}(\vec{r}) = FT[\vec{\mathcal{B}_e}(\vec{r},t)] = \vec{\mathtt{B}}_{e}(\vec{r}) \, e^{\mathrm{i} \xi(\vec{r})}$, $\delta(\omega)$ is the Dirac delta function, and the sign function is defined as
\begin{eqnarray}
\operatorname{sgn}(x) & \equiv &
\begin{cases}
-1 & \text{if } x < 0, \\
+1  & \text{if } x > 0.
\end{cases}
\end{eqnarray}
The current density $\vec{J}_{a_{u}}$ describes the up-conversion term at angular frequency $\omega = \omega_e + \omega_a$, while $\vec{J}_{a_{d}}$ describes the down-conversion term at $\omega = |\omega_e - \omega_a|$. The formalism presented above is valid for both processes; however, for conciseness, the remainder of this work will focus on the up-conversion case.

\subsection{Detected power in heterodyne detection}

To derive the signal power extracted from axion-photon up-conversion in the readout mode, we follow the standard resonant mode treatment adopted in Refs.~\cite{Superconducting_upconversion,Upconversion_annalen,Lasenby_2020}. Retaining only the up-conversion term of the axion-induced current density in Eq. \eqref{eq:current_density_up}, and setting $\varphi=0$, the source term of the up-conversion term can be rewritten as
\begin{equation}
\vec{J}_{a_u}(\vec r)
=
-\mathrm{i}\,\frac{1}{2}\,g_{a\gamma\gamma}\,
\sqrt{\frac{\varepsilon_0}{\mu_0}}\,
a_0 \, \omega_a\,\vec{B}_p(\vec r).
\label{eq:J_res_2}
\end{equation}
Here, $\vec B_p \equiv \vec{B}_e$ (we rename, accordingly, $\omega_p = \omega_e$) denotes the magnetic field complex phasor of the pump mode, and the amplitude of the axion field is given by $a_0 = \sqrt{2\rho_a \hbar c^3 / \omega_a^2}$, where the dark matter density is given by $\rho_a\approx0.4\,\mathrm{GeV/cm^3}$. Since only the component of the source overlapping with the readout electric field $\vec{E}_r$ contributes to the excitation of that mode, we define the complex source overlap amplitude \cite{Upconversion_annalen}
\begin{equation}
\Upsilon_{rp} \, 
\equiv \,
2\int_V \vec{E}_r^*(\vec{r})\cdot \vec{J}_{a_u}(\vec r)\,dV,
\label{eq:Ipr_def}
\end{equation}
where the super-index $^*$ denotes complex conjugation. After substitution of Eq.~\eqref{eq:J_res_2}, this becomes
\begin{equation}
\Upsilon_{rp}
=
-\mathrm{i}\,g_{a\gamma\gamma}\,
\sqrt{\frac{\varepsilon_0}{\mu_0}}\,
a_0 \, \omega_a
\int_V \vec{E}_r^*\cdot \vec{B}_p\,dV.
\label{eq:Ipr_final}
\end{equation}
Accordingly,
\begin{equation}
|\Upsilon_{rp}|^2
=
g_{a\gamma\gamma}^2\,
\frac{\varepsilon_0}{\mu_0}\,
a_0^2\, \omega_a^2
\left|
\int_V \vec{E}_r^*\cdot \vec{B}_p\,dV
\right|^2.
\label{eq:Ipr_sq}
\end{equation}
The integral in Eq.~\eqref{eq:Ipr_sq} plays the role of the geometric coupling factor in the up-conversion detection, as it quantifies the overlap between the electric field of the readout mode and the magnetic field of the pump mode. At the resonant angular frequency $\omega_r$, the electric $U_e$ and magnetic $U_m$ time-average energies are equal ($U_e=U_m$), and the stored electromagnetic energy of the readout mode ($U_r$) is
\begin{equation}
U_r \, = \, 2 \, U_e \, = \, 2 \, U_m \, = \,
\frac{\varepsilon_0}{2}\int_V \,||\vec E_r||^2\,dV.
\label{eq:Ur_res}
\end{equation}
At this point, we introduce losses in the pump and readout modes to consider a more realistic scenario, characterized by loaded quality factors $Q_{L,p}$ and $Q_{L,r}$, respectively. For a time-harmonic volumetric current source driving the readout mode exactly on resonance, the steady-state generated power is equal to the total power dissipated in that mode, and therefore, $P_{\mathrm{gen},r}=P_{\mathrm{loss}}$. Using the standard resonant response of a cavity mode, the generated power can be written as
\begin{equation}
P_{\mathrm{gen},r}
=
\frac{Q_{L,r}}{4\,\omega_r U_r}\,
|\Upsilon_{rp}|^2,
\label{eq:Prcav_general}
\end{equation}
where $Q_{L,r}$ is the loaded quality factor of the readout mode. Then, substituting \eqref{eq:Ipr_sq} into \eqref{eq:Prcav_general}, it yields
\begin{equation}
P_{\mathrm{gen},r}
=
\frac{1}{4}\,
\frac{Q_{L,r}}{\omega_r U_r}\,
g_{a\gamma\gamma}^2\,
\frac{\varepsilon_0}{\mu_0}\,
a_0^2 \, \omega_a^2
\left|
\int_V \vec{E}_r^*\cdot \vec{B}_p\,dV
\right|^2.
\label{eq:Prcav_final}
\end{equation}

Only a fraction of this internally generated power is extracted through the readout port. Denoting by $\beta_r$ the coupling coefficient of the readout mode to the readout port, and assuming that any residual coupling of the pump port to the readout mode and from the pump mode to the readout port is negligible, the extracted power in the readout port is given by
\begin{equation}
P_{\mathrm{out},r}
=
\frac{\beta_r}{1+\beta_r}\,
P_{\mathrm{gen},r}.
\label{eq:Prout_general}
\end{equation}
We now express the result in terms of the incident pump power $P_{\mathrm{inc},p}$. Let $\beta_p$ be the coupling coefficient of the pump port to the pump mode. On resonance, the accepted pump power is
\begin{equation} 
P_{\mathrm{acc},p}
=
\frac{4\beta_p}{(1+\beta_p)^2}\,
P_{\mathrm{inc},p}
\label{eq:Ppacc}
\end{equation}
and the electromagnetic energy stored in the pump mode at resonance is then
\begin{equation}
U_p=
\frac{Q_{0,p}}{\omega_p}\,
P_{\mathrm{acc},p}
=
\frac{Q_{L,p}}{\omega_p}\,
\frac{4\beta_p}{1+\beta_p}\,
P_{\mathrm{inc},p}.
\label{eq:Up_final}
\end{equation}
This energy can also be written as
\begin{equation}
U_p=
\frac{1}{2 \, \mu_0}\int_V \, ||\vec B_p||^2\,dV.
\label{eq:Up_B}
\end{equation}

To isolate the purely geometric part of the coupling, we define the dimensionless pump-readout overlap factor as
\begin{equation}
C_{rp}=\frac{\varepsilon_0}{\mu_0}\frac{
\left|
\int_V \vec E_r^* \cdot \vec B_p\,dV
\right|^2
}{
4\,U_r U_p
}= \frac{\left|
\int_V \vec E_r^*\cdot \vec B_p\,dV
\right|^2}{\int_V||\vec E_r||^2\,dV \int_V ||\vec B_p||^2\,dV}.
\label{eq:Cpr_def}
\end{equation}

This quantity is the natural analogue, in the up-conversion method, of the form factor used in conventional haloscope searches: it measures the overlap between the magnetic field of the pump mode and the electric field of the readout mode over the cavity volume. We now use Eq.~\eqref{eq:Cpr_def} and the definition of the axion field amplitude in order to express Eq.~\eqref{eq:Prcav_final} in terms of the dark matter density and the overlap factor
\begin{equation}
P_{\mathrm{gen},r}
= \frac{2}{\omega_r}
\,
g_{a\gamma\gamma}^2\,\rho_a\,\hbar \,c^3\,Q_{L,r}\,
U_p\,C_{rp},
\label{eq:Pgen_Cpr}
\end{equation}
and, after substitution into Eq.~\eqref{eq:Prout_general} together with Eq.~\eqref{eq:Up_final}, the extracted power at the readout port becomes
\begin{equation}
P_{\mathrm{out},r}
=
2\frac{\beta_r}{1+\beta_r}\,
\frac{4\beta_p}{1+\beta_p}\,
\frac{Q_{L,r} Q_{L,p}}{\omega_r\omega_p}\,
g_{a\gamma\gamma}^2\,\rho_a\,\hbar \,c^3\,
C_{rp}\,
P_{\mathrm{inc},p}.
\label{eq:Prout_final_1}
\end{equation}
For compactness, we define
\begin{equation}
\kappa_{rp}
\equiv
\frac{\beta_r}{1+\beta_r}\,
\frac{\beta_p}{1+\beta_p},
\label{eq:kappapr}
\end{equation}
and since $P_{\mathrm{out},r}$ denotes the extracted power at the resonance peak, the frequency dependence of the detected signal can therefore be described by a Lorentzian profile normalized to unity at resonance, whose spectral bandwidth is given by $Q_{L,r}$ as
\begin{equation} \label{eq:lorentzian}
    f_L(\omega) = \frac{1}{1+\left( 2\,Q_{L,r}\left(\frac{\omega-\omega_r}{\omega_r}\right)\right)^2}
\end{equation}
and therefore, 
\begin{equation}
P_{\mathrm{out},r}(\omega)
=
8\,\kappa_{rp}\,
\frac{Q_{L,r} Q_{L,p}}{\omega_r\omega_p}\,
g_{a\gamma\gamma}^2\,\rho_a\,c^3\,\hbar\,
C_{rp}\,
P_{\mathrm{inc},p}\,f_L(\omega).
\label{eq:Prout_final_2}
\end{equation}
%
%
From the expression of the extracted power, an interesting point worth highlighting is the apparent absence of an explicit cavity volume ($V$) scaling in heterodyne experiments. This is not a contradiction with the usual Sikivie scaling, but rather a consequence of expressing the signal in terms of a fixed circulating pump power. Under this assumption, and for a fixed pump frequency and comparable normalized mode profiles, the pump magnetic field amplitude scales as $V^{-1/2}$. Since the axion-induced signal power scales as $B_p^2 V$, this decrease in field amplitude exactly compensates the explicit volume factor. A direct volume dependence is recovered whenever the experimentally relevant constraint is instead a maximum sustainable pump field rather than a fixed pump power. Nevertheless, this does not imply that the cavity volume becomes irrelevant for the design. Even if the signal power is formally volume-independent at a fixed circulating power, a larger cavity can achieve the same signal level with lower field amplitudes, reducing the probability of electron discharge effects to occur, such as multipactor effect \cite{Hatch1958,Vaughan1988,Woode1989,Kishek1998,Kishek1998,ECSS2003,Perez2009,GonzalezIglesias2016,Semenov2018Multipactor,Vague2018,GonzalezIglesias2024} and RF Breakdown phenomena \cite{Kilpatrick1957,Germain1968,Bane2005,Grudiev2009,Wuensch2017,MartinezReviriego2023}. It also applies for the magnetic field amplitude and the possibility of reaching the critical value when SRF cavities are used. In addition, another dependence on the cavity volume arises in the range of axion masses to be scanned, since the modal resonance frequencies depend on the cavity dimensions (in classical cavities). As a result, the role of volume in heterodyne detection experiments is shifted: rather than entering as an explicit signal-enhancement factor, it becomes an important parameter through its impact on  the cryostat size, target axion masses and achievable pump fields. Moreover, another inherent limitation of the heterodyne detection comes from the relationship $P_{\mathrm{out},r} \propto \frac{1}{\omega_r\,\omega_p}$, which implies a significant reduction in the detected power when operating at high frequencies. This limitation has a direct impact on the cavity volume, since lower volumes result in higher resonance frequencies. Therefore, a compromise must be reached between cavity volume $-$constrained by the cryostat available space$-$ and mode frequencies, since the detected power benefits from operating at low frequencies. \\

An important advantage worth discussing of this method is that, since it does not rely on the application of a strong external magnetostatic field, it becomes possible to exploit SRF cavities with much higher quality factors than those achievable with conventional haloscopes. The concept of utilizing SRF cavities for the detection of dark matter axions was initially introduced in \cite{Sikivie_SRF}, and subsequent experiments have expanded upon this concept \cite{Superconducting_upconversion, SRF_Borogad, SRF_Janish}. This can improve the achievable sensitivity, since the Signal-to-Noise ratio ($S/N$) in resonant searches benefits directly from large cavity quality factors. In this regard, because superconducting materials reduce Joule losses, these cavities can be kept cold with a lower cryogenic heat load than a conventional (copper) cavity. In this sense, as said before, a limitation of the maximum pump signal power is set by the superconductor critical magnetic field. Type-II superconductors are particularly well-suited for use in high field SRF cavities due to their high critical field ($H_c$) values. These values correspond to the field value at which vortices arise, leading to increased dissipation. Some examples include Nb with $H_c \simeq 0.2$ T and $\textrm{Nb}_3\textrm{Sn}$ with $H_c \simeq 0.41$ T \cite{NbSn_paper}.  \\

Nevertheless, the critical magnetic field does not represent the only limitation to the achievable quality factor in a practical SRF cavity. The surface resistance of niobium can depend on the RF field amplitude \cite{Romanenko2017,Ciovati2006}. In particular, the high-field $Q$ drop in bulk niobium typically becomes relevant for peak surface magnetic fields of order $B_{\rm pk}\sim 90$--$100$ mT \cite{Ciovati2006}. In the operating regime considered here, RF field vortex penetration and high-field $Q$ degradation are therefore not expected to constitute the dominant limitation, provided that the peak surface field remains well below this regime. Other mechanisms, however, may reduce the quality factor even at low RF fields. Ambient magnetic flux trapped during the superconducting transition produces vortex-related residual dissipation \cite{Gonnella2016}, while fabrication tolerances, gaps, mechanical joints, and resistive interfaces can substantially degrade the quality factor of multipart or mechanically tunable SRF cavities \cite{Sharping2020}.

Other technical limitations are also discussed in \cite{Lasenby_2020}. Beyond thermal noise, heterodyne experiments may also be affected by mechanical vibrations, field-emitted charges, and pump power leakage into the readout chain. Mechanical deformations of the cavity walls can parametrically couple the pump and signal modes inducing spurious power in the readout channel, while also producing slow variations of the mode frequencies and their frequency split \cite{Heterodyne_2021}. However, for the example presented here in section IV, this effect is not expected to be a leading limitation: vibration-induced noise is estimated to exceed thermal electromagnetic noise only below axion frequencies of order $\sim 300$ kHz, well below our target range of $1$--$35$ MHz. Nevertheless, microphonic detuning may still shift the ultranarrow SRF resonance and should therefore be characterized and mitigated in the final cryogenic setup. Field-emitted electrons may also constitute a background source if they deposit energy into the signal mode. In favorable cavity-mode configurations, this contribution can be strongly reduced by keeping the electric field at the cavity surface sufficiently small, although the importance of this effect remains geometry dependent. A further important systematic is pump drive leakage, since phase noise from the pump source, imperfect mode orthogonality, fabrication tolerances, or parasitic couplings can transfer a fraction of the pump power directly into the readout channel. Given the extremely small axion-induced signal, this leakage must be carefully monitored experimentally and suppressed, as we propose in Sec. \ref{sec:Results}. 
\subsection{Analysis for non-monochromatic sources}\label{sub:Qequiv}
\begin{figure}[ht]
    \centering
    \includegraphics[width=0.85\linewidth]{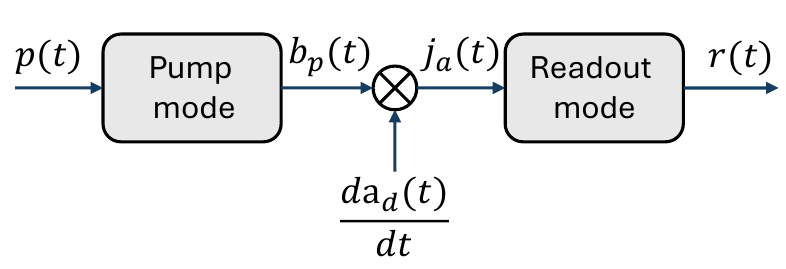}
    \caption{Signal processing from the pump mode excitation to the readout mode detection in up-conversion. The signal $p(t)$ is the pump signal injected by the RF signal generator, $b_p(t)$ the pump mode signal resulted from the interaction between $p(t)$ and the pump resonant mode, $d \mathtt{a}_d/d t$ the derivative of the dispersive axion field, $j_a(t)$ the axion current density and $r(t)$ the signal resulting from the interaction between $j_a(t)$ and the readout resonant mode.}
    \label{fig:scheme_upconversion}
\end{figure}

Since the axion and the pump signal are not strictly monochromatic, their finite spectral width must be taken into account when evaluating the detected power, as it limits the extent to which the cavity quality factors can enhance the signal. These quality factors also have an impact on the bandwidth of the detected signal, which must be considered when choosing the detector bandwidth. Thus, the detected spectrum is shaped by three successive processes, as Fig. \ref{fig:scheme_upconversion} shows. We develop here this process only for the up-converted signal for simplicity and without loss of generality. 

\begin{figure*}[ht]
    \centering
    \includegraphics[width=0.98\linewidth]{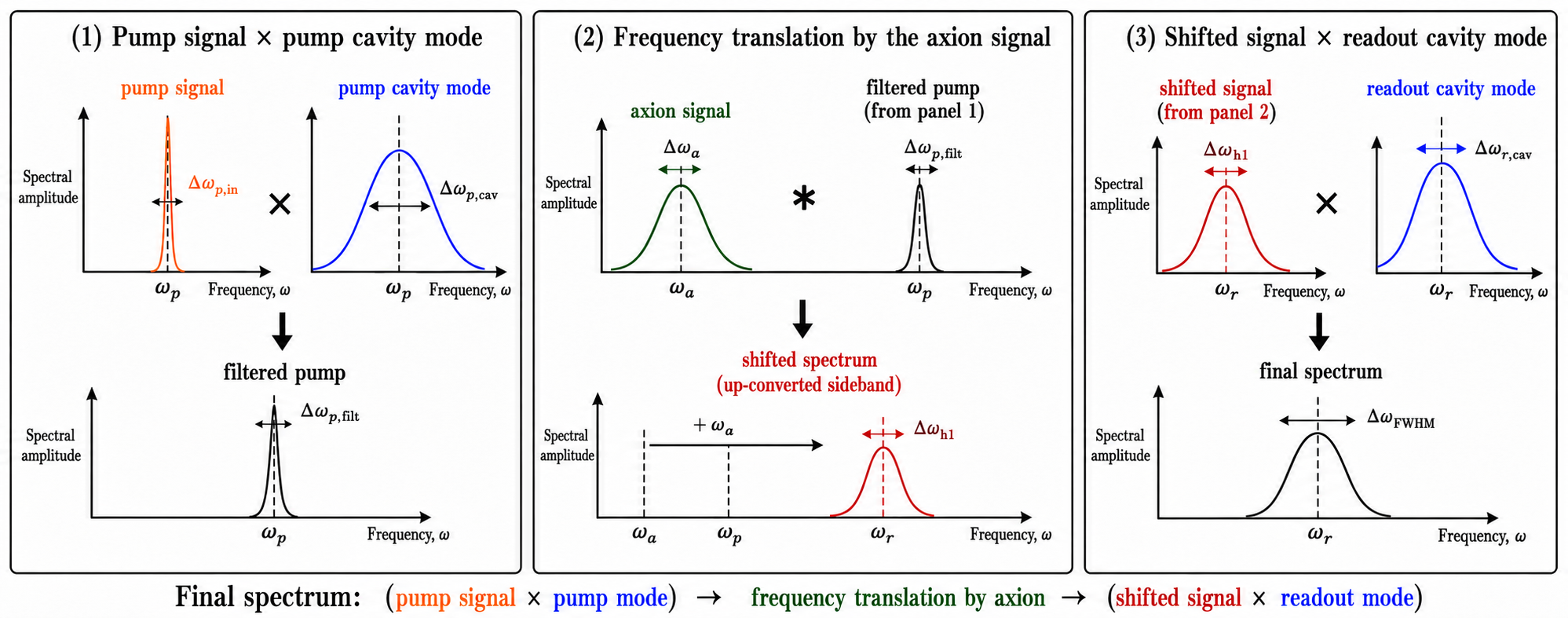}
    \caption{Schematic representation of the up-conversion process in the frequency domain. Left, \textit{Stage 1}: the injected pump signal (orange), with linewidth $\Delta\omega_{p,\mathrm{in}}$, is assumed to be narrower or in the order of the pump mode linewidth $\Delta\omega_{p,\mathrm{cav}}$ (blue), depending on the cavity material. This signal is filtered by the resonant response of the pump cavity, resulting in a filtered pump spectrum with linewidth $\Delta\omega_{p,\mathrm{filt}}$. Middle, \textit{Stage~2}: mixing with the axion signal centered at $\omega_a<\omega_p$ translates the pump-filtered spectrum to the up-converted sideband centered at $\omega_r=\omega_p+\omega_a$. Under the narrow-pump approximation considered here, the linewidth of the mixed signal is $\Delta\omega_{\mathrm{h1}}=\omega_r/Q_{h,1}\simeq\omega_a/Q_a$. Right, \textit{Stage 3}: the shifted spectrum is subsequently filtered by the readout-mode response, so that the final detected spectrum is given by the product of the up-converted signal and the Lorentzian response of the readout mode, with final linewidth $\Delta\omega_{\mathrm{FWHM}}=\omega_r/Q_{h,2}$. The figure is schematic and not to scale.
}
    \label{fig:signal_interactions}
\end{figure*}

According to  Fig. \ref{fig:scheme_upconversion}, the pump signal injected by the generator, centered at $\omega_p$, is spectrally filtered by the resonant response of the pump cavity mode, so that the effective spectrum is given by the product of both Lorentzian profiles (\textit{Stage 1}). Next, this filtered pump signal mixes with the axion signal, which has a finite bandwidth, producing a frequency translation toward the readout band centered at $\omega_r = \omega_p + \omega_a$, with a linewidth modified by the spectral contribution of the axion (\textit{Stage 2}). Finally, the shifted signal is filtered by the resonant readout mode of the cavity, so that the detected spectrum is determined by the product of the up-converted sideband and the Lorentzian response of the readout mode (\textit{Stage 3}). A schematic view of this process describing the transformations of the signals spectrum is shown in Fig.~\ref{fig:signal_interactions}. These three stages are now explained in detail:  \\
%

\textit{Stage 1}: The pump mode is excited by a signal composed of a very pure tone centered at $\omega_p$ and a residual phase noise contribution. For the calculation of the axion signal, the pump excitation can be treated as monochromatic, particularly in the copper implementation, where the source linewidth is much narrower than the pump mode linewidth. In practice, the relevant quantity for frequency locking is the instantaneous detuning between the generator frequency and the pump mode resonance, $\delta f_p(t)=f_{p, \rm in}(t)-f_{p,\rm cav}(t)$. To keep the pump within the $-3$ dB bandwidth of the cavity resonance, the feedback system must satisfy $|\delta f_p|<\Delta f_{p,\rm cav}/2=f_p/(2Q_{L,p})$. Therefore, particularly for very high-$Q$ implementations, an active phase-locked loop (PLL) or a resonance tracking RF feedback scheme is required to compensate generator drift, cavity pump frequency variation due to the tuning procedure (which affects both readout and pump modes) and other possible fluctuations induced by mechanical vibrations, microphonics, or thermal drifts, ensuring that the instantaneous detuning remains within the $-3$ dB bandwidth of the pump resonance \cite{Bernard2000PACO}. 

\noindent
The pump mode response can be evaluated at $\omega_p$ as
\begin{equation} \label{eq:bpt1}
    b_p(t) = \mathtt{B}_{p0}\,\mathrm{cos}(\omega_p t)\, ,
\end{equation}
where $\mathtt{B}_{p0}$ is a positive real number that represents the magnetic amplitude of the pump mode. Thus, the time dependence remains harmonic at $\omega_p$, while the resonator response only fixes the stored energy through $Q_{L,p}$ and the coupling coefficient. However, for the estimation of pump leakage into the readout channel (Sec. \ref{sec:electrical_response}), the finite spectral purity of the RF source, i.e, the phase noise, must be included in this analysis. \\
%
%
%
%
%
%
%

\textit{Stage 2}: The axion field in Eq. \eqref{axion_field_timedomain} was assumed to be monochromatic. However, the axion has a finite bandwidth and therefore an associated quality factor $Q_a \sim 10^6$. Again, given that $\vec{k}_a \approx \vec{0}$ and $\varphi = 0$, the axion field can be expressed by approximating to a Lorentzian spectrum as
\begin{equation} \label{eq:adt}
    \mathtt{a}_d(t) = a_0\,e^{-\frac{\omega_a}{2Q_a}t}\mathrm{cos}(\omega_at)
\end{equation}
where the sub-index $d$ indicates that this axion field is dispersive. Therefore,
\begin{equation}
    \frac{d  \mathtt{a}_d}{d t} = -a_0\,e^{-\frac{\omega_a}{2Q_a}t}\left( \frac{\omega_a}{2Q_a}\mathrm{cos}(\omega_a t)+\omega_a\mathrm{sin}(\omega_a t)\right),
\end{equation}
which, for $Q_a>>1$, can be approximated as
\begin{equation}
    \frac{d  \mathtt{a}_d}{d t} \approx -a_0\,e^{-\frac{\omega_a}{2Q_a}t}\omega_a\mathrm{sin}(\omega_a t).
\end{equation}
Then, the resulting axion current density is given by
\begin{equation}
\begin{aligned}
\vec{\mathtt{J}}_{a}(\vec r,t) ={}&
\,- g_{a\gamma\gamma}\sqrt{\frac{\varepsilon_0}{\mu_0}}\,a_0\,
\omega_a\,\mathtt{B}_{p0}\,
e^{-\left(\frac{\omega_a}{2Q_a}\right)t}
\\[3pt]
& \sin(\omega_a t)\cos(\omega_p t) \, \Re{(\vec{B}_p (\vec{r}))}. 
\end{aligned}
\label{eq:Ja_full} 
\end{equation}
The time-dependent part of $\vec{\mathtt{J}}_{a}(\vec r,t)$ is
\begin{align}\label{eq:ja_mix}
j_{a}(t) 
& = e^{-\left(\frac{\omega_a}{2Q_a}\right)t}
\,\omega_a \sin(\omega_a t)\cos(\omega_p t)= \nonumber
\\
& = \frac{\omega_a}{2}\,
e^{-\left(\frac{\omega_a}{2Q_a}\right)t} \nonumber  \\
&  \left(
\sin\big((\omega_a+\omega_p)t)
+
\sin\big((\omega_a-\omega_p)t)
\right).
\end{align}
The exponential decay of this expression can be written in terms of the readout angular frequency and the quality factor $Q_{h,1}$ at that frequency as follows
\begin{equation}
    e^{-\left(\frac{\omega_a}{2Q_a}\right)t}=e^{-\frac{\omega_r}{2Q_{h,1}}t},
\end{equation}
and so $Q_{h,1}$ is given by
\begin{equation} \label{eq:Qh1}
    Q_{h,1}=\frac{\omega_r Q_a}{\omega_a}.
\end{equation}
\textit{Stage 3}: The response of the readout mode to the up-converted axion signal is obtained as the linear convolution of the input temporal profile and the impulse response of the readout mode, which may be written as
\begin{equation} \label{eq:rt}
    r(t) = \left(e^{-\left( \frac{\omega_r}{2Q_{h,1}}\right)t}\mathrm{sin}(\omega_r t)\right) \, * \, \left(e^{-\left(\frac{\omega_r}{2Q_{L,r}}\right)t}\mathrm{sin}(\omega_r t)\right),
\end{equation}
where the symbol $*$ denotes the linear convolution operator defined as,
\begin{eqnarray}
f(t) * g(t) \, \equiv \, \int_{-\infty}^{+\infty} f(\tau) \, g(t-\tau)\,d\tau
\end{eqnarray}
for the functions $f(t)$ and $g(t)$. \\

In the frequency domain, the filtered up-converted spectrum is proportional to the product of the Lorentzian profile associated with the finite axion linewidth and the Lorentzian response of the readout mode,

\begin{equation} \label{eq:cov_product}
\frac{1}{1+\left[2Q_{h,1}\left(\frac{\omega-\omega_r}{\omega_r}\right)\right]^2}
\frac{1}{1+\left[2Q_{L,r}\left(\frac{\omega-\omega_r}{\omega_r}\right)\right]^2}.
\end{equation}

We define the effective detection quality factor \(Q_{h,2}\) from the full width at half maximum of this product, in the same way as for a single Lorentzian. That is, \(Q_{h,2}\) is defined by

\begin{equation}
    \Delta \omega_{\rm FWHM}=\frac{\omega_r}{Q_{h,2}} .
\end{equation}

Let

\[x \equiv \frac{\omega-\omega_r}{\omega_r}.
\]

The half-maximum condition for the product in Eq. \eqref{eq:cov_product} is

\[\left(1+4Q_{h,1}^2 x^2\right)\left(1+4Q_{L,r}^2 x^2\right)=2 .\]

Writing \(u=x^2\), this condition becomes

\[16Q_{h,1}^2Q_{L,r}^2u^2+4\left(Q_{h,1}^2+Q_{L,r}^2\right)u-1=0 .\]

The physically relevant positive solution is

\[u =\frac{-\left(Q_{h,1}^2+Q_{L,r}^2\right)+\sqrt{\left(Q_{h,1}^2+Q_{L,r}^2\right)^2+4Q_{h,1}^2Q_{L,r}^2}}{8Q_{h,1}^2Q_{L,r}^2}.\]

Since the half-width at half maximum is

\[\frac{\Delta \omega_{\rm FWHM}}{2\omega_r}=\sqrt{u}=\frac{1}{2Q_{h,2}},\]
the exact effective quality factor associated with the bandwidth of the product of the two Lorentzian responses is

\begin{equation} \label{eq:Qh2}
Q_{h,2}=\left[\frac{Q_{h,1}^2+Q_{L,r}^2+\sqrt{\left(Q_{h,1}^2+Q_{L,r}^2\right)^2+4Q_{h,1}^2Q_{L,r}^2}}{2}\right]^{1/2}.
\end{equation}

This expression can be approximated by the simpler 

\begin{equation} \label{eq:Qh2_simplif}
Q_{h,2}\approx\sqrt{Q_{h,1}^2+Q_{L,r}^2}. 
\end{equation}
The accuracy of this approximation as a function of the ratio $Q_{h,1}/Q_{L,r}$ is quantitatively assessed in Appendix~\ref{app:Qh2_accuracy}. If \(Q_{L,r}\gg Q_{h,1}\), the readout resonance is much narrower than the up-converted axion spectrum and \(Q_{h,2}\simeq Q_{L,r}\). Conversely, if \(Q_{h,1}\gg Q_{L,r}\), the axion-induced sideband is narrower than the readout resonance and \(Q_{h,2}\simeq Q_{h,1}\). Therefore, the bandwidth of the detected spectrum is controlled by the narrower of the two spectral responses. In the parameter regime relevant for the SRF implementation considered here, typically \(Q_{L,r}\gg Q_{h,1}\), and hence \(Q_{h,2}\simeq Q_{L,r}\), which sets the detector bandwidth in the receiver.\\

%
%
%
%
Finally, for a finite-linewidth axion field, the relevant power quantity is not the monochromatic peak power, but the signal power integrated over the up-converted axion spectrum after filtering by the readout resonance. We therefore define \(P_d\) as this effective detected signal power within the analysis bandwidth. The finite spectral overlap between the axion-induced sideband, characterized by \(Q_{h,1}\), and the readout resonance, characterized by \(Q_{L,r}\), replaces the readout power enhancement factor \(Q_{L,r}\) in \eqref{eq:Prout_final_1} by

\begin{equation} \label{eq:QeffP}
Q_{\rm eff}^{(P)} =\frac{Q_{h,1}Q_{L,r}}{Q_{h,1}+Q_{L,r}} .
\end{equation} \\

Accordingly, the detected signal power becomes

\begin{equation} \label{eq:Pd1}
P_d =8\kappa_{rp}\frac{Q_{L,p}}{\omega_r\omega_p}\frac{Q_{h,1}Q_{L,r}}{Q_{h,1}+Q_{L,r}}g_{a\gamma\gamma}^2\rho_a c^3\hbar C_{rp}P_{\mathrm{inc},p}.
\end{equation}

An important consequence of $Q_{h,1}<<Q_{L,r}$ is that it sets the detected power as

\begin{equation}  \label{eq:Pdd}
    P_d
\approx
8\,\kappa_{rp}\,
\frac{Q_{L,p}\, Q_{h,1}}{\omega_r\omega_p}\,
g_{a\gamma\gamma}^2\,\rho_a\,c^3\,\hbar\,
C_{rp}\,P_{\mathrm{inc},p}.
\end{equation}

Comparing \eqref{eq:Pdd} with \eqref{eq:Prout_final_1} it is clear that taking into account the dispersive nature of the axion and the use of SRF cavities produce a reduction in the detected power by a factor $(Q_{L,r}\,\omega_a) / (Q_a\,\omega_r)$. For realistic cases this can be a reduction factor around  $10 - 1000$, depending on the distance between $\omega_a$ and $\omega_r$.

\subsection{Figure of merit in the heterodyne detection}
In haloscope axion detection experiments, the scanning rate $df_a/dt$ is commonly used as a figure of merit. Maximizing this quantity requires determining the optimal values of the port couplings $(\beta_p, \beta_r)$ which depend on other experimental parameters, such as the relevant quality factors of the detection system and the system noise temperature $T_\mathrm{sys}=T_\mathrm{cav}+T_\mathrm{add}$, with $T_\mathrm{cav}$ the cavity and $T_\mathrm{add}$ the readout chain noise temperature, respectively. The scanned axion frequency per frequency step can be expressed as
\begin{equation} \label{eq:faQa}
    \Delta f_a=\frac{\omega_r}{2\pi\,Q_{h,1}}= \frac{\omega_a}{2\pi\,Q_a},
\end{equation}
where \eqref{eq:Qh1} has been used. The Dicke's radiometer equation in a microwave detector system is \cite{pozar}
\begin{equation}
    \frac{S}{N}=\frac{P_d}{P_N}=\frac{P_d}{k_B\,T_{sys}}\sqrt{\frac{\Delta t}{\Delta f}}
\end{equation}
where $k_B$ is the Boltzmann constant, the noise power is given by $P_N = k_B \, T_{sys} \, \Delta f$, and $\Delta f$  is the detection frequency bandwidth. The integration time can be obtained from it as
\begin{equation}
    \Delta t = \Delta f \left(\frac{\frac{S}{N}k_B T_{sys}}{P_d}\right)^2
\end{equation}
and the scanning rate is therefore
%
%
%
\begin{equation}
    \frac{df_a}{dt}=\frac{\Delta f_a}{\Delta t}=\frac{\omega_a}{2\pi\,Q_a\,\Delta f}\left(\frac{P_d}{\frac{S}{N}k_B T_{sys}}\right)^2.
\end{equation}
Taking $\Delta f = \frac{\omega_r}{2\pi\,Q_{h,2}}$ as the effective detection bandwidth matched to the filtered signal spectrum, and considering the effects of mismatching in the noise \cite{CAPP_revisiting}, we obtain
\begin{equation}
    \frac{df_a}{dt}=\frac{P_d^2\,\omega_a\,Q_{h,2}}{\left(\frac{S}{N}k_B \left(\left( \frac{4\beta_r}{(1+\beta_r)^2}\right)T_\mathrm{cav}+T_\mathrm{add}\right)\right)^2\omega_r\,Q_a}.
\end{equation}
And finally, the scanning rate dependence on the quality factors in the heterodyne detection is given by
\begin{equation} \label{eq:dma_final}
\frac{d f_a}{d t}
\propto
\frac{
\left(
\frac{4\beta_p\beta_r}{(1+\beta_p)(1+\beta_r)}\,
\frac{Q_{L,p}Q_{h,1}Q_{L,r}}{Q_{h,1}+Q_{L,r}}
\right)^2
\omega_a Q_{h,2}
}{
\left(
\frac{4\beta_r}{(1+\beta_r)^2}\,T_{\mathrm{cav}}+T_{\mathrm{add}}
\right)^2
\omega_r Q_a
}.
\end{equation}

\section{Application of the BI-RME 3D method for the heterodyne detection}\label{sec:birmecircuit}

\subsection{Formulation}

The analytical model derived above is mainly aimed at describing the resonant axion-induced signal and its scaling with the relevant cavity parameters. However, a realistic assessment of the heterodyne technique also requires an accurate description of the spectral response away from resonance, as well as of the direct pump leakage into the readout port, which can mask the axion-induced contribution. This motivates the use of the BI-RME 3D (Boundary Integral--Resonant Mode Expansion) method, which provides the full frequency-domain response of the cavity and therefore enables a more complete and accurate evaluation of the signal spectra.

Originally developed in the 1980s and 1990s at the Università degli Studi di Pavia (Italy), this technique can be used to determine the modal electromagnetic field distributions inside an arbitrarily shaped cavity coupled to a given number of ports. In particular, it provides access to the full complex response of the system, including both amplitude and phase information for the relevant input and output signals over a broad frequency range, rather than only at the resonant frequencies, as is the case for conventional figure of merit estimates of the detected power. The BI-RME 3D method has been presented and extended in a number of works \cite{conciauro,birme3d_3Dcavities,birme3d_3Dcavities_ports,birme3d_angel_multipactor,birme3d_angel_posts,birme3d_fermin,birme3d_jordi_MTT,birme3d_jordi_MWCL,birme3d_overview,birme3d_pavia_MTT}; here, we restrict ourselves to those aspects that are directly relevant to the analysis of up- and down-conversion axion detection. It is worth emphasizing that BI-RME 3D is formulated in the frequency-domain.

The BI-RME 3D method was first applied to the electromagnetic analysis of haloscopes immersed in a static magnetic field \cite{bi-rme3d_axions}, corresponding to the conventional inverse Primakoff effect. In that formulation, the starting point is the expression for the magnetic field inside a lossless microwave resonator excited by volumetric and surface electric and magnetic current densities, denoted by $\vec{J}$, $\vec{J}_S$, $\vec{M}$, $\vec{M}_S$, respectively,
\begin{eqnarray}  \label{birme3D_H_1}
	\vec{H}(\vec{r}) & = & \frac{1}{\mathrm{i} k \eta} \nabla \int_V g^m
	(\vec{r},\vec{r}^{\, \prime}) \, \nabla' \cdot \vec{M}(\vec{r}^{\,
		\prime}) \, dV' \, + \nonumber \\
  & &  \frac{1}{\mathrm{i} k \eta} \nabla \int_S g^m
	(\vec{r},\vec{r}^{\, \prime}) \, \nabla_S' \cdot \vec{M}_s(\vec{r}^{\,
		\prime}) \, dS' \, - \nonumber \\        
       & & \frac{\mathrm{i} k}{\eta} \int_V \mathbf{\vec{G}^{\rm F}}
	(\vec{r},\vec{r}^{\, \prime}) \cdot \vec{M} (\vec{r}^{\, \prime})
	\, dV '\,  - \nonumber \\
    & & \frac{\mathrm{i} k}{\eta} \int_S \mathbf{\vec{G}^{\rm F}}
	(\vec{r},\vec{r}^{\, \prime}) \cdot \vec{M}_S (\vec{r}^{\, \prime})
	\, dS'  \, + \nonumber \\ 
     & & \int_V \nabla \times \mathbf{\vec{G}^{\rm A}}
	(\vec{r},\vec{r}^{\, \prime}) \cdot \vec{J}(\vec{r}^{\, \prime})   
	\, dV'  \, + \nonumber \\
    & &  \int_S \nabla \times \mathbf{\vec{G}^{\rm A}}
	(\vec{r},\vec{r}^{\, \prime}) \cdot \vec{J}_S(\vec{r}^{\, \prime})   
	\, dS'. 
\end{eqnarray}
where $\eta = \sqrt{\mu_0/\varepsilon_0} \approx 377 \, \Omega$ is the vacuum impedance; $\vec{n}$ is the inward unitary normal vector to the cavity surface ($||\vec{n}|| = 1$); $\nabla_s$ is the surface divergence operator \cite{chentotai}; $g^e(\vec{r},\vec{r}^{\, \prime})$ and $g^m(\vec{r},\vec{r}^{\, \prime})$ are the electric and magnetic static scalar potentials Green's functions of the cavity under Coulomb gauge, respectively; and $\mathbf{\vec{G}^{\rm A}}
(\vec{r},\vec{r}^{\, \prime})$ and $\mathbf{\vec{G}^{\rm F}} (\vec{r},\vec{r}^{\, \prime})$ are the electric and magnetic dyadic potentials Green's functions of the cavity under Coulomb gauge, respectively. In the present BI-RME 3D formulation, fictitious magnetic currents are introduced only to account for the coupling between the cavity and the input/output waveguide ports. Accordingly, the volumetric magnetic current density is set to zero, $\vec{M} = \vec{0}$, which simplifies Eq.~\eqref{birme3D_H_1},
\begin{eqnarray}  \label{birme3D_H_2}
	\vec{H}(\vec{r}) & = &  \frac{1}{\mathrm{i} k \eta} \nabla \int_S g^m
	(\vec{r},\vec{r}^{\, \prime}) \, \nabla_S' \cdot \vec{M}_s(\vec{r}^{\,
		\prime}) \, dS' \, - \nonumber \\        
    & & \frac{\mathrm{i} k}{\eta} \int_S \mathbf{\vec{G}^{\rm F}}
	(\vec{r},\vec{r}^{\, \prime}) \cdot \vec{M}_S (\vec{r}^{\, \prime})
	\, dS'  \, + \nonumber \\ 
     & &  \int_V \nabla \times \mathbf{\vec{G}^{\rm A}}
	(\vec{r},\vec{r}^{\, \prime}) \cdot \vec{J}(\vec{r}^{\, \prime})   
	\, dV'  \, + \nonumber \\
    & &  \int_S \nabla \times \mathbf{\vec{G}^{\rm A}}
	(\vec{r},\vec{r}^{\, \prime}) \cdot \vec{J}_S(\vec{r}^{\, \prime})   
	\, dS'. 
\end{eqnarray}
The last integral in Eq.~\eqref{birme3D_H_2} requires separate treatment because of the singular behavior of the term $\nabla \times \mathbf{\vec{G}^{\rm A}}(\vec{r},\vec{r}^{\, \prime})$. Its treatment is detailed in Appendix~\ref{Appendix_singular_surface_integral}. The resulting expression is
\begin{eqnarray}  \label{birme3D_H_3}
	\vec{H}(\vec{r}) & = &  \frac{1}{\mathrm{i} k \eta} \nabla \int_S g^m
	(\vec{r},\vec{r}^{\, \prime}) \, \nabla_S' \cdot \vec{M}_s(\vec{r}^{\,
		\prime}) \, dS' \, - \nonumber \\        
    & & \frac{\mathrm{i} k}{\eta} \int_S \mathbf{\vec{G}^{\rm F}}
	(\vec{r},\vec{r}^{\, \prime}) \cdot \vec{M}_S (\vec{r}^{\, \prime})
	\, dS'  \, + \nonumber \\ 
     & &  \int_V \nabla \times \mathbf{\vec{G}^{\rm A}}
	(\vec{r},\vec{r}^{\, \prime}) \cdot \vec{J}(\vec{r}^{\, \prime})   
	\, dV'  \, + \nonumber \\
    & &  \frac{-1}{2} \left(\vec{n} \times \vec{J}_S(\vec{r})\right).
\end{eqnarray}
The last term in this equation must be retained only when the magnetic field $\vec{H}(\vec{r})$, evaluated through Eq.~\eqref{birme3D_H_3}, is computed at a point lying on the inner cavity surface. In fact, we want to emphasize that this singular surface integral is not requested in the analysis of haloscopes embedded in an external static magnetic field because the discontinuity of the magnetostatic field occurs on the inner wall of the magnet (for example, the inner wall of a solenoid), and therefore the RF electromagnetic energy radiated cannot penetrate into the haloscope, since it has been designed and constructed as a Faraday cage. Therefore, the corresponding surface integral does not appear in the standard BI-RME 3D formulation for haloscopes with magnetostatic fields, as reported in \cite{bi-rme3d_axions}.

\begin{figure}[ht]
    \centering
    \includegraphics[width=0.7\linewidth]{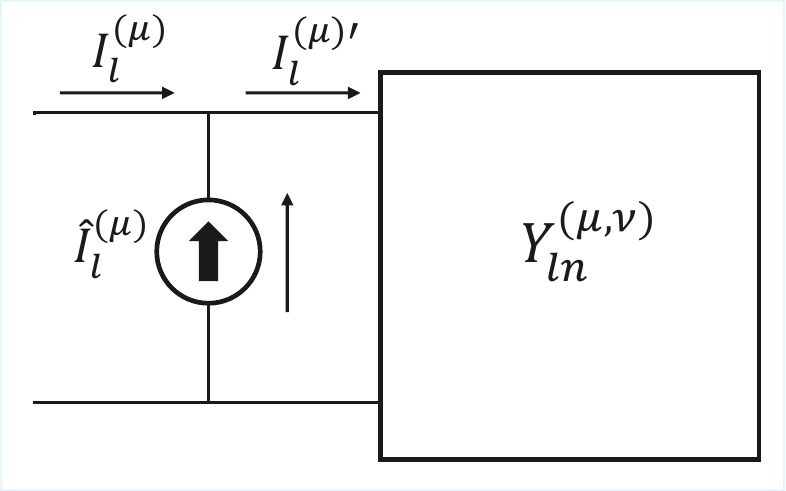}
    \caption{Multimode equivalent network representation of a cavity resonator excited by an axion field. We have represented the port $\mu$ and the waveguide mode $l$.}
    \label{fig:birme_port}
\end{figure}

Following the derivation of Sec.~2.1 of Ref.~\cite{bi-rme3d_axions}, but now starting from Eq.~\eqref{birme3D_H_3}, Eq.~(7) of Ref.~\cite{bi-rme3d_axions} can be rewritten as
\begin{eqnarray}  \label{currents_extraction}
	I_l^{(\mu)} \,  =  \, I_l^{(\mu) \, \prime}  \,   -  \,  \hat{I}_l^{(\mu)}
\end{eqnarray}
where a new set of RF axion current sources $\hat{I}_l^{(\mu)}$ are defined as,
\begin{eqnarray}   
\label{driven_source_birme3d}
	\hat{I}_l^{(\mu)} \, & \equiv & \,  
	- \sum_{m=1}^M  \, F_{ml}^{(\mu)} \, \frac{k_m}{k_m^2 - k^2}  \,  \int_ V \vec{E}_m(\vec{r}^{\, \prime}) \cdot \vec{J_a}(\vec{r}^{\, \prime}) \, dV' \, + \nonumber \\ & + & \, \frac{1}{2} \, \int_ {S(\mu)}  
   \vec{e}_l^{(\mu)}(\vec{r}^{\, \prime}) \cdot \vec{J}_{a_S}(\vec{r}^{\, \prime}) \, dS' \, \, ; \, \, \mu \in \left\{1, 2...P \right\}. \nonumber \\
\end{eqnarray}	
This set of RF axion current sources is associated with the $P$ waveguide ports, as shown in Fig.~\ref{fig:birme_port} for one port. In this expression, we have used the relation between the normalized electric ($\vec{e}_l^{(\mu)}$) and magnetic ($\vec{h}_{l}^{(\mu)}$) modal vector functions of the waveguide ports.

\noindent
Obviously, the axion-induced surface current density $\vec{J}_{a_S}$ has also been included, together with the following surface-coupling integral, which accounts for the coupling between each excited cavity resonance and the corresponding waveguide port, as described in Ref.~\cite{bi-rme3d_axions},
\begin{equation}
    F_{ml}^{(\mu)} \equiv \int_{S(\mu)}\vec{H}_{m}\left(\vec{r}\right)\cdot \vec{h}_{l}^{(\mu)}\left(\vec{r}\right)\ dS,
\end{equation}
being $\vec{H}_{m}$ the normalized magnetic field of the $m$th resonant mode of the cavity. 

\subsection{Heterodyne detection analysis of a resonant two-ports haloscope with the BI-RME 3D method}

\subsubsection{Complete network including all sources}

In order to apply this formulation to the heterodyne conversion technique in a conventional two-ports haloscope, we consider a cavity with two ports ($P=2$), both connected to standard coaxial transmission lines supporting only the fundamental TEM mode ($l=1$). The pump mode is injected through the port $\mu=1$, while the readout signal is extracted from the port $\mu=2$. Under these assumptions, the normalized electric and magnetic vector mode functions of the TEM mode at both ports, expressed in cylindrical coordinates $(r,\varphi,z)$ of a reference frame centered in the coaxial line, are given by
\begin{equation}
    \vec{e}_{\mathrm{TEM}} \, = \, \frac{1}{\sqrt{2 \, \pi \, \mathrm{ln}(R_{\mathrm{out}}/R_{\mathrm{in}})}} \, \frac{1}{r} \, \hat{r}\,;
\end{equation}
\begin{equation}
     \vec{h}_{\mathrm{TEM}} \, = \, \frac{1}{\sqrt{2 \, \pi \, \mathrm{ln}(R_{\mathrm{out}}/R_{\mathrm{in}})}} \, \frac{1}{r} \, \hat{\varphi}
\end{equation}
where $R_{\mathrm{out}}$ and $R_{\mathrm{in}}$ denote the outer and inner radii of the coaxial line, respectively, and $\varepsilon_r$ is the relative electric permittivity of the dielectric medium between the conductors. The modal impedance of the TEM coaxial mode is given by $Z_w=\sqrt{\mu_0/(\varepsilon_0\varepsilon_r)}$, and it should be distinguished from the characteristic impedance of the coaxial line \footnote{The characteristic impedance of the coaxial line is given by $Z_0 = \frac{Z_w}{2\pi}\ln\!\left(\frac{R_{\mathrm{out}}}{R_{\mathrm{in}}}\right)$}. The corresponding modal admittance is $Y_w=1/Z_w$. The orthonormalization condition adopted here is
\begin{equation}
	\int_{CS} \, \vec{e}_{\mathrm{TEM}} \cdot  (\vec{h}_{\mathrm{TEM}} \times \hat{z}) \, dS \, = \, 1 \, ,
\end{equation}
where $CS$ is the coaxial transmission line surface cross-section. 

\begin{figure}[t]
    \centering
        \includegraphics[width=0.95\linewidth]{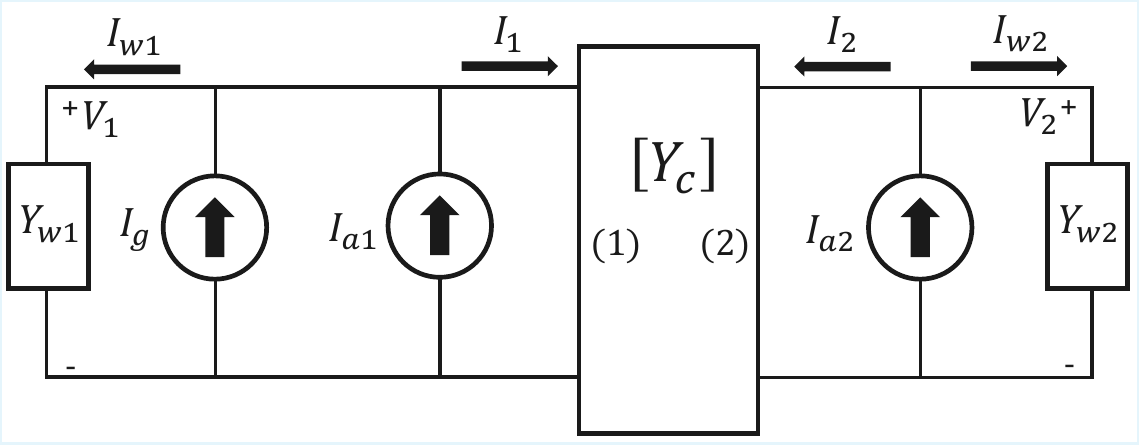}
        
    \caption{BI-RME 3D network representation: The pump mode is generated with the current source $I_g$ which is injected to the cavity through the port $(1)$, whereas the readout port corresponds with the port $(2)$; the axion excitation is represented by the current sources $I_{a_1}$ and $I_{a_2}$ coupled to both ports $(1)$ and $(2)$, respectively; the ports are loaded with the modal admittances $Y_{w_1}$ and $Y_{w_2}$.}
    \label{fig:circuit_BIRME3D_general}
\end{figure}

Fig.~\ref{fig:circuit_BIRME3D_general} shows the two-port network representation of the BI-RME 3D formulation for the present case. The driven current source $I_g$, which excites the pump mode, is connected to port $\mu=1$, whereas the readout signal is extracted at port $\mu=2$. Since both ports are coupled to identical coaxial lines, their modal admittances are the same, $Y_{w_1}=Y_{w_2}=Y_w$. For the numerical evaluation of the current sources defined in Eq.~\eqref{driven_source_birme3d}, we use the up-conversion expression given in Eq.~\eqref{eq:J_res_2}. 

To solve the circuit of Fig.~\ref{fig:circuit_BIRME3D_general} we apply the Kirchhoff laws, obtaining the relationships between voltages and currents,
\begin{equation}
    I_g + I_{a_1} = I_{w_1} + I_1 ; I_{a_2} = I_{w_2} + I_2 ; 
    V_1 =  I_{w_1}/Y_w ;  V_2 = I_{w_2}/Y_w. \nonumber
\end{equation}
The relationship between currents and voltages in the cavity ports is given by $\overline{I}=\overline{\overline{Y}}_c \cdot \overline{V}$, where $\overline{I}$ and $\overline{V}$ denote the current and voltage vectors, respectively, and $\overline{\overline{Y}}_c$ is the cavity admittance matrix,
\begin{eqnarray}
\overline{I} \equiv \begin{pmatrix}
  I_1  \\
  I_2  
\end{pmatrix}  \, ; \, 
\overline{V} \equiv \begin{pmatrix}
  V_1  \\
  V_2  
\end{pmatrix}  \, ; \, 
\overline{\overline{Y_c}} \equiv 
\begin{pmatrix}
  Y_{11} & Y_{12} \\
  Y_{21} & Y_{22}
\end{pmatrix}.  \nonumber
\end{eqnarray}
In the present implementation, $\overline{\overline{Y}}_c$ is evaluated numerically from the scattering parameters, following the standard microwave network theory described in Ref.~\cite{pozar}. Finally we obtain 
\begin{align}
     V_1 \, & = \, \frac{I_{a_2}}{Y_{21}} - \frac{Y_{w_2} + Y_{22}}{Y_{12} Y_{21} - \gamma} \left(I_g + I_{a_1} - \frac{Y_{w_1} + Y_{11}}{Y_{21}} I_{a_2}\right)  \\
      V_2 \,&=\, \frac{Y_{21}}{Y_{12} Y_{21} - \gamma} \left( I_g + I_{a_1} - \frac{Y_{w_1} + Y_{11}}{Y_{21}} I_{a_2} \right)
\end{align}
where $\gamma$ is defined as $\gamma \equiv (Y_{w_1} + Y_{11})\, (Y_{w_2} + Y_{22})$. 

The average power supplied by the pump source, $P_g$, the average power dissipated in the cavity, $P_c$, the average power injected by the axion in the haloscope, $P_a$, and the average powers delivered to the pump and readout ports, $P_{w_1}$ and $P_{w_2}$, respectively, are then given by
\begin{align}
P_g
&= \frac{1}{2}\,\Re\!\left(V_1 I_g^*\right)  
\label{eq:Pg}
\\[4pt]
P_c
&= \frac{1}{2}\,\Re\!\left(V_1 I_1^*\right)
 + \frac{1}{2}\,\Re\!\left(V_2 I_2^*\right)  
\label{eq:Pc}
\\[4pt]
P_a
&= \frac{1}{2}\,\Re\!\left(V_1 I_{a_1}^*\right)
 + \frac{1}{2}\,\Re\!\left(V_2 I_{a_2}^*\right)  
\label{eq:Pa}
\\[4pt]
P_{w_1}
&= \frac{1}{2}\,\Re\!\left(V_1 I_{w_1}^*\right)
 = \frac{1}{2}\,|V_1|^2\,\Re\!\left(Y_{w_1}^*\right)  
\label{eq:Pw1}
\\[4pt]
P_{w_2}
&= \frac{1}{2}\,\Re\!\left(V_2 I_{w_2}^*\right)
 = \frac{1}{2}\,|V_2|^2\,\Re\!\left(Y_{w_2}^*\right)  
\label{eq:Pw2}
\end{align}

As expected, Energy Conservation Principle is satisfied,
\begin{equation}
    P_g + P_a = P_c + P_{w_1} + P_{w_2},
\end{equation}
which states that the energy entering into the network by both the pump current source ($P_g$) and the axion field coupled to the cavity ($P_a$), is dissipated by Joule effect within the cavity ($P_c$), and carried out to the waveguide ports ($ P_{w_1}$ and $ P_{w_2}$).

\subsubsection{Pump network}

\begin{figure}[t]
    \centering
        \includegraphics[width=0.95\linewidth]{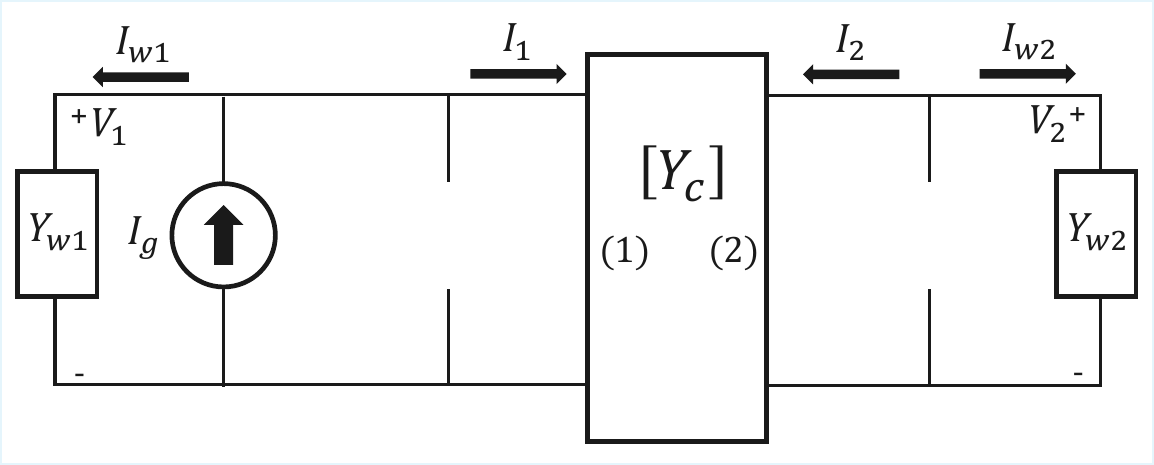}
        
    \caption{BI-RME 3D network representation of the pump circuit, which has only the current source $I_g$; the axion currents $I_{a_1}$ and $I_{a_2}$ have been removed because they are open circuited.}
    \label{fig:circuit_BIRME3D_pump}
\end{figure}

For the present application, it is important to establish the relationship between the pump current source $I_g$ and the power generated by the current source $P_g \equiv P_{\mathrm{inc},p}$.
Therefore, now we have to analyze the network of Fig.~\ref{fig:circuit_BIRME3D_pump}, where the axion current sources have been removed because they are open-circuited. In this case, the Kirchhoff laws are again applied, obtaining
\begin{equation}
    I_g = I_{w_1} + I_1 ; I_{w_2} = - I_2 ; 
    V_1 =  I_{w_1}/Y_w ;  V_2 = I_{w_2}/Y_w. \nonumber
\end{equation}

For this case the solution of the circuit is simpler than in the previous case, obtaining
\begin{align}
     V_1 \, = \, \frac{Y_{w_2} + Y_{22}}{\gamma - Y_{12} \, Y_{21}} \, I_g \, \, ; \, \,  V_2 \, = \, \frac{Y_{21}}{Y_{12} \, Y_{21} \, - \, \gamma} \, I_g.  
\end{align}
The power generated by the pump current source can be written as
\begin{equation}
    P_g \, = \, \frac{1}{2} \Re{(V_1 I_g^*)} \, = \, |I_g|^2 \, \chi(\omega_p) , 
\end{equation}
where the function $\chi(\omega)$ is defined as
\begin{eqnarray}
\chi(\omega) \, \equiv \, \frac{1}{2} \, \Re{\left( \frac{Y_{w_2} + Y_{22}}{\gamma - Y_{12} \, Y_{21}}\right)},  
\end{eqnarray}
which allows to calculate the amplitude of the pump current source as a function of the power generated by the microwave source employed in the experimental setup. In this case we will assume that the phase of $I_g$ is zero: $I_g = |I_g| \, e^{\mathrm{i} \, 0^{\circ}}$.

%
%
%
%
%
%
%

%
\section{Application of the heterodyne detection to a conventional resonant haloscope}\label{sec:Results}
Different low-frequency resonant haloscopes have been proposed in the technical literature, and some of them operated so far: ADMX \cite{ADMX_LF}, FLASH \cite{FLASH} and RADES-BabyIAXO \cite{RADES_BabyIAXO}. Among these, we choose the RADES-BabyIAXO largest cavity as the reference design for this study because it provides a particularly suitable platform for heterodyne axion detection since its quasi-cylindrical geometry gives rise to an interesting modal structure, including higher-order and degenerate modes that are especially favorable for pump-readout combinations. In addition, its tuning system based on rotating plates allows to vary the resonant frequency of the modes, and therefore, to search at different axion frequencies with a relative tuning of $\sim18$\%. These features make the RADES-BabyIAXO cavity a well-motivated and realistic design to explore the application of the heterodyne detection.

\subsection{The RADES-BabyIAXO cavity} \label{sub:cavity}

\begin{figure}[ht]
    \centering
    \includegraphics[width=0.99\linewidth]{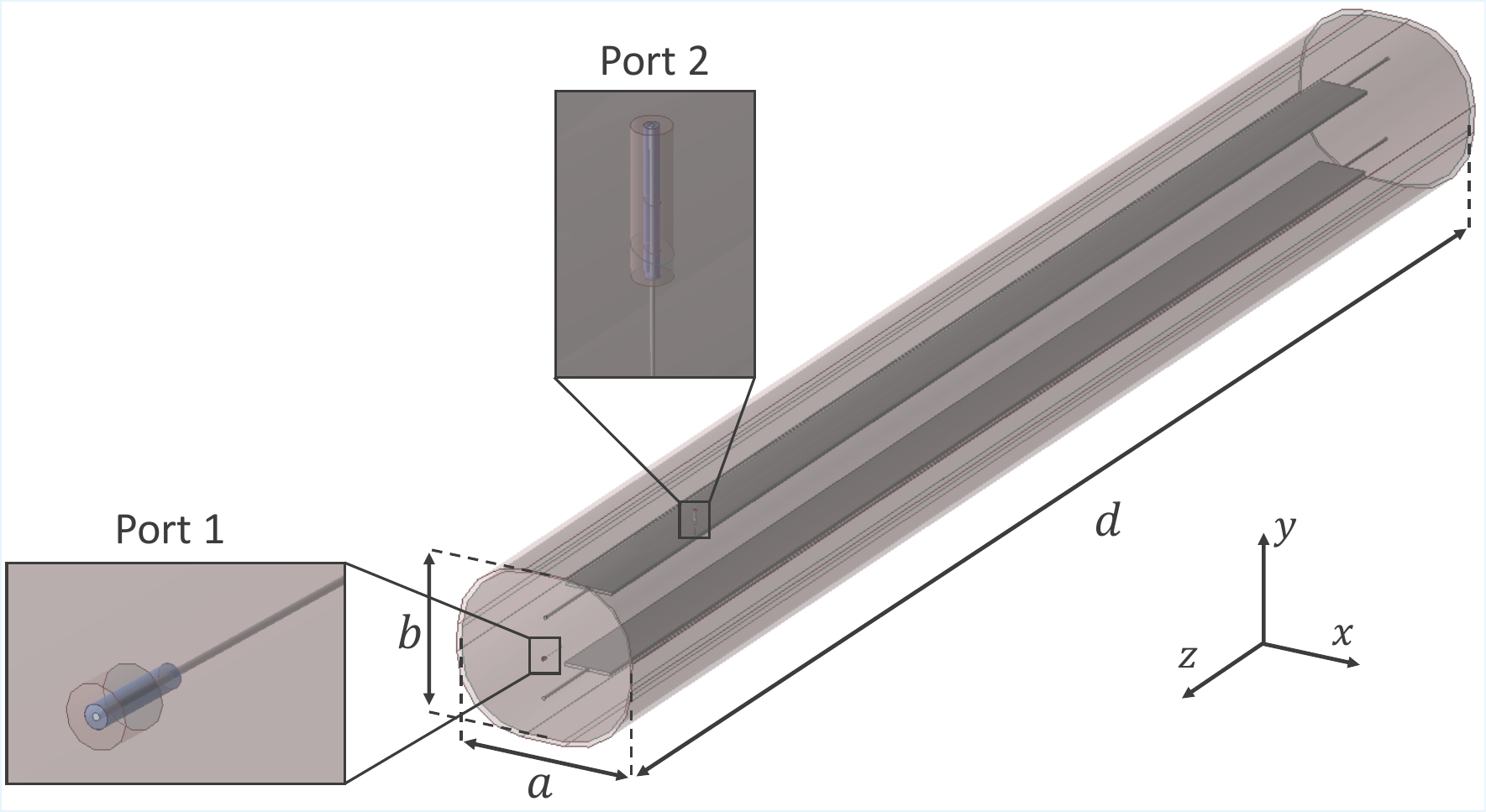}
    \caption{3D schematic of the quasi-cylindrical RADES-BabyIAXO cavity. The cavity chosen for the study has dimensions $a=56$ cm, $b=50.4$ cm, and $d=500$ cm. The aspect ratio $b/a=0.9$ is chosen to lift the degeneracy between the $\mathrm{qTE}_{111-\mathrm{y}}$ and $\mathrm{qTE}_{111-\mathrm{x}}$ modes. The two tuning plates are shown in dark grey. For the heterodyne operation the cavity is equipped with two coaxial probes (external radius $R_{out} = 2.05$ mm, internal radius $R_{in} = 0.635$ mm, and relative electrical permittivity $\varepsilon_r=2$) located at the cavity walls. Port 1 is placed at the center of the end cap to couple to $\mathrm{qTM}_{0mp}$ modes, while port 2 is positioned on the upper part of the cavity, at a distance of 62.5 cm from the end cap, to couple to $\mathrm{qTE}_{0mp}$ modes. The orthogonal orientation of the ports helps to reduce leakage from the pump mode to the readout mode to some extent.}
    \label{fig:cavity}
\end{figure}

The RADES-BabyIAXO cavities were proposed for DM axion searches within the magnet of the future BabyIAXO helioscope \cite{BabyIAXO2021}. Their design, described in \cite{RADES_BabyIAXO}, was developed to optimize the space within the dipole magnet bore. In order to prevent the $\mathrm{TE}_{111}$ mode degeneracy and to provide space for cables and instrumentation on the sides, the devices were designed with a quasi-cylindrical shape. The cavities also incorporate a tuning system based on rotating metallic plates along the longitudinal axis which can turn from $\alpha=0^\circ$ to $90^\circ$. This mechanism allows the frequency shifting of different electromagnetic modes by varying their field distribution. The up-conversion study was performed for the largest of the four cavities designed for BabyIAXO, the geometry and dimensions of which are shown in Fig. \ref{fig:cavity}.

The initial idea was to implement the heterodyne detection method using the two $\mathrm{qTE}_{111}$ \footnote{Since the cavity geometry is not canonical and incorporates tuning mechanisms, the resonant modes are renamed as quasi-TE and quasi-TM, abbreviated as qTE and qTM, respectively.} polarizations (-x and -y) of the RADES-BabyIAXO cavity, with one mode acting as the pump mode and the other as the readout mode. This option was particularly appealing because the tuning plates affect both resonances in opposite ways: as the plates are rotated, one mode shifts to lower frequency while the other moves upward, naturally allowing the mode frequency separation to be scanned over a significant range. This behavior has been explicitly reported for the two $\mathrm{qTE}_{111}$ modes of the RADES-BabyIAXO cavity: under plate tuning from $\alpha=0^\circ$ to $90^\circ$, the $\mathrm{qTE}_{111-\mathrm{y}}$ mode shifts downward from $296.2$ to $252.8$ MHz, whereas the $\mathrm{qTE}_{111-\mathrm{x}}$ mode shifts upward from $303.4$ to $333.2$ MHz, thereby yielding a potentially accessible axion-frequency range of $7.3$--$80.4$ MHz \cite{BabyIAXO_HFGW}. However, despite this attractive tunability, the value of the overlap factor was found to be $C_{rp}\sim10^{-3}$. The reason is that both modes are TE-like, so the magnetic field of each mode is not favorably aligned with the electric field of the other one, leading to a low mode overlap and therefore to a suppressed axion-induced coupling. For this reason, although the $\mathrm{qTE}_{111}$ pair provides a natural tunable frequency splitting in this cavity, it is not an efficient choice for heterodyne axion detection. As suggested by \cite{Upconversion_australianos}, mixed TM–TE mode pairs are expected to provide more favorable conditions for heterodyne detection, with mode overlap factors significantly larger.

\subsection{Coupling factor study}\label{sub:couplingfactor}

\begin{figure*}[ht]
    \centering
    \includegraphics[width=1\linewidth]{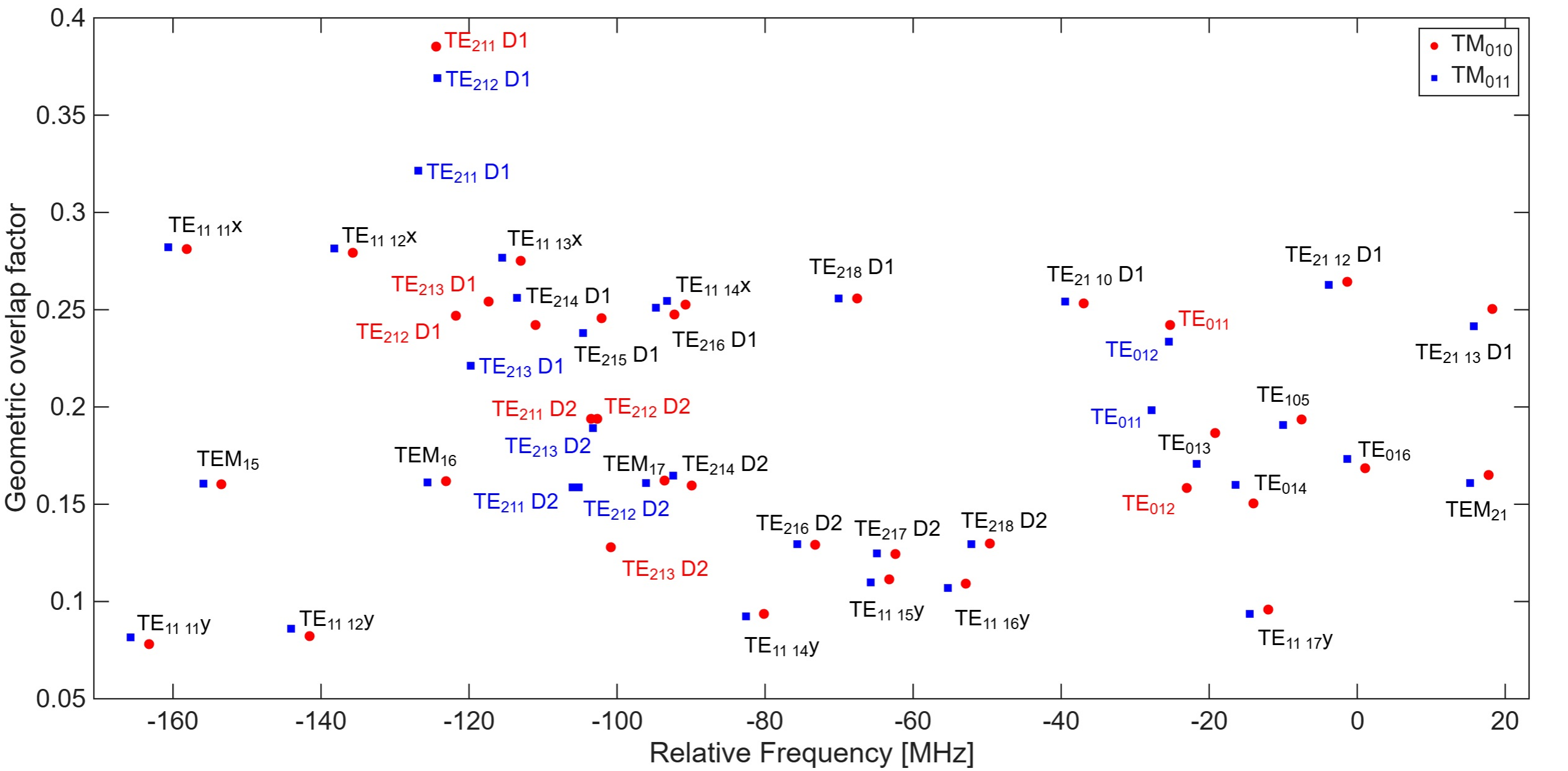}
    \caption{Geometric overlap factor results for $\alpha = 0^{\circ}$ plates angle for all possible combinations between 39 different TE modes as readout modes with $\mathrm{TM_{010}}$ and $\mathrm{TM_{011}}$ as pump modes. The number of modes shown correspond to all the modes that can be found in this frequency range. All modes are quasi-TM/TE, although the ‘q’ has been omitted in this plot to alleviate the notation. Four $\mathrm{TEM}$ modes have also been considered for this calculation, $\mathrm{TEM_{15}}$, $\mathrm{TEM_{16}}$, $\mathrm{TEM_{17}}$ and $\mathrm{TEM_{21}}$. The sub-indexes `D1' and `D2' denote degenerate modes with orthogonal polarizations. Black labels correspond to pairs of red (TE-$\mathrm{TM_{010}}$) and blue (TE-$\mathrm{TM_{011}}$) data, close enough in the plot to be labeled simultaneously. Data labeled with its corresponding color (e.g. $\mathrm{TE_{211}\,D1}$ in red at the top of the plot) implies that the complementary color is not close enough in the chart, being labeled individually. The readout modes selected from now on for the analysis are $\mathrm{TE_{011}}$, $\mathrm{TE_{012}}$ and $\mathrm{TE_{014}}$. Negative frequencies make reference to down-conversion, and positive frequencies to up-conversion.}
    \label{fig:general_plot}
\end{figure*}

The study of mode coupling begins taking as pump the $\mathrm{qTM}_{010}$ and $\mathrm{qTM}_{011}$ modes, since these are the lowest order TM modes. For this cavity, their resonant frequencies are 604.86 and 607.34 MHz, respectively (with a plate angle of $\alpha=0^{\circ}$). According to the results in \cite{Upconversion_australianos}, one might expect that a mode combination likely to yield a good value of the overlap factor is the $\mathrm{TE}_{011}$-$\mathrm{TM}_{010}$. However, due to the complex geometry of the RADES-BabyIAXO cavity, a study of the values of $C_{rp}$ for different mode combinations has been carried out. The study was extended to all modes lying up to $180~\mathrm{MHz}$ below the pump mode because the resulting mode separations correspond to axion frequencies for which conventional resonant cavities would become prohibitively large. However, the computational cost limited the search for modes above the pump frequency to a range of approximately \(20\,\mathrm{MHz}\). To this end, the cavity was simulated for different plate angles from $\alpha=0^\circ$ to $\alpha = 90^\circ$ with a step of $5^\circ$ using the CST Studio Suite \cite{CST} (Eigensolver), and the fields were extracted to determine the value of $C_{rp}$. Fig. \ref{fig:general_plot} shows the computed overlap factors for all modes at a distance of $\lesssim 200$ MHz from the pump modes only for the case $\alpha=0^\circ$.

\begin{figure*}[ht]
\centering

\subfloat[]{
    \includegraphics[width=0.32\textwidth]{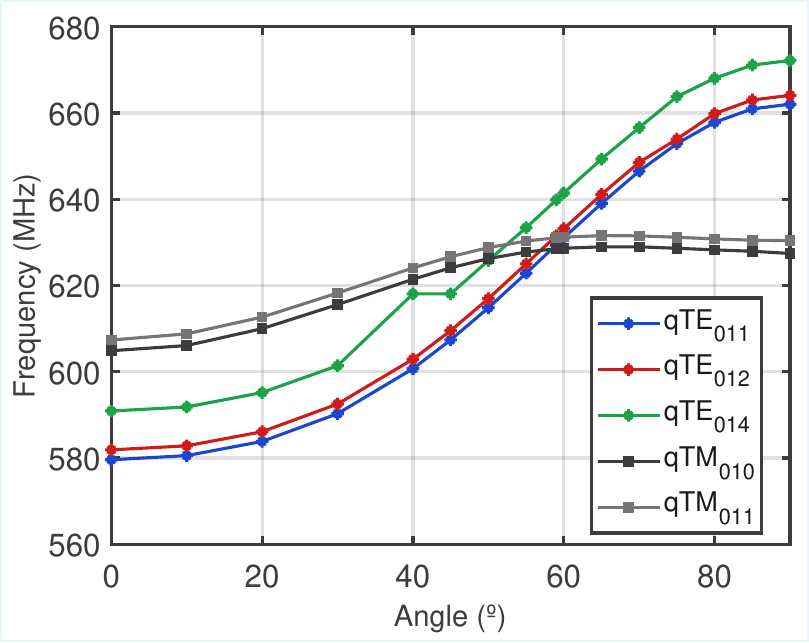}
}
\hfill
\subfloat[]{
    \includegraphics[width=0.32\textwidth]{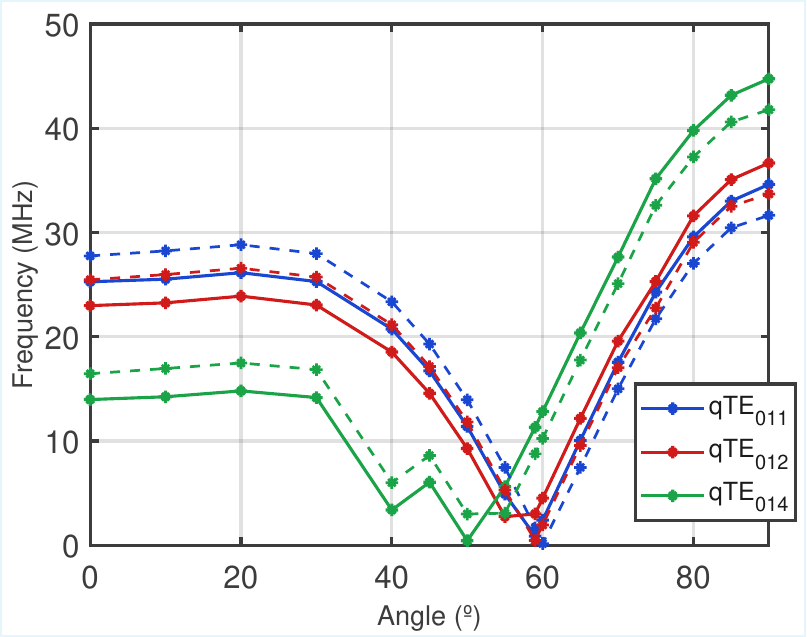}
}
\hfill
\subfloat[]{
    \includegraphics[width=0.32\textwidth]{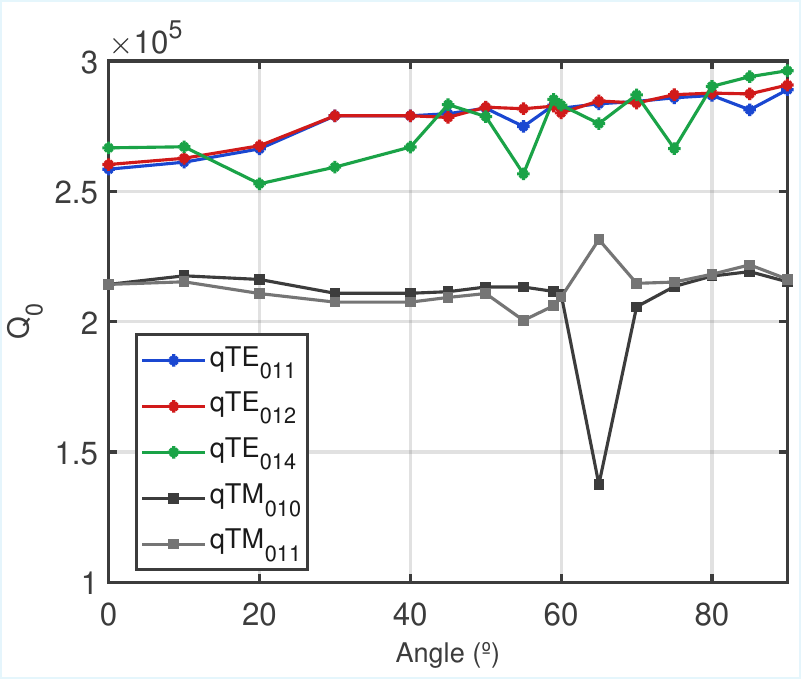}
}

\caption{(a) Variation of the resonant frequencies of the selected qTE ($f_r = \omega_r/(2 \pi)$) and qTM ($f_p = \omega_p/(2 \pi)$) modes, (b) Frequency difference ($f_a=|f_p-f_r|$) between the qTE modes and the qTM modes; solid curves denote the separation from $\mathrm{qTM_{010}}$ whereas dashed curves denote the separation from $\mathrm{qTM_{011}}$, and (c) Unloaded quality factor of all analyzed modes as a function of the tuning plates angle $\alpha$.}
\label{fig:freqs_and_Q0}
\end{figure*}

As can be seen, a large number of possible pump--readout mode pairs can be formed in this frequency range, owing to the presence of numerous higher-order and degenerate cavity modes. Therefore, in order to facilitate mode identification and the analysis, and imposing the selection criterion $C_{rp}~\geq~0.15$, we restrict our attention to the following mode combinations: $\mathrm{qTE}_{011},\,\mathrm{qTE}_{012},\,\mathrm{qTE}_{014}-\mathrm{qTM}_{010}$, and $\mathrm{qTE}_{011},\,\mathrm{qTE}_{012},\,\mathrm{qTE}_{014}-\mathrm{qTM}_{011}$. Following a series of parametric simulations with a step of $5^\circ$ of plates rotation, the absolute and relative tuning range, as well as the unloaded quality factor (cryogenic OFHC copper with electric conductivity $\sigma = 10^9$ S/m), were extracted across the entire range of plates rotation for the selected modes (see Fig. \ref{fig:freqs_and_Q0}). 

From Fig. \ref{fig:freqs_and_Q0} (a), it is evident that the tuning system has a significant impact on all selected modes, with a more pronounced effect on qTE modes compared to qTM modes. It has also been noted that there is a mode crossing point at the region $50^\circ<\alpha < 60^\circ$ for all modes, which is of particular interest since this allows to scan for very low axion masses, always limited by the minimum power established to detect. Furthermore, Fig. \ref{fig:freqs_and_Q0} (b) demonstrates that there is a wider explorable range of axion masses between the angles $\alpha \approx  58^\circ$ and $\alpha = 90^\circ$. With regard to the quality factor, this varies only slightly for all modes except at a specific point for the $\mathrm{qTM_{010}}$ mode. This is due to a mode crossing close to that point. 

The next step is to calculate the overlap factors for the selected mode combinations across the entire tuning range. To do so, the electric fields of the qTE modes and the magnetic fields of the qTM modes were extracted from the simulations, and the definition of $C_{rp}$ in Eq.~\eqref{eq:Cpr_def} was applied. 

\begin{figure*}[ht]
\centering

\subfloat[]{
    \includegraphics[width=0.45\textwidth]{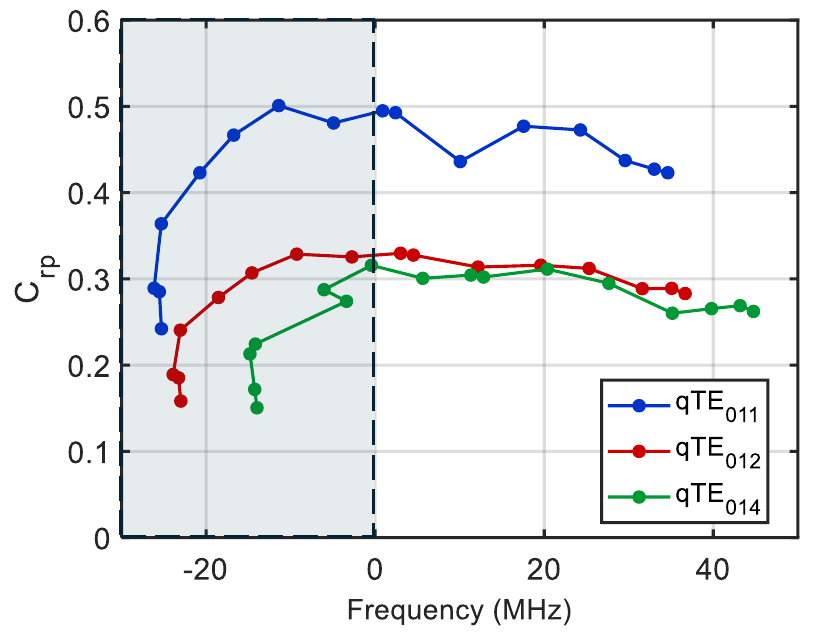} 
}
\hfill
\subfloat[]{
    \includegraphics[width=0.45\textwidth]{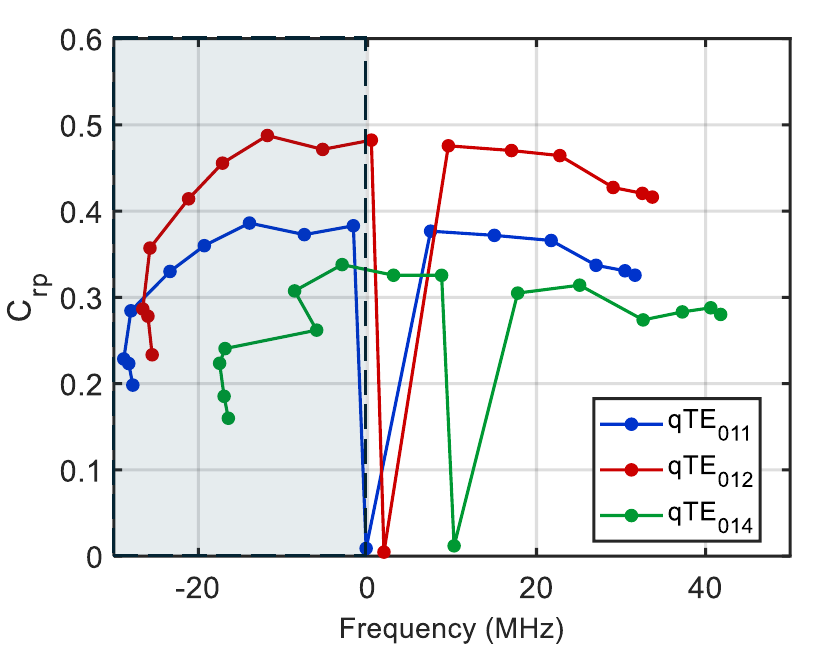}
}

\caption{Geometric overlap factors between the selected qTE modes with (a) $\mathrm{qTM}_{010}$ mode, and (b) $\mathrm{qTM}_{011}$ mode as a function of the relative frequency during tuning. While mode overlap occurs within the whole interval, the coupling strength varies significantly, with pronounced local suppressions at specific tuning points. In the following, we restrict the analysis to the configurations of positive frequencies denoted with a white background (up-conversion), where the coupling remains consistently large across the tuning range, discarding those with reduced coupling (negative frequencies correspond to down-conversion).}
\label{fig:form_factor}
\end{figure*}

As illustrated in Fig. \ref{fig:form_factor}, the values of $C_{rp}$ for the $\mathrm{qTE}-\mathrm{qTM_{010}}$ and $\mathrm{qTE}-\mathrm{qTM_{011}}$ combinations are indicated. When using the $\mathrm{qTM_{010}}$ mode as the pump mode, the readout mode that provides the highest overlap factors is $\mathrm{qTE_{011}}$. Similarly, when using the $\mathrm{qTM_{011}}$ mode as the pump mode, the $\mathrm{qTE_{012}}$ mode yields better values than the other ones. Therefore, these are two combinations that can be used as a pump-readout pair for the same range of axion masses. Furthermore, the values of $C_{rp}$ calculated in the range prior to the mode crossing (vertical black dashed line) are lower than those calculated in the range beyond the mode crossing for all cases, due to less deformation of the electric field at these plates rotation angles. An interesting feature is the ability to switch mode pairs to extend the scanning range slightly. For instance, using the pair $\mathrm{qTE_{011}}-\mathrm{qTM_{010}}$ permits scanning axion masses from a minimum set at 0.9 MHz up to 34.62 MHz. Then, the pair $\mathrm{qTE_{014}}-\mathrm{qTM_{010}}$ could be used for scanning, despite its lower value of the overlap factor, in order to cover an additional $10$ MHz range up to $44.75$ MHz by dedicating more integration time. Therefore, we choose the pair $\mathrm{qTE_{011}}-\mathrm{qTM_{010}}$ for the application of the BI-RME 3D method, corresponding to an axion frequency range of $f_a = \omega_a/(2 \pi) = 0.9-34.62$ MHz, and plates rotation angle range of $\alpha \approx 58^\circ$ to $\alpha=90^\circ$ \footnote{Therefore, after selecting these modes, the position of the readout port can be optimized to couple this pair of modes and further reduce the leakage of the pump mode.}. 


\subsection{Study of the electrical response: detected power and pump leakage} \label{sec:electrical_response}

In this section, the results from the power estimates at each port are shown. The BI-RME 3D method has been applied to two specific cases:
\begin{itemize}
    \item Case 1: corresponds to the minimum chosen separation between the pump mode and the readout mode ($\alpha \approx 58^\circ ,\ f_a=0.9$ MHz).
    
    \item Case 2: corresponds to the maximum possible separation ($\alpha = 90^\circ, \ f_a=34.62$ MHz).
\end{itemize}
At first glance, in terms of restrictions, Case 1 is the most stringent as the pump and readout modes are closer in frequency. Consequently, the interference from the pump power will, in principle, be more significant than in Case 2. 

First of all, the BI-RME 3D formulation described in the previous section is used to calculate the wide-band electrical response of the haloscope. We model the incident pump signal used in BI-RME 3D as the output of a RF signal generator including the phase noise contribution, in this case, taking the specifications for the R\&S SMA100B signal generator~\cite{RohdeSchwarz_SMA100B_2026}. The scattering parameters of the RADES-BabyIAXO cavity were first obtained from full-wave CST simulations assuming cryogenic copper walls with conductivity $\sigma_\textrm{Cu}=10^{9}$ S/m. However, a direct electromagnetic simulation of the cavity considering a superconducting material, in this case niobium (with $R_s = 1\,\mathrm{n\Omega}$ \footnote{The value $R_s=1\,\mathrm{n}\Omega$ should be understood as an optimistic SRF benchmark; actual values will depend on material treatment, RF field amplitude, temperature, trapped flux, and residual losses.}), becomes impractical since the resulting loaded quality factors are much larger, leading to extremely narrow resonances that require a prohibitively dense frequency sampling and very high numerical error in simulations. To overcome this limitation, the scattering parameters computed for copper were fitted using a rational function model in which each scattering parameter is represented as the sum of a smooth non-resonant polynomial background, and a set of resonant pole contributions associated with the cavity modes. This approach follows standard fitting procedures for high-Q microwave resonators \cite{Petersan1998Q,Gustavsen1999VF,Ramella2021HighQ}, in particular the formalism used in Eq.~(13) of \cite{Ramella2021HighQ}, where the polynomial term accounts for the slowly varying non-resonant response and the resonant terms that describe the modes contributions. Once the copper response was fitted, the superconducting response was extrapolated by rescaling the intrinsic quality factor of each mode according to the ratio between the cryogenic copper and niobium surface resistances, while keeping the external coupling coefficients fixed. The corresponding loaded quality factors were then recomputed and substituted back into the rational model. Nevertheless, this approach has certain limitations, such as not accounting for the small shifts in the resonant frequencies of the modes and changes in the non-resonant polynomial background arising from changes in the quality factor. Then, the admittance parameters needed in BI-RME 3D are obtained as a function of the scattering parameters \cite{pozar}. This procedure and the scattering parameters fitting for both cases can be consulted in Appendix \ref{Appendix_scattering_parameters}. The parameters of the modes are extracted from the simulations for the case of cryogenic copper. These are shown in Table \ref{tab:variables}.

\begin{table}[h]
\centering
\begin{tabular}{| c | c | c |}
\hline
  & \textbf{Case 1 (Cu)} & \textbf{Case 2 (Cu)}\\
\hline
$f_a$ (MHz) & 0.9 & 34.62 \\
\hline
$f_p$ (MHz) & 627.12 & 627.41 \\
\hline
$f_r$ (MHz) & 628.05 & 662.03 \\
\hline
$Q_{L,p}$ & $5.872\cdot10^4$ & $5.938\cdot10^4$ \\ 
\hline
$Q_{L,r}$ & $7.856\cdot10^4$ & $7.970\cdot10^4$ \\
\hline
$\beta_p$ & 0.59 & 0.55 \\
\hline
$\beta_r$ & 2.12 & 2.15 \\
\hline
\end{tabular}
\caption{Frequencies, loaded quality factors and port couplings for the pump and readout modes in the two studied cases. The values of the port couplings are not the optimal ones and should be recalculated to optimize the experiment performance.}
\label{tab:variables}
\end{table}

In the case of niobium, the mode frequencies and port couplings have been assumed to be the same as for the cryogenic copper. The loaded quality factors have been calculated by rescaling using equations \eqref{eq:q0_srf_scaling} and \eqref{eq:ql_srf} in Appendix \ref{Appendix_scattering_parameters}. The values are $Q_{L,p}=1.82\cdot10^{10}$ and $Q_{L,r} = 1.52\cdot10^{10}$ for Case 1, and $Q_{L,p} =1.87\cdot10^{10}$ and $Q_{L,r}=1.92\cdot10^{10}$ for Case 2. \\

The phase noise specification of the RF signal generator is given in dBc/Hz relative to the pump carrier. In the BI-RME 3D calculation, this spectral density is integrated over the effective detection bandwidth of the readout signal, $B_{\rm det}=f_r/Q_{h,2}$. For the cryogenic copper cases, the corresponding bandwidths are $1.18$ Hz and $1.25$ Hz for Cases 1 and 2, respectively, and $0.48$ Hz and $0.51$ Hz for the corresponding niobium cases. \\

For the same configurations, the $-3$ dB lock criterion introduced in Sec.~\ref{sub:Qequiv}, $|\delta f_p|<f_p/(2Q_{L,p})$, corresponds to maximum pump--cavity detunings of approximately $5.34$ kHz and $5.28$ kHz for the two copper cases, and $17.2$ mHz and $16.8$ mHz for the corresponding niobium cases. These values therefore set the frequency tracking requirement that the PLL or equivalent feedback system must satisfy. \\

\begin{figure*}[t]
    \centering

    \subfloat[]{%
        \includegraphics[width=0.45\textwidth]{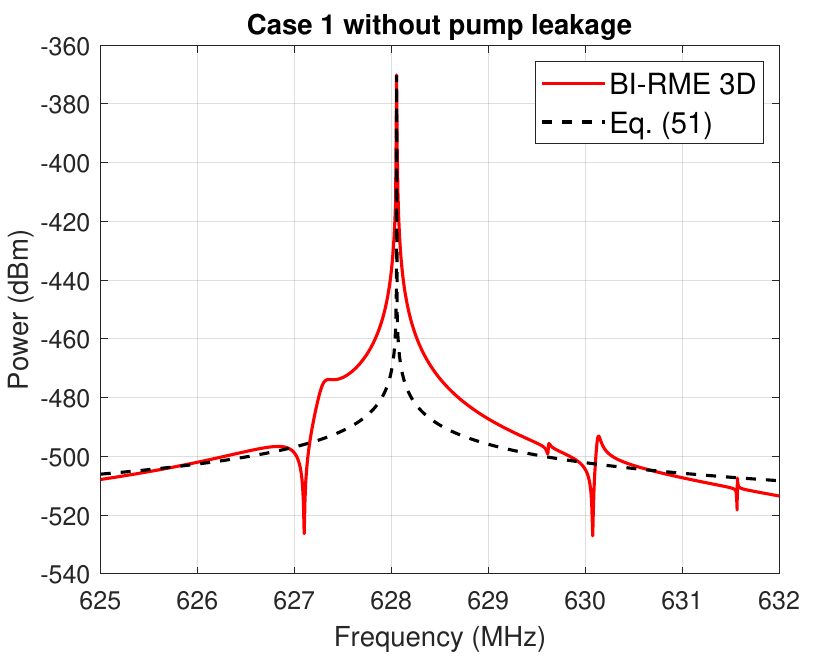}
        \label{fig:case1_a_c}
    }\hfill
    \subfloat[]{%
        \includegraphics[width=0.45\textwidth]{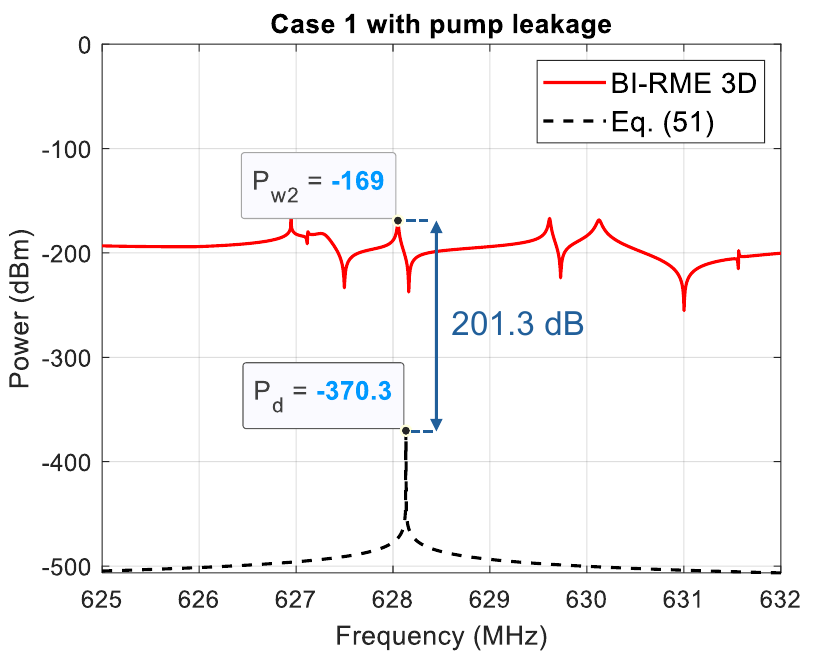}
        \label{fig:case1_b_c}
    }

    \vspace{0.4cm}

    \subfloat[]{%
        \includegraphics[width=0.45\textwidth]{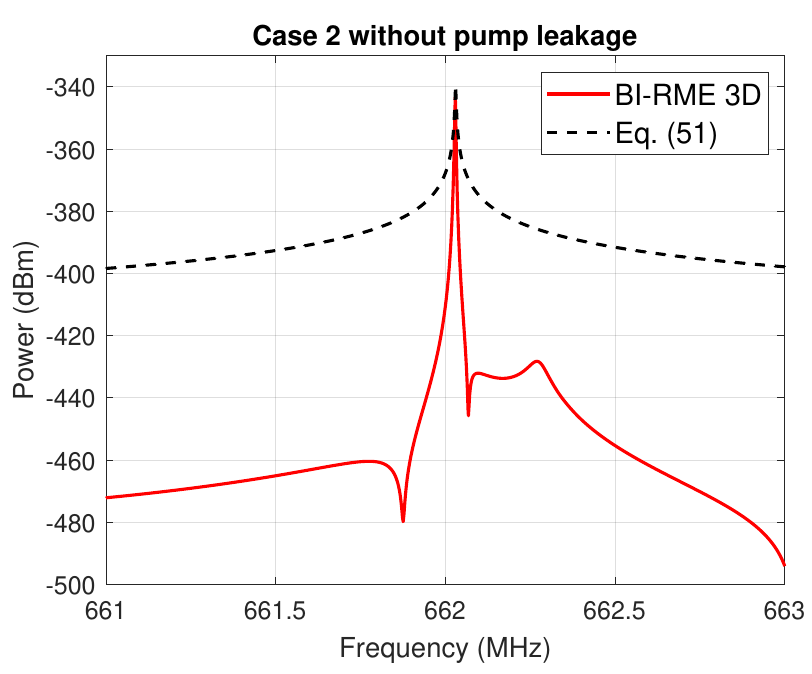}
        \label{fig:case2_a_c}
    }\hfill
    \subfloat[]{%
        \includegraphics[width=0.45\textwidth]{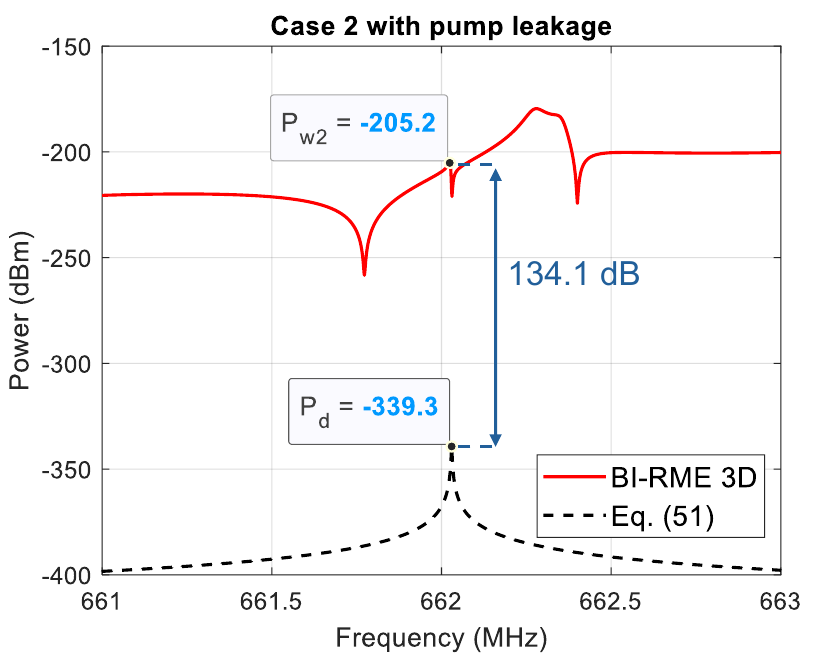}
        \label{fig:case2_b_c}
    }

    \caption{
    Comparison between the theoretical prediction (Eq. \eqref{eq:Pd1}) and the BI-RME 3D results for the two tuning cases of the RADES-BabyIAXO cavity assuming cryogenic copper. Panels (a) and (b) correspond to the same readout frequency point (Case 1), while panels (c) and (d) correspond to the second readout frequency condition (Case 2). In panels (a) and (c), the axion-induced power generated at port 2 is compared between the theoretical expression (black dashed line) and the BI-RME 3D result (red line), without including the leakage of the pump mode into the readout port. In panels (b) and (d), the total power detected at port 2 obtained from BI-RME 3D (red line) is shown. The blue double arrow indicates the power difference between the pump leakage evaluated at the readout frequency and the expected axion signal; it does not represent an absolute leakage or isolation level. In these panels, the theoretical axion detected power is also displayed for reference (black dashed line), in order to illustrate how pump leakage can mask the signal generated by axion-photon conversion. For both cases, the calculations assume the KSVZ axion coupling, ($g_{a\gamma\gamma}=-0.38\,m_a$), together with a local DM density of $\rho_a=0.4,\mathrm{GeV,cm^{-3}}$ and therefore, $g_{a \gamma \gamma} a_0 = -9.42\cdot 10^{-22}$. The incident pump power is $P_{\mathrm{inc},p} = 4$ mW.}
    \label{fig:BIRME_results_copper}
\end{figure*}

\begin{figure*}[t]
    \centering

    \subfloat[]{%
        \includegraphics[width=0.45\textwidth]{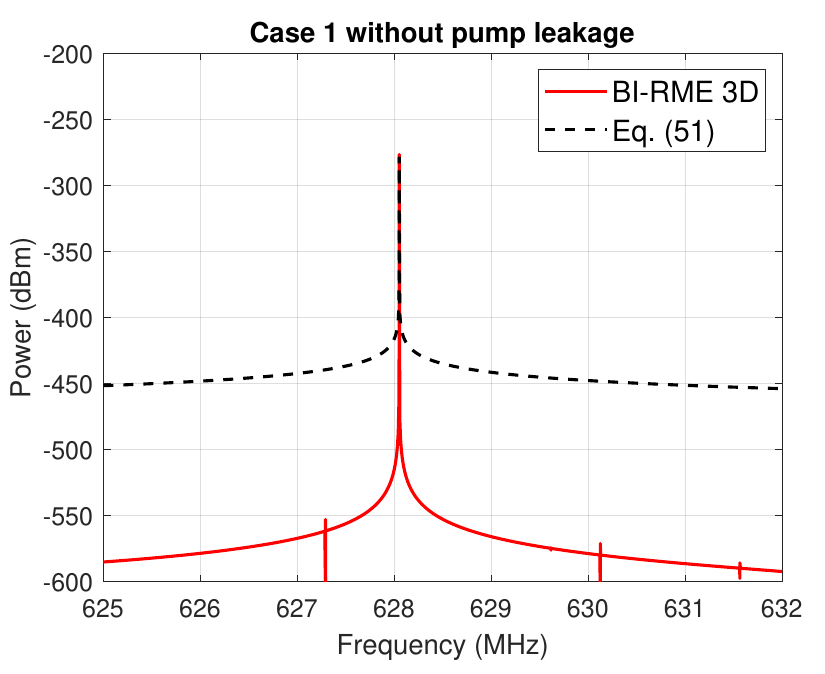}
        \label{fig:case1_a_s}
    }\hfill
    \subfloat[]{%
        \includegraphics[width=0.45\textwidth]{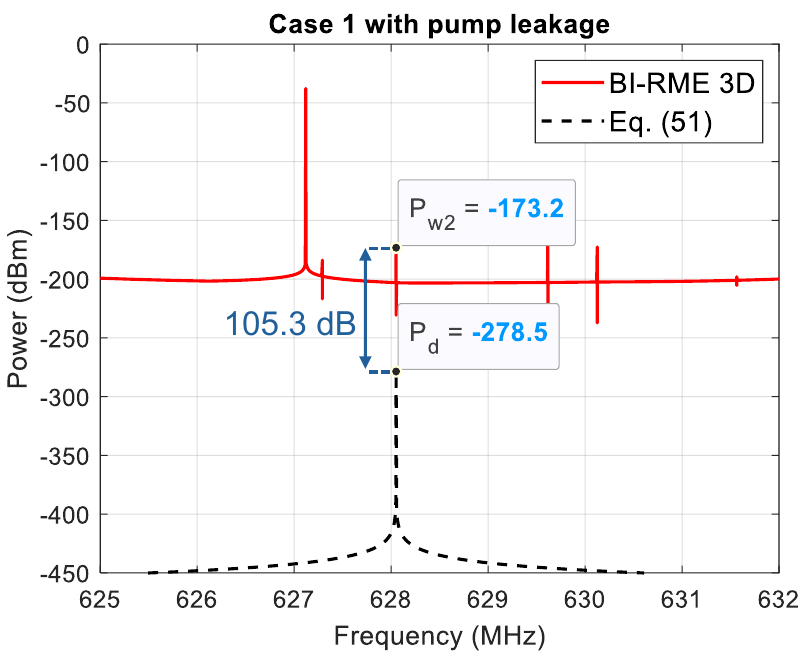}
        \label{fig:case1_b_s}
    }

    \vspace{0.4cm}

    \subfloat[]{%
        \includegraphics[width=0.45\textwidth]{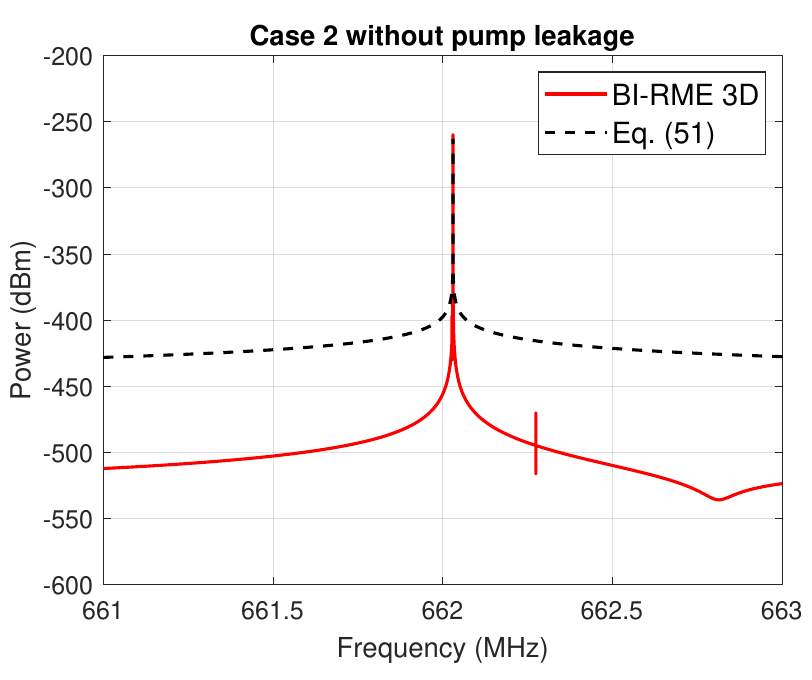}
        \label{fig:case2_a_s}
    }\hfill
    \subfloat[]{%
        \includegraphics[width=0.45\textwidth]{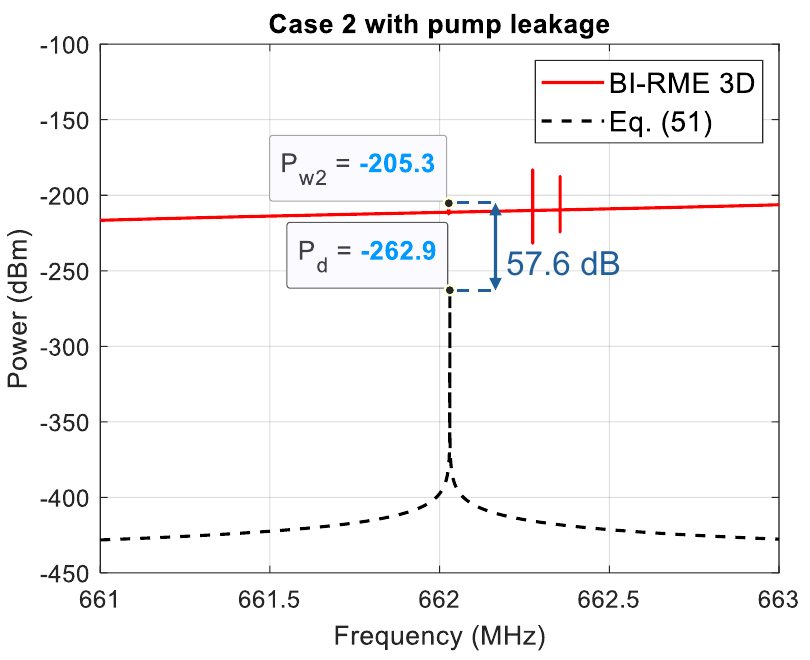}
        \label{fig:case2_b_s}
    }
    \caption{
    Comparison between the theoretical prediction (Eq. \eqref{eq:Pd1}) and the BI-RME 3D results for the two tuning cases of the RADES-BabyIAXO cavity assuming niobium. Panels (a) and (b) correspond to the same readout frequency point (Case 1), while panels (c) and (d) correspond to the second readout frequency condition (Case 2). In panels (a) and (c), the axion-induced power generated at port 2 is compared between the theoretical expression (black dashed line) and the BI-RME 3D result (red line), without including the leakage of the pump mode into the readout port. In panels (b) and (d), the total power detected at port 2 obtained from BI-RME 3D (red line) is shown. The blue double arrow indicates the power difference between the pump leakage evaluated at the readout frequency and the expected axion signal; it does not represent an absolute leakage or isolation level. In these panels, the theoretical axion detected power is also displayed for reference (black dashed line), in order to illustrate how pump leakage can mask the signal generated by axion-photon conversion. For both cases, the calculations assume the KSVZ axion coupling, ($g_{a\gamma\gamma}=-0.38\,m_a$), together with a local DM density of $\rho_a=0.4,\mathrm{GeV,cm^{-3}}$ and therefore, $g_{a \gamma \gamma} a_0 = -9.42\cdot 10^{-22}$. The incident pump power is $P_{\mathrm{inc},p} = 4$ mW. }
    \label{fig:BIRME_results_sc}
\end{figure*}

The results of the BI-RME 3D method application are shown in Figs. \ref{fig:BIRME_results_copper} and \ref{fig:BIRME_results_sc} for the copper and SRF cavities, respectively. For both cases, the analytical prediction is shown as a reference against which the BI-RME 3D result is compared. For visualization purposes, this analytical curve is obtained by normalizing the effective detected power in Eq. \eqref{eq:Pd1} with the readout Lorentzian profile. This approximation is adequate near resonance in the regimes considered here; the complete spectral shaping is captured by the BI-RME 3D calculation.
The left-hand panels (subfigures (a) and (c)) compare the power generated at port 2 for the two cases. In BI-RME 3D results, the theoretical axion-induced power has been evaluated from Eq. \eqref{eq:Pw2}

The comparison shows that the BI-RME 3D approach reproduces the frequency response more accurately away from resonance, while preserving an excellent agreement with the analytical expression at the resonance peak. Specifically, BI-RME 3D allows to accurately define the pump source from the signal generator specifications, and it takes into account the two different leakage sources: the coupling of the pump signal with the readout mode through the pump port, and the coupling of the pump mode to the readout port, both at all the simulated frequencies. This allows to properly analyze the problem of the pump leakage in the readout signal.

As expected, the axion generated power is larger in the niobium case (Fig. \ref{fig:BIRME_results_sc}), owing to its higher quality factor. The detected power spectra at the readout port are shown in subfigures (b) and (d). These panels are used to quantify accurately the leakage of the pump signal into the readout channel at different frequencies, including the readout frequency. At the readout frequency, the residual pump leakage exceeds the expected axion power by $201.3$ dB for cryogenic copper and $105.3$ dB for niobium. These values quantify the pump leakage level at $f_r$ relative to the expected axion signal. It can be seen that the use of the superconductor further reduces leakage of the pump signal thanks to improved cavity filtering due to its high quality factor. In both cases, however, the residual leakage remains well above the axion-induced contribution and therefore masks the axion-photon signal. By contrast, in Case 2 the leakage level is substantially reduced, obtaining $134.1$ dB and $57.6$ dB for cryogenic copper and niobium, respectively. This behavior is explained by the larger frequency separation between the pump and readout resonances, for which the pump spectral amplitude at \(f_r\) is significantly lower.

 It can be seen that the use of the superconductor further reduces the pump leakage thanks to the stronger cavity filtering associated with its higher quality factor.  By contrast, in Case 2 the leakage level is substantially reduced, exceeding the expected axion signal by 134.1 dB and 57.6 dB for cryogenic copper and niobium, respectively.

These results indicate that Case 1 represents the most unfavorable configuration from the point of view of pump leakage, severely compromising the detectability of the axion-photon conversion signal. Furthermore, the use of superconductors is practically essential for this type of experiment, as it substantially improves the detected power.

Several strategies can be adopted to mitigate the coupling of the pump signal into the readout port. A first possibility is to optimize the relative position of the pump and readout ports so that the readout port is placed close to an electric field minimum of the pump mode. In the present design, the ports location was chosen following this criterion, leading to an isolation level of $|S_{21}| \approx-30$ dB. Additional suppression may be achieved by inserting a band-reject filter in the input pump line. In this case, the rejected band should be centered at the readout frequency covering the operational range from the minimum achievable readout frequency to the maximum. Commercial solutions provide attenuations of up to 50 dB \cite{BandReject,Wainwright_Tunable} for tunable filters and up to 80 dB for fixed filters \cite{Wainwright_Fixed}. In Case 1, as the leakage from the pump mode exceeds 100 dB, two filters connected in cascade could be used. A separate filtering requirement concerns the strong pump carrier itself, since the residual carrier at $f_p$ entering the readout chain can be much larger and may saturate the first low noise amplifier. Therefore, a slightly tunable notch filter centered at the pump frequency should be installed in the readout chain before the first amplifier. Its stop-band must be sufficiently narrow to provide strong carrier rejection while maintaining negligible insertion loss at the readout frequency.

In addition to these measures, coherent pump leakage can be further suppressed using active feed-forward cancellation. A reference copy of the pump signal, taken after the main noisy RF components, can be amplitude- and phase-adjusted and recombined destructively with the readout signal. This technique suppresses only the component correlated with the pump reference and must be recalibrated at each tuning point. Residual leakage can be further identified in the analysis by monitoring its coherence with the pump reference, by repeating measurements under Local Oscillator/pump/readout retunings, and by applying null tests with detuned modes or reduced overlap. Candidate axion signals are required to be stationary in axion-frequency space, to have the expected linewidth, and to be incoherent with the pump reference.

\subsection{Optimization of the scanning rate and projected upper-limit sensitivity}
To optimize the figure of merit related to the scanning rate, we now determine the optimal coupling values for the pump and readout ports. Since the most promising implementation of the up-conversion technique is expected to rely on SRF cavities, we restrict the present study to this case. To this end, Eq.~\eqref{eq:dma_final} is evaluated assuming the niobium cavity with the corresponding unloaded quality factors for the pump and readout modes, together with the corresponding quality factors of the up-conversion process $(Q_{h,1},Q_{h,2})$. The cavity temperature $T_\mathrm{cav}=300$ mK is assumed, obtained by extrapolating from a commercial specification of $450~\mu{\rm W}$ at 100 mK using the relationship $P_{\rm cool}\propto T^2$ for the cooling power of the dilution mixture for the cryostat model Bluefors LD350/LD450 \cite{Cryostat}. In practice, the actual value must be verified using the load curve of the specific cryostat. We emphasize that this extrapolation only provides an estimate of the available cooling power and does not constitute a complete thermal budget. Using the same $P_{\rm cool}\propto T^2$ scaling, the nominal cooling power at 300 mK would be approximately 4.05 mW. Since the pump mode is  critically coupled, the accepted pump power is $P_{\rm acc,p}=P_{\rm inc,p}$, so that a continuous incident power of 4 mW would by itself consume almost all of this estimated cooling capacity. Additional static and parasitic heat loads from mechanical supports, RF lines, instrumentation, the tuning system, and thermal radiation are not included in this simplified estimate. Therefore, $P_{\rm inc,p}=4$ mW should be interpreted here as a reference pump power for the electromagnetic sensitivity calculation. A practical implementation will require a complete thermal budget using the measured load curve of the selected cryostat. Reduced continuous pump power, increased cryogenic capacity or improved thermal anchoring may be required.

\begin{figure}[ht]
\centering
\includegraphics[width=0.45\textwidth]{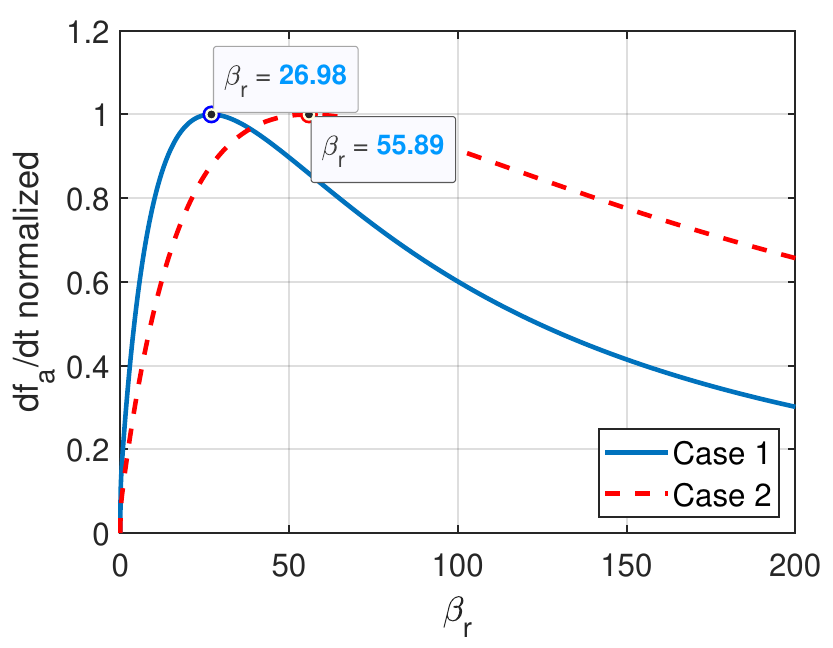}
\caption{Plot of the scanning rate with normalized units for Case 1 and Case 2, depending on $\beta_r$, assuming $T_\mathrm{cav}=300\,\mathrm{mK}$, $T_\mathrm{add}=20$ mK. The unloaded quality factors of the pump and readout modes are $Q_{0,p}=2.74\cdot 10^{10}$ , $Q_{0,r}=4.63\cdot 10^{10}$ for Case 1; and $Q_{0,p}=2.87\cdot 10^{10}$, $Q_{0,r}=5.92\cdot 10^{10}$ for Case 2.}
\label{fig:scanning_rate_opt}
\end{figure}

An added noise of $T_\mathrm{add}=20\,\mathrm{mK}$ is assumed for the readout chain, mainly produced by a SQUID amplifier. We assume here that pump leakage is sufficiently suppressed, filtered, or calibrated so that it does not dominate over the thermal/noise-limited readout. Fig.~\ref{fig:scanning_rate_opt} shows the resulting normalized scanning rate versus $\beta_r$ values for the cases 1 and 2, respectively. It is obvious from \eqref{eq:dma_final} that the optimal value in any case for the pump port is $\beta_p=1$ (critical coupling, i.e, all the incident power is accepted by the cavity).

The existence of an optimum in the readout port is a consequence of several effects present in the scanning rate expression, such as the enhancement of the signal in the cavity, the frequency step in the experiment, and the cavity noise and the noise added by the readout chain.

For the present configuration, the optimal values are found to be $\beta_r = 26.98$ for Case 1, and $\beta_r = 55.89$ for Case 2. These results indicate that the readout port should be significantly overcoupled in order to optimize the scanning rate. From an experimental point of view, this implies that the port geometry (in the present design, implemented through coaxial antennas) must be chosen so as to reproduce these target couplings as closely as possible. Moreover, since the coupling factors generally vary during the tuning process, a practical implementation should incorporate a re-coupling mechanism capable of keeping the system close to the optimal working point throughout the scan. Finally, at these optimal port couplings, we evaluate the scanning rate assuming \(\mathrm{SNR}=1.28\) and the KSVZ axion–photon coupling corresponding to the axion mass of each case. For detection bandwidths of \(\Delta f=0.379\,\mathrm{Hz}\) in Case 1 and \(\Delta f=0.636\,\mathrm{Hz}\) in Case 2, the resulting maximum scanning rates are \(df_a/dt=6.70\cdot10^{-15}\,\mathrm{Hz/s}\) and \(df_a/dt=7.98\cdot10^{-10}\,\mathrm{Hz/s}\), respectively. These extremely low values highlight the need for substantial improvements before practical broadband data taking can be achieved.
 \\

\begin{figure*}[ht]
    \centering
    \includegraphics[width=0.95\linewidth]{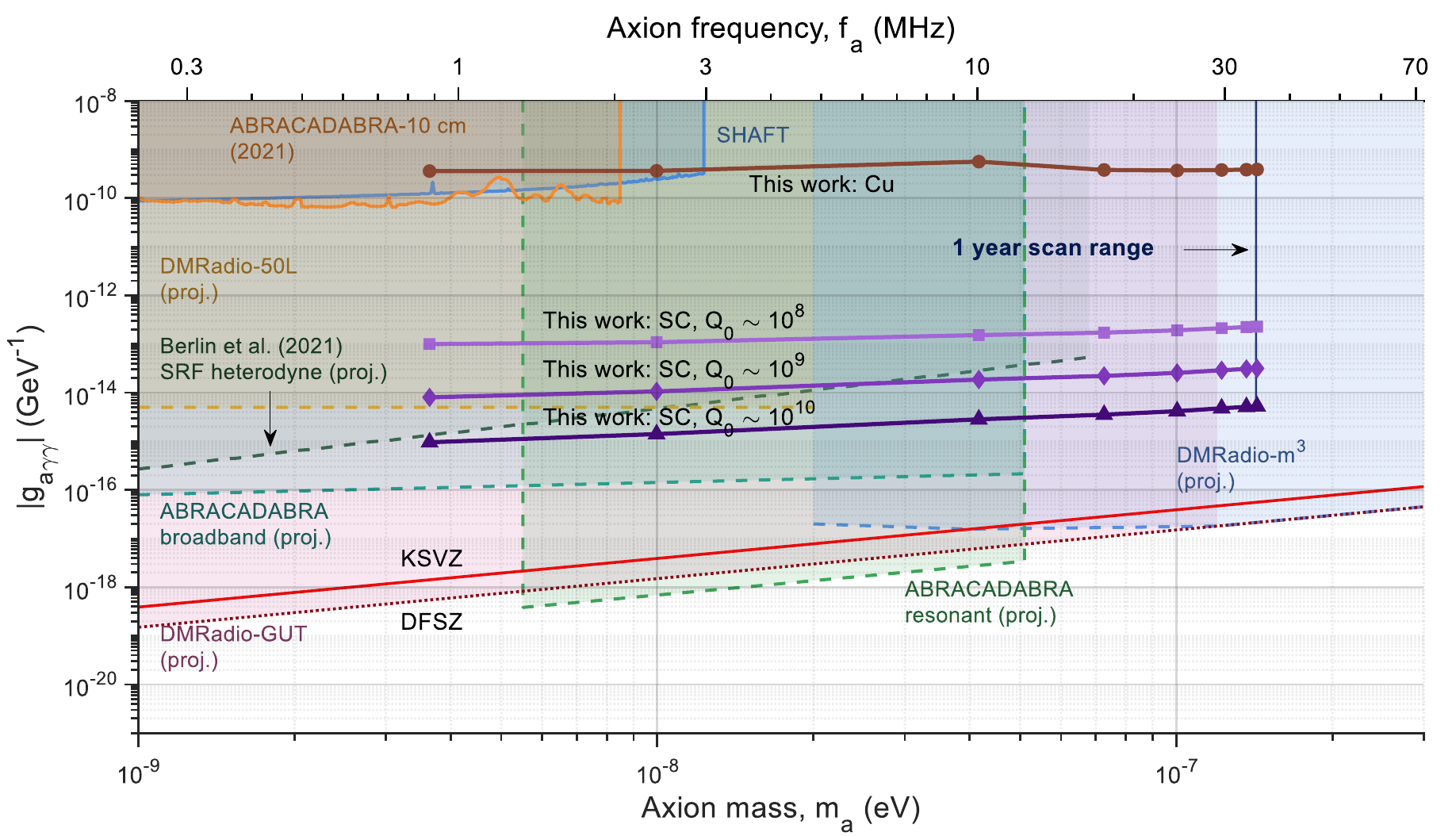}
\caption{
Projected upper-limit sensitivity to the axion-photon coupling
$|g_{a\gamma\gamma}|$ at 90\% CL (Confidence Level) for the proposed up-conversion experiment showing other experiments in this axion mass range \cite{LC_resonator2, Heterodyne_2021, Salemi2021ABRACADABRA, Rapidis2023DMRadio, Brouwer2022DMRadiom3, Brouwer2022DMRadioGUT}. The plot compares a cryogenic copper cavity with three superconducting scenarios corresponding to different assumptions for the unloaded quality
factor. For the superconducting cases,
$(Q_{0,p},Q_{0,r})=(2.87,5.92)\cdot10^8,\cdot10^9,\cdot10^{10}$ and assuming the corresponding optimized readout couplings
$\beta_r=21.87$, $39.72$, and $55.89$, respectively.
The lower horizontal axis gives the equivalent axion mass, while the upper axis shows the corresponding axion frequency. The calculation assumes
$\mathrm{SNR}=1.28$, $\Delta t=3600~\mathrm{s}$ per frequency point,
$T_{\mathrm{cav}}=300~\mathrm{mK}$,
$T_{\mathrm{add}}=20~\mathrm{mK}$,
$\beta_r\mathrm{(Cu)}=2$, $\beta_p=1$,
$P_{\mathrm{inc},p}=4~\mathrm{mW}$,
$\rho_a=0.4~\mathrm{GeV\,cm^{-3}}$, and $Q_a=10^6$.
The achievable tuning range is
$f_a=0.9$--$34.62~\mathrm{MHz}$, corresponding to
$m_a=3.64\cdot10^{-3}$--$0.143~\mu\mathrm{eV}$.
The full sensitivity curves represent the local sensitivity achievable at
each tunable frequency and should therefore not be interpreted as the
frequency interval that can be scanned within a fixed total acquisition
time. The shaded regions above each curve indicate the values of
$|g_{a\gamma\gamma}|$ excluded at $\ge$ 90\% CL under the assumed experimental conditions. For the representative superconducting
case with $Q_0\sim10^{10}$, the dark-blue band indicates the realistically
scannable range for one year of data taking. Assuming one hour of
integration per frequency point, corresponding to $N=8760$ points, and
frequency steps $\Delta f_a=f_a/Q_a$, the covered
interval is approximately $f_a=34.32$--$34.62~\mathrm{MHz}$,
i.e. $0.3~\mathrm{MHz}$. These sensitivity projections assume that
pump leakage is properly treated so that it does not dominate over the
thermal/noise-limited readout.
}
    \label{fig:sensitivity_plot}
\end{figure*}

Assuming that pump leakage is sufficiently suppressed, filtered, or calibrated so that it does not dominate over the thermal/noise-limited readout, Fig. \ref{fig:sensitivity_plot} shows the projected sensitivity to the axion-photon coupling obtained for the proposed up-conversion experiment over the achievable axion frequency range $f_a=0.9–34.62~\mathrm{MHz}$, corresponding to an axion mass interval $m_a=3.64\cdot 10^{-3}–1.43\cdot 10^{-1}~\mu\mathrm{eV}$. The sensitivity is evaluated at $90\%$ confidence level by imposing $\mathrm{SNR}=1.28$, with an integration time of $\Delta t=3600~\mathrm{s}$ per frequency point. The lower horizontal axis gives the equivalent axion mass, while the upper axis gives the corresponding axion frequency.


Fig.~\ref{fig:sensitivity_plot} includes existing experimental limits and future projections from relevant lumped-element searches, together with the projected sensitivity of a SRF heterodyne detection proposal of Berlin et al.~\cite{Heterodyne_2021}. This allows a direct comparison of the present up-conversion approach with low-mass axion searches. In the copper case, the sensitivity remains approximately flat over most of the tunable range, with values in the range $|g_{a\gamma\gamma}| \simeq 3.56\cdot10^{-10}-5.58\cdot10^{-10}~\mathrm{GeV}^{-1}$. The superconducting implementation provides a substantially improved reach, although its performance strongly depends on the achievable quality factor. For the representative case with $Q_0\sim10^{10}$, the projected sensitivity ranges from $|g_{a\gamma\gamma}| \simeq 9.47\cdot10^{-16}$ to $5.21\cdot10^{-15}~\mathrm{GeV}^{-1}$. For the more conservative scenarios with $Q_0\sim10^{9}$ and $Q_0\sim10^{8}$, the corresponding ranges are $|g_{a\gamma\gamma}| \simeq 7.94\cdot10^{-15}$--$3.13\cdot10^{-14}~\mathrm{GeV}^{-1}$ and $9.89\cdot10^{-14}$--$2.26\cdot10^{-13}~\mathrm{GeV}^{-1}$, respectively. Compared with cryogenic copper, the superconducting scenarios provide sensitivity improvements of about \(10^{5}\), \(10^{4}\), and a few \(10^{3}\) for \(Q_0\sim10^{10}\), \(10^{9}\), and \(10^{8}\), respectively, depending on the tuning point. This enhancement results from the combined effect of the larger quality factors achievable in the superconducting cavity and the optimization of the readout coupling. In particular, the copper case is evaluated with $\beta_r=2$, whereas the superconducting scenarios use $\beta_r=55.879$, $39.718$, and $21.876$ for $Q_0\sim10^{10}$, $10^9$, and $10^8$, respectively. Therefore, the improvement with respect to copper should not be attributed solely to the increase of the quality factor, but to the combined effect of higher quality factors and optimized readout coupling.

The solid curves in Fig.~\ref{fig:sensitivity_plot} represent the local sensitivity achievable at each frequency within the $0.9$--$34.62~\mathrm{MHz}$ tuning range and should not be interpreted as the frequency interval that can be covered within a fixed acquisition time. For example, for the $Q_0\sim10^{10}$
scenario, assuming $\Delta t=3600~\mathrm{s}$ per frequency point and one year of continuous data taking, corresponding to $N=8760$ points, with a frequency step of $\Delta f_a=f_a/Q_a$ according to Eq.~\eqref{eq:faQa}, the realistically scannable interval is approximately $34.32$--$34.62~\mathrm{MHz}$,
corresponding to a bandwidth of $0.3~\mathrm{MHz}$, that is, around 1\% of relative bandwidth. Nevertheless, even under the more conservative superconducting assumptions considered here, the use of an SRF cavity provides a substantial sensitivity improvement over cryogenic copper, with the $Q_0\sim10^{10}$ scenario reaching the $10^{-15}~\mathrm{GeV}^{-1}$ level.

\section{Conclusions}
In this work, we have studied the application of the heterodyne detection technique to a low-frequency haloscope, in this case the RADES-BabyIAXO haloscope, as a possible strategy to search for low-mass dark matter axions using microwave cavity modes. Starting from axion electrodynamics, we derived the effective axion-induced current density and the corresponding expression for the power extracted from the readout port. A key point of the formulation is the derivation of the effective quality factors governing the up-conversion process. The finite linewidth of the axion signal was explicitly included, instead of assuming a purely monochromatic axion. The interaction between the pump signal and the pump resonant mode, the axion field and the pump mode, and finally, the filtering of the up-converted signal by the readout mode, were taken into account, leading to the definitions of new quality factors that set the detection bandwidth and impact the extracted power. 

This treatment was also implemented in the BI-RME 3D analysis by considering an axion signal with finite bandwidth. The method was applied to the largest quasi-cylindrical RADES-BabyIAXO cavity. A study has been carried out on various combinations of TE-TM modes, finding that the pair $\mathrm{qTE}_{011}$-$\mathrm{qTM}_{010}$ provides a good overlap factor enhancing the exploration of axion frequencies from approximately $0.9$ to $34.62$ MHz. The analytical model was compared with the BI-RME 3D full-wave formulation, showing good agreement at the readout resonance and confirming the validity of the derived power expression. In addition, the BI-RME 3D technique allowed us to include realistic two-port effects, especially the electromagnetic leakage from the pump port to the readout port at the readout frequency, which can become a critical limitation that must be minimized experimentally. 


Finally, the scanning rate expression was derived including the relevant quality factors of the up-conversion process, and the optimal coupling coefficients were determined to obtain the maximum the scanning rate. Sensitivity estimates for cryogenic copper and superconducting scenarios with \(Q_0\sim10^{8}\)–\(10^{10}\) show that, depending on the achievable quality factor and optimized readout coupling, the projected upper-limit sensitivity can reach the \(10^{-15}\,\mathrm{GeV}^{-1}\) level over part of the potential scannable range. Overall, the results indicate that the RADES-BabyIAXO cavity is a promising platform for heterodyne axion searches in the low MHz range.

Future work will focus on improving the reachable sensitivity and the scanning rate. Possible strategies include determining the maximum pump power that can be sustained in the pump mode $-$as constrained by the critical magnetic field of the superconducting material$-$ and the available cooling power of the cryostat employing a pulsed pump signal. Pulsed operation could reduce the average RF heat load, thereby limiting the increase in the cavity temperature and the associated thermal noise. The pulse duration and duty cycle should nevertheless be optimized by accounting for the cavity filling and decay times. In addition, another objective is to perform thermal simulations of the cavity heating caused by the pump signal and to identify an optimal cooling distribution depending on the readout mode current density, in order to reduce the readout noise. We also plan to develop an experimental setup for validation of the present results.

\section{\label{acknowledgments} Acknowledgments}

This work was performed within the RADES group; we thank our colleagues for their support. We also thank Juan Monzó Cabrera, Camilo Garcia-Cely, Valerie Domcke, Sung Mook Lee and Álvaro Hernández Zamorano for their valuable comments and discussions. The research leading to these results has received funding from the Spanish Ministry of Science and Innovation with the projects PID2022-137268NB-C53, PID2022-137268NA-C55 and PID2025-171343NB-C53, funded by MICIU/AEI/10.13039/501100011033/ and by “ERDF/EU”. This research was also supported by the Horizon Europe programme (ERC-2023-SyG DarkQuantum, grant agreement No. 101118911). J. R. Navarro-Madrid acknowledges the financial support received from the Universidad Politécnica de Cartagena through the DEBUTA Project. J. Reina-Valero has the support of "Plan de Recuperación, Transformación y Resiliencia (PRTR) 2022 (ASFAE/2022/013)", funded by Conselleria d'Innovació, Universitats, Ciència i Societat Digital from Generalitat Valenciana, and NextGenerationEU from European Union. This publication has received financial support from the Severo Ochoa project CEX2023-001292-S funded by Ministerio de Ciencia, Innovación y Universidades/Agencia Estatal de Investigación.

\appendix

\section{Orthogonality relationship of the TE and TM modes of a microwave cavity} \label{Appendix_modes_cavity}

Here it is demonstrated that for any TE or TM solenoidal eigenmode of an empty cavity, the electric and magnetic fields are orthogonal at every point of the resonator \cite{DarkPhoton_JR}. The cavity is assumed to be vacuum-filled, with permittivity $\varepsilon_0$ and permeability $\mu_0$.

We first consider a TM mode, whose electric and magnetic fields are denoted by $\vec{E}^{\,\mathrm{TM}}_m$ and $\vec{H}^{\,\mathrm{TM}}_m$, respectively. In this case, the fields can be decomposed into transverse and axial components as
\[
\vec{E}^{\,\mathrm{TM}}_m=\vec{E}^{\,\mathrm{TM}}_{t,m}+E^{\,\mathrm{TM}}_{z,m}\,\hat z,
\qquad
\vec{H}^{\,\mathrm{TM}}_m=\vec{H}^{\,\mathrm{TM}}_{t,m},
\]
so that
\[
\vec{E}^{\,\mathrm{TM}}_m\cdot \vec{H}^{\,\mathrm{TM}}_m
=
\vec{E}^{\,\mathrm{TM}}_{t,m}\cdot \vec{H}^{\,\mathrm{TM}}_{t,m}.
\]
Using Eq.~(8.76) of Ref.~\cite{jackson}, one finds that
\[
\vec{E}^{\,\mathrm{TM}}_{t,m}\propto \nabla_t E^{\,\mathrm{TM}}_{z,m},
\qquad
\vec{H}^{\,\mathrm{TM}}_{t,m}\propto \hat z\times \nabla_t E^{\,\mathrm{TM}}_{z,m},
\]
and therefore
\[
\vec{E}^{\,\mathrm{TM}}_{t,m}\propto \hat z \times \vec{H}^{\,\mathrm{TM}}_{t,m}.
\]
It follows immediately that the transverse electric and magnetic fields are orthogonal, and hence
\[
\vec{E}^{\,\mathrm{TM}}_m\cdot \vec{H}^{\,\mathrm{TM}}_m=0.
\]

The same argument applies to TE modes. Let $\vec{E}^{\,\mathrm{TE}}_n$ and $\vec{H}^{\,\mathrm{TE}}_n$ denote the electric and magnetic fields of the $n$-th TE mode. These fields may be written as
\[
\vec{E}^{\,\mathrm{TE}}_n=\vec{E}^{\,\mathrm{TE}}_{t,n},
\qquad
\vec{H}^{\,\mathrm{TE}}_n=\vec{H}^{\,\mathrm{TE}}_{t,n}+H^{\,\mathrm{TE}}_{z,n}\,\hat z.
\]
From Eq.~(8.77) of Ref.~\cite{jackson}, one obtains
\[
\vec{E}^{\,\mathrm{TE}}_{t,n}\propto \hat z\times \nabla_t H^{\,\mathrm{TE}}_{z,n},
\qquad
\vec{H}^{\,\mathrm{TE}}_{t,n}\propto \nabla_t H^{\,\mathrm{TE}}_{z,n},
\]
which implies
\[
\vec{H}^{\,\mathrm{TE}}_{t,n}\propto \hat z\times \vec{E}^{\,\mathrm{TE}}_{t,n}.
\]
Therefore, the transverse electric and magnetic fields are again orthogonal, and one concludes that
\[
\vec{E}^{\,\mathrm{TE}}_n\cdot \vec{H}^{\,\mathrm{TE}}_n=0.
\]

\section{Accuracy of the effective quality factor approximation}
\label{app:Qh2_accuracy}

In Sec.~\ref{sub:Qequiv}, the effective quality factor associated with the
FWHM of the product of the two Lorentzian responses was obtained
exactly as

\begin{equation} \label{eq:Qh2_exact}
Q_{h,2}^\mathrm{exact}=
\left[
\frac{
Q_{h,1}^2+Q_{L,r}^2+
\sqrt{\left(Q_{h,1}^2+Q_{L,r}^2\right)^2
+4Q_{h,1}^2Q_{L,r}^2}
}{2}
\right]^{1/2},
\end{equation}

and approximated by

\begin{equation} \label{eq:Qh2_approx}
Q_{h,2}^{\mathrm{app}}
=
\sqrt{Q_{h,1}^2+Q_{L,r}^2}.
\end{equation}

To quantify the accuracy of this approximation, we introduce the
dimensionless ratio

\begin{equation}
r_t=\frac{Q_{h,1}}{Q_{L,r}}.
\end{equation}

Both expressions can then be normalized to $Q_{L,r}$, so that their
relative difference depends only on $r_t$. We define the relative error as

\begin{equation}
\epsilon(r_t)=
\frac{
\left|Q_{h,2}^{\mathrm{app}}-Q_{h,2}^{\mathrm{exact}}\right|
}{
Q_{h,2}^{\mathrm{exact}}
}\cdot100\%.
\end{equation}

\begin{figure}[ht]
    \centering
    \includegraphics[width=0.9\linewidth]{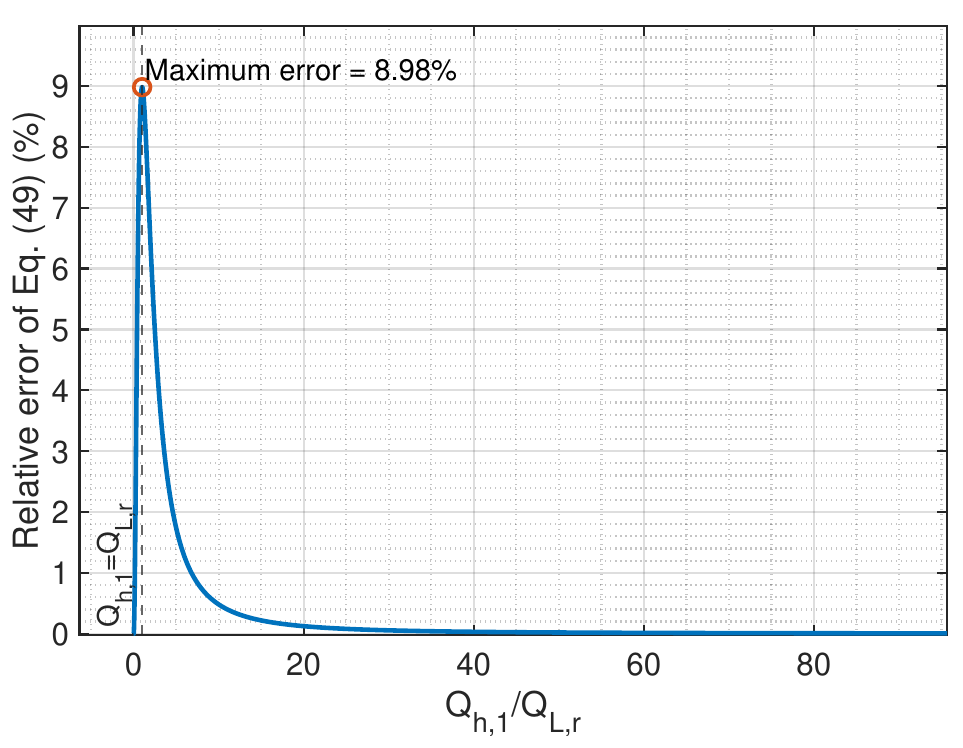}
    \caption{Relative error introduced by the approximation in Eq.~\eqref{eq:Qh2_approx}
with respect to the exact effective quality factor in Eq.~\ref{eq:Qh2_exact}, as a
function of the ratio $Q_{h,1}/Q_{L,r}$.}
    \label{fig:Qh2_error}
\end{figure}

Figure~\ref{fig:Qh2_error} shows $\epsilon$ as a function of
$Q_{h,1}/Q_{L,r}$. The maximum deviation occurs when the two quality
factors are equal, $Q_{h,1}=Q_{L,r}$, for which the exact and
approximate values are

\begin{equation}
Q_{h,2}^{\mathrm{exact}}=1.5538\,Q_{L,r},
\qquad
Q_{h,2}^{\mathrm{app}}=1.4142\,Q_{L,r},
\end{equation}

and therefore, corresponding to a relative error of approximately $8.98\%$.

The approximation rapidly improves when the two quality factors become
well separated. For example, the error is about $6.35\%$ for
$Q_{h,1}/Q_{L,r}=0.5$ (or $2$), $1.74\%$ for $0.2$ (or $5$),
and falls below $0.5\%$ for ratios smaller than $0.1$ or larger than
$10$. Therefore, Eq.~\eqref{eq:Qh2_approx} provides an accurate approximation in the asymptotic regimes $Q_{L,r}\gg Q_{h,1}$ or $Q_{h,1}\gg Q_{L,r}$. When both quality factors are comparable, the deviation increases but remains bounded by $8.98\%$, with the maximum occurring at $Q_{h,1}=Q_{L,r}$. Thus, the approximation remains useful for analytical estimates

\section{Evaluation of a singular surface integral}    \label{Appendix_singular_surface_integral}

\begin{figure}[H]   
    \centering
    \includegraphics[width=0.65\linewidth]{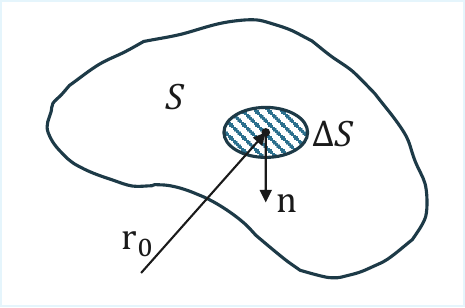}
    \caption{Scheme used for the analytical calculation of the singular surface integral \ref{singular_surface_integral}; $S$ is the cavity surface; $\Delta S$ is a very small circle of radius $\varepsilon$, with $\varepsilon \rightarrow 0$; $\vec{r}_0$ is the position vector where we want to calculate the surface integral (the circle $\Delta S$ is centered around $\vec{r}_0$); $\vec{n}$ is a unitary vector pointing from the cavity surface towards the cavity.}
    \label{fig:birme_surface_singular_1}
\end{figure}

We now evaluate the singular surface integral defined as
\begin{equation}     \label{singular_surface_integral}
\vec{I}(\vec{r}) \equiv 
\int_S 
\nabla \times \mathbf{\vec{G}^{\rm A}}
(\vec{r},\vec{r}^{\,\prime}) \cdot 
\vec{J}_S(\vec{r}^{\,\prime}) \, dS' .
\end{equation}
We focus on the case in which the observation point lies on the cavity surface, namely $\vec{r}=\vec{r}_0\in S$, as shown in Fig.~\ref{fig:birme_surface_singular_1}. In this case, the integral $\vec{I}(\vec{r}_0)$ contains a singularity that must be extracted and
integrated analytically in the Cauchy Principal Value (CPV) sense. To this end, the integration surface $S$ is divided into two regions: a small circular surface $\Delta S$ of characteristic radius $\varepsilon$, with $\varepsilon \ll 1$, surrounding the singular point $\vec{r}_0$, and the remaining surface $S-\Delta S$. Therefore, the integral can be written as
\begin{align}  \label{integral_I_general}
\vec{I}(\vec{r}) =
&\int_{S-\Delta S}
\nabla \times \mathbf{\vec{G}^{\rm A}}
(\vec{r},\vec{r}^{\,\prime}) \cdot 
\vec{J}_S(\vec{r}^{\,\prime}) \, dS'
\nonumber \\
&+
\int_{\Delta S}
\nabla \times \mathbf{\vec{G}^{\rm A}}
(\vec{r},\vec{r}^{\,\prime}) \cdot 
\vec{J}_S(\vec{r}^{\,\prime}) \, dS' .
\end{align}
The first term is interpreted as the CPV contribution, since the singular neighborhood around $\vec{r}_0$ has been excluded from the integration domain. It is therefore defined as
\begin{equation}
\operatorname{CPV}\!\left[\vec{I}(\vec{r}_0)\right]
\equiv
\lim_{\varepsilon \to 0}
\int_{S-\Delta S}
\nabla \times \mathbf{\vec{G}^{\rm A}}
(\vec{r}_0,\vec{r}^{\,\prime}) \cdot 
\vec{J}_S(\vec{r}^{\,\prime}) \, dS' .
\end{equation}
\noindent
It is simple to probe that this integral is zero using the general expression of the electric dyadic potentials Green’s function and the relationship $\nabla~\times~\vec{E}_m~=~k_m~\vec{H}_m$~\cite{conciauro},
\begin{eqnarray}
\int_{S - \Delta S} \nabla \times \mathbf{\vec{G}^{\rm A}}
	(\vec{r},\vec{r}^{\, \prime}) \cdot \vec{J}_S(\vec{r}^{\, \prime}) \, dS'  & = & \nonumber   \\
\int_{S - \Delta S} \nabla \times \left(\sum_{m=1}^{+\infty} \frac{\vec{E}_m(\vec{r}) \vec{E}_m(\vec{r}^{\, \prime})}{k_m^2 - k^2}\right) \cdot \vec{J}_S(\vec{r}^{\, \prime}) \, dS'  & = & \nonumber   \\  
\int_{S - \Delta S}  \left(\sum_{m=1}^{+\infty} \frac{(\nabla \times \vec{E}_m(\vec{r})) \vec{E}_m(\vec{r}^{\, \prime})}{k_m^2 - k^2}\right) \cdot \vec{J}_S(\vec{r}^{\, \prime}) \, dS'  & = & \nonumber   \\
\sum_{m=1}^{+\infty}  \frac{k_m}{k_m^2 - k^2} \, \vec{H}_m(\vec{r}) \int_{S - \Delta S}  \vec{E}_m(\vec{r}^{\, \prime}) \cdot \vec{J}_S(\vec{r}^{\, \prime}) \, dS'  & = & \vec{0} \nonumber  
\end{eqnarray}
considering the boundary condition that states that the electric field parallel to a perfect conductor is zero ($\vec{E}_m(\vec{r}^{\, \prime})|_{S} = \vec{0}$). As a consequence: $\mathrm{CPV}[\vec{I}(\vec{r}_0)] = \vec{0}$.

Next step is to extract the singularity of the electric dyadic potential $\mathbf{\vec{G}^{\rm A}}$ given by $\frac{1}{8 \pi R} \left(\mathbf{\vec{I}} + \frac{\vec{R} \vec{R}}{R^2} \right)$ in the second integral of (\ref{integral_I_general}), as reported in \cite{conciauro},
\begin{eqnarray}    \label{integral_I_deltaS_general}
\vec{I}(\vec{r}) \, & = & \, \int_{\Delta S} \nabla \times \mathbf{\vec{G}^{\rm A}}
 \cdot \vec{J}_S(\vec{r}^{\, \prime})   \, dS' \, =  \nonumber  \\ 
   &  &  \int_{\Delta S} \nabla \times \left(\mathbf{\vec{G}^{\rm A}} - 
   \frac{1}{8 \pi R} \left(\mathbf{\vec{I}} + \frac{\vec{R} \vec{R}}{R^2}\right)\right)  \cdot \vec{J}_S(\vec{r}^{\, \prime})  \, dS' \, +  \nonumber  \\ 
    &  &  \int_{\Delta S} \nabla \times \left( \frac{1}{8 \pi R} \left(\mathbf{\vec{I}} + \frac{\vec{R} \vec{R}}{R^2}\right)\right)  \cdot \vec{J}_S(\vec{r}^{\, \prime})  \, dS'  
\end{eqnarray}
where the unitary dyadic is defined as: $\mathbf{\vec{I}} \equiv \hat{x} \hat{x} + \hat{y} \hat{y} + \hat{z} \hat{z}$. It is important to remark that such singularity does not depend on the frequency. The first integral of (\ref{integral_I_deltaS_general}) is zero because the integrand is regular (the singularity has been extracted), and the surface $\Delta S$ tends to zero ($\varepsilon \rightarrow 0$). Finally, we have to integrate analytically the singularity of the second integral of (\ref{integral_I_deltaS_general}). Dyadic algebra \cite{chentotai}, \cite{hanson_yakovlev} allows to express the curl of the singularity as follows,
\begin{eqnarray}   
 \nabla \times \left( \frac{1}{8 \pi R} \left(\mathbf{\vec{I}} + \frac{\vec{R} \vec{R}}{R^2}\right)\right) = - \frac{(\vec{R}/R) \times \mathbf{\vec{I}}}{4 \pi R^2}   \nonumber
\end{eqnarray}
so the singular integrand is,
\begin{eqnarray}   
- \frac{(\vec{R}/R) \times \mathbf{\vec{I}}}{4 \pi R^2} \cdot \vec{J}_S(\vec{r}^{\, \prime}) \, = \, \frac{- 1}{4 \pi R^3} \left( \vec{R} \times  \vec{J}_S(\vec{r}^{\, \prime}) \right),   \nonumber
\end{eqnarray}
where we have employed this property: $(\vec{a} \times \mathbf{\vec{I}}) \cdot \vec{b} = \vec{a} \times \vec{b}$. By assuming that the surface electric current density $\vec{J}_S(\vec{r}^{\, \prime})$ is constant on the small circle $\Delta S$, we find
\begin{eqnarray}   \label{integral_I_deltaS_pre}
\vec{I}(\vec{r}_0) \, & = & \,  \int_{\Delta S} \nabla \times \left( \frac{1}{8 \pi R} \left(\mathbf{\vec{I}} + \frac{\vec{R} \vec{R}}{R^2}\right)\right)\biggm|_{\vec{r}_0}  \cdot \, \vec{J}_S(\vec{r}^{\, \prime})  \, dS'  \, = \, \nonumber \\
&  &  \,  \int_{\Delta S} \frac{- 1}{4 \pi R^3} \left( \vec{R} \times  \vec{J}_S(\vec{r}^{\, \prime})   \, dS' \right)\biggm|_{\vec{r}_0} \, = \,   \nonumber   \\
&  & \frac{-1}{4 \pi}  \, \left( \int_{\Delta S} \frac{\vec{R}}{R^3}\biggm|_{\vec{r}_0} dS' \right) \times \vec{J}_S(\vec{r}_0)
\end{eqnarray}
In order to solve analytically such surface integral, we take a local cylindrical reference system $(\rho, \varphi,z)$ centered on the center of the circle $\Delta S$; in this reference framework the vector $\vec{R}$ is expressed as $\vec{R} = - \rho' \, \hat{\rho}' + \delta \, \vec{n}$ with $\delta \rightarrow 0$; obviously $R =\sqrt{\rho'^2 + \ \delta^2}$ and $dS' = \rho' d\rho' \varphi'$. Thus, the integral (\ref{integral_I_deltaS_pre}) is analytically evaluated,
\begin{eqnarray}   \label{integral_I_deltaS_final}
\vec{I}(\vec{r}_0) & = & \frac{-1}{4 \pi}  \left( \int_{\Delta S} \frac{- \rho' \, \hat{\rho}' + \delta \, \vec{n}}{(\rho'^2 + \ \delta^2)^{3/2}}\biggm|_{\vec{r}_0} \rho' d\rho' d\varphi' \right) \times \vec{J}_S(\vec{r}_0) \, =  \nonumber \\
&  &  \frac{-1}{4 \pi}  \left( \int_{\Delta S} \frac{\delta \, \vec{n}}{(\rho'^2 + \ \delta^2)^{3/2}}\biggm|_{\vec{r}_0} \rho' d\rho' d\varphi' \right) \times \vec{J}_S(\vec{r}_0) \, =  \nonumber \\
&  &  \frac{-1}{4 \pi} \left[\frac{-\delta}{\sqrt{\rho'^2 + \ \delta^2}}\right]_0^{\varepsilon} \left[ \oint_{0}^{2 \pi}  d\varphi'\right] \, \left[\vec{n} \times \vec{J}_S(\vec{r}_0)\right] \, = \nonumber  \\
&  & - \frac{1}{2} \left[\vec{n} \times \vec{J}_S(\vec{r}_0)\right]
\end{eqnarray}
where we have considered $\oint_{0}^{2 \pi} \,  \hat{\rho}' \, d\varphi' = \oint_{0}^{2 \pi} \  (\cos \varphi' \hat{x} + \sin \varphi' \hat{y}) \, d\varphi' = \vec{0}$ and $\delta/|\delta| = 1$ within the cavity.  

\section{Rational fit in the frequency domain of the scattering parameters for a SRF cavity} \label{Appendix_scattering_parameters}

In order to obtain a compact and physically interpretable representation of the simulated scattering parameters, the frequency response of the two-port cavity is described by means of a rational modal model. The underlying idea is to separate the response into two contributions: a slowly varying non-resonant background, and a set of resonant terms associated with the resonant modes of the cavity.

For each scattering parameter $S_{ij}$, with $i,j=1,2$, the fitted response in the frequency domain is written as \cite{Ramella2021HighQ}
\begin{equation}
S_{ij}(f)=\sum_{n=0}^{P} b_{ij,n}\Gamma^n+
\sum_{m=1}^{M} \frac{a_{ij,m}}{D_m(f)},
\label{eq:rational_model}
\end{equation}
where
\begin{equation}
D_m(f)=1+2jQ_{L,m}\left(\frac{f}{f_{0,m}}-1\right),
\label{eq:modal_denominator}
\end{equation}
and
\begin{equation}
\Gamma \, \equiv  \, \frac{f-f_c}{\Delta f/2}.
\label{eq:normalized_frequency}
\end{equation}
Here, $f$ is the operation frequency, $f_c$ is the center of the fitted frequency interval, and $\Delta f$ is the total frequency span. Therefore, $\Gamma$ is a normalized frequency variable, which maps the fitted band into a dimensionless interval. This normalization improves the numerical conditioning of the polynomial expansion allowing to fit the non-resonant part of the frequency response to be represented in a stable way. The first term of Eq.~\eqref{eq:rational_model},
\begin{equation}
\sum_{n=0}^{P} b_{ij,n}\Gamma^n,
\end{equation}
represents the smooth non-resonant background. This contribution accounts for direct electromagnetic coupling between ports, slowly varying mismatches, feedline effects, and any broad-band response that does not originate from a high-$Q$ resonance. The order of the polynomial is $P$, and the coefficients $b_{ij,n}$ are complex numbers which are independently fitted for each scattering parameter. Their complex nature allows to simulate the model in order to reproduce both the magnitude and the phase of the background response. The second term of Eq.~\eqref{eq:rational_model},
\begin{equation}
\sum_{m=1}^{M} \frac{a_{ij,m}}{D_m(f)},
\end{equation}
contains the resonant contribution of the cavity modes. Each mode $m$ is characterized by a resonant frequency $f_{0,m}$ and a loaded quality factor $Q_{L,m}$. The denominator $D_m(f)$ has the standard form of a Lorentzian response around $f_{0,m}$. At resonance, $f=f_{0,m}$, the denominator is equal to unity, whereas away from resonance its magnitude increases according to the loaded linewidth of the mode. Thus, $Q_{L,m}$ determines the spectral width of the resonance, while $f_{0,m}$ fixes its central frequency. The coefficients $a_{ij,m}$ are complex residues which determine how strongly each resonant mode contributes to a particular scattering parameter, and also include the phase associated with the excitation and the detection paths. Therefore, the same physical pole can appear with different amplitudes and phases in $S_{11}$, $S_{21}$, $S_{12}$, and $S_{22}$. This is particularly important in a two-port cavity, where a given eigenmode may couple strongly to one port and weakly (or even negligible) to the other.

A key aspect of the model is that the resonant denominators are common to all four scattering parameters. This reflects the fact that the poles are properties of the cavity itself, not of a specific measurement channel. The different $S_{ij}$ parameters share the same set of resonant frequencies and loaded quality factors, but they differ in their residues and in their non-resonant backgrounds. This structure provides a physically consistent description of the multi-port response and avoids identifying the same cavity mode independently in each scattering parameter.

The model also naturally accounts for transmission zeros and antiresonances. In particular, deep minima in $S_{21}$ or $S_{12}$ do not necessarily correspond to additional physical modes. They may arise from destructive interference between the resonant terms and the non-resonant background. In the rational representation, these features appear as zeros of the numerator rather than as poles of the denominator. This distinction is important when fitting responses with Fano-like lineshapes, where resonant and direct signals interfere and produce asymmetric peaks and notches \cite{Bachbauer2025Fano, Kancleris2020Fano, steshenko2025inclinedFano}. 

\begin{figure*}[ht]
\centering

\subfloat[]{
    \includegraphics[width=0.47\textwidth]{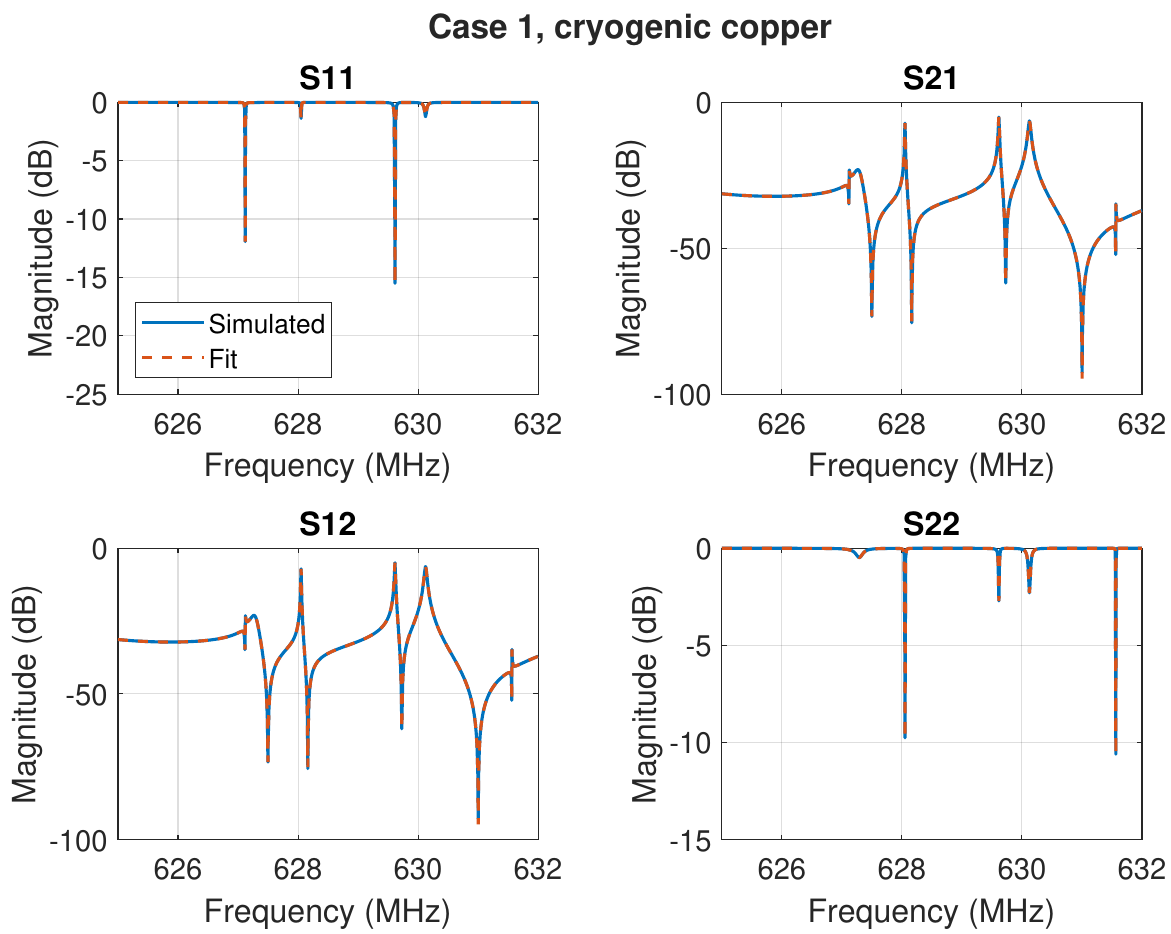} 
}
\hfill
\subfloat[]{
    \includegraphics[width=0.49\textwidth]{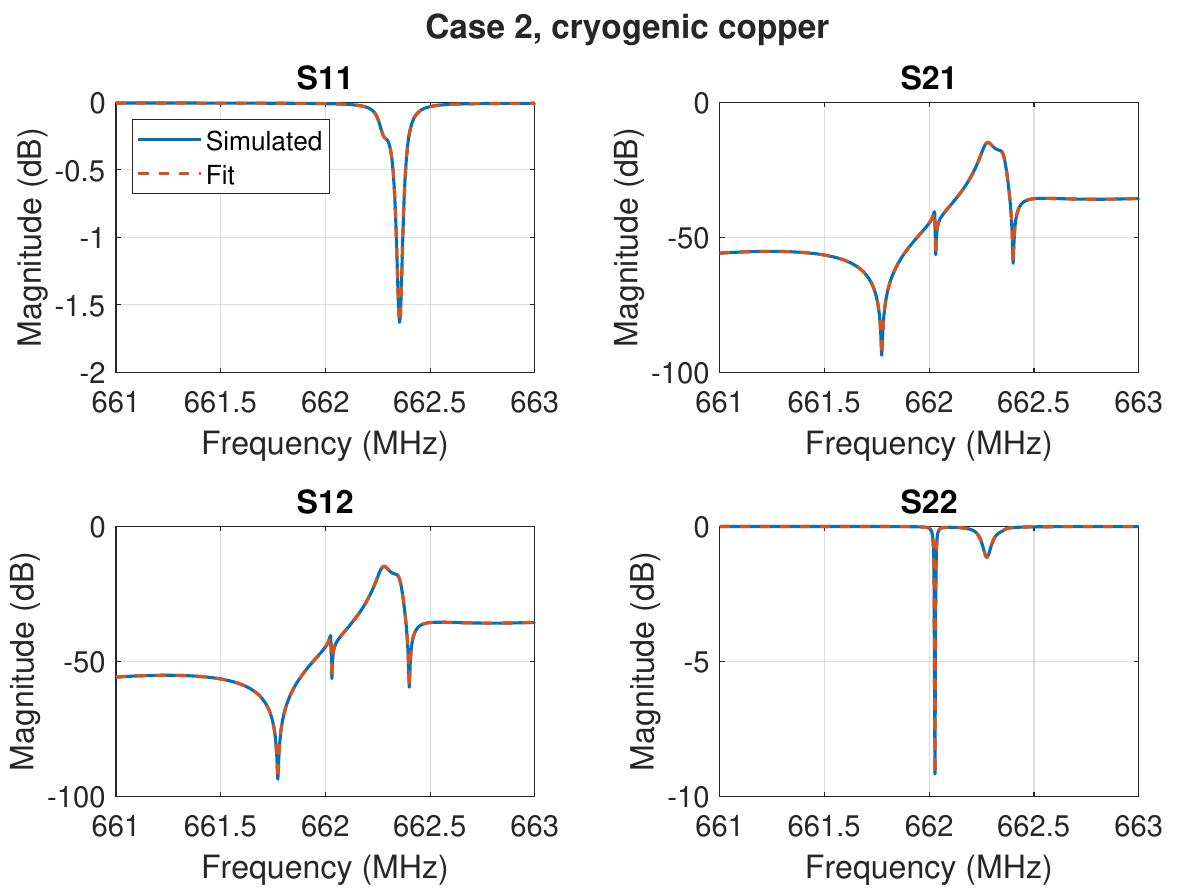}
}
    \vspace{0.4cm}
\subfloat[]{
    \includegraphics[width=0.45\textwidth]{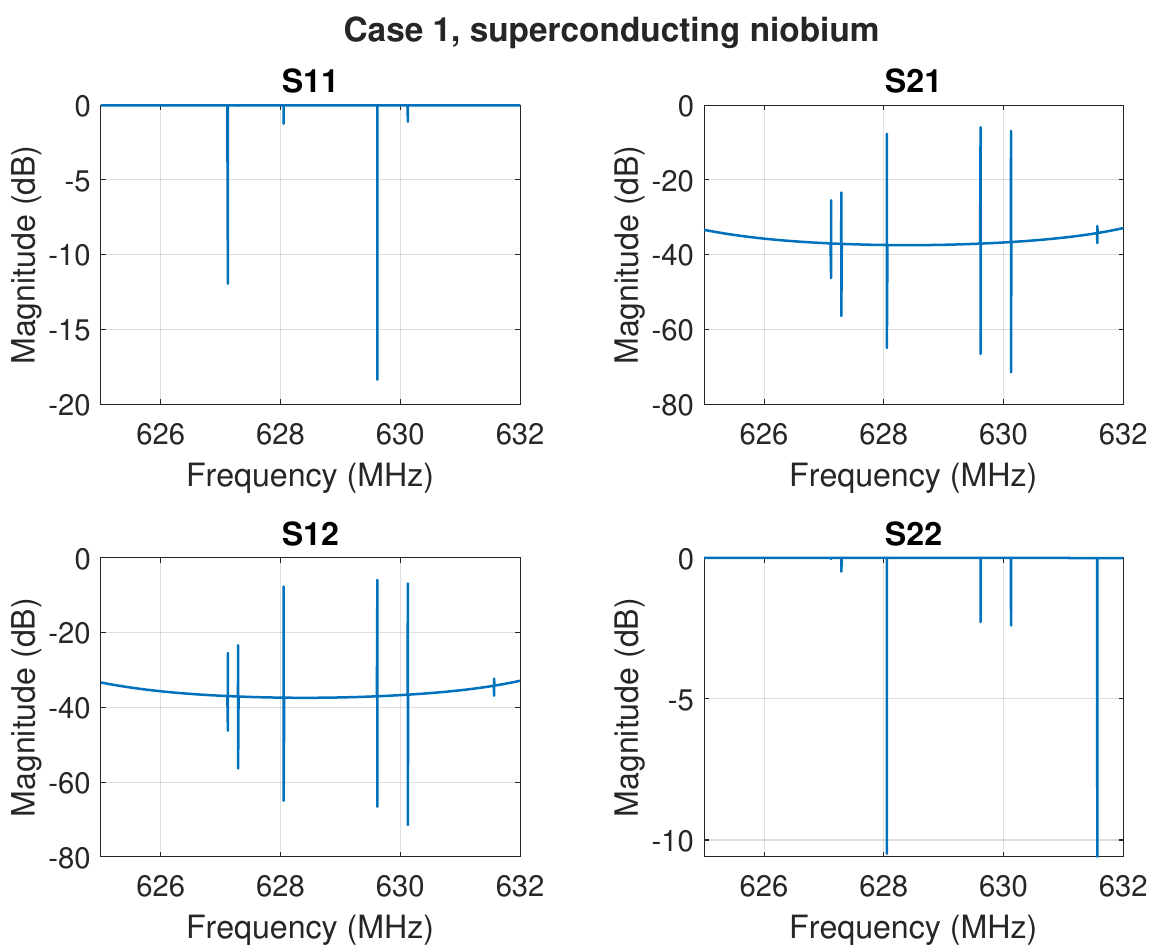} 
}
\hfill
\subfloat[]{
    \includegraphics[width=0.49\textwidth]{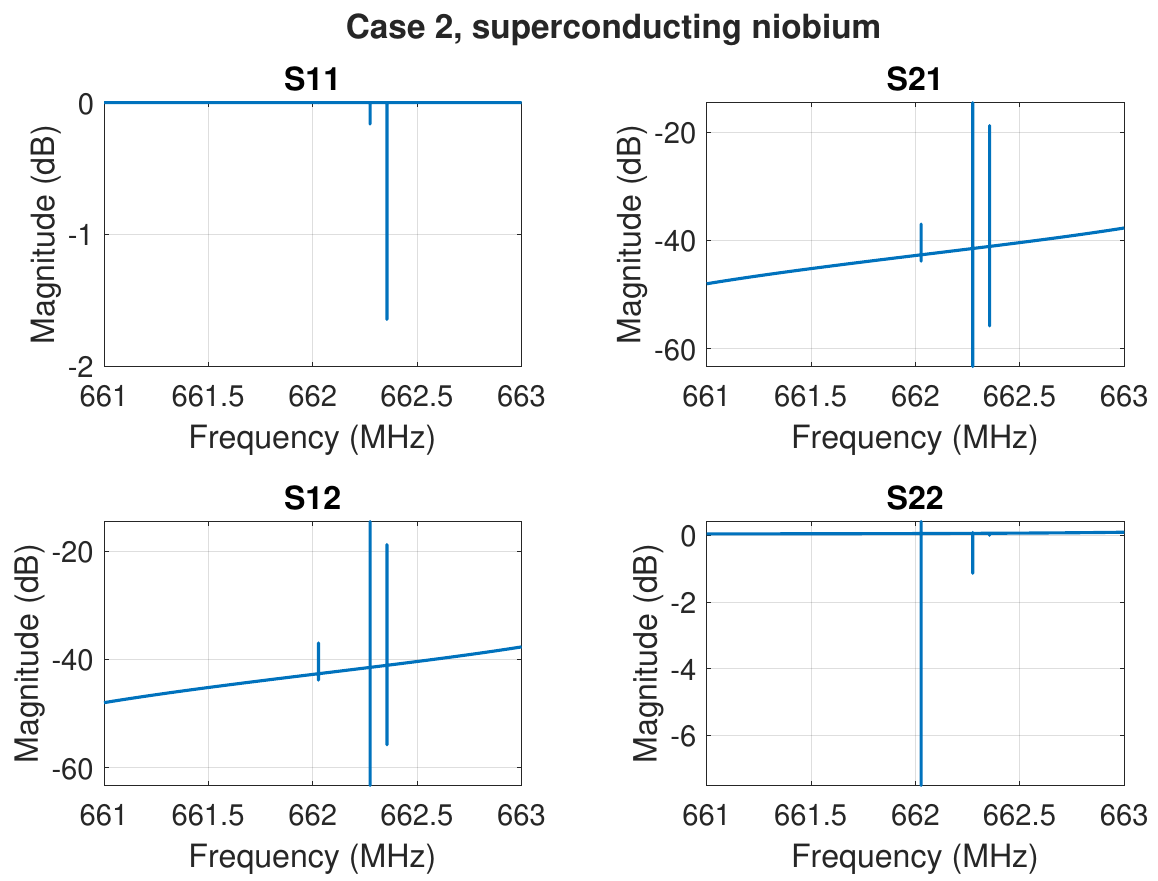}
}

\caption{Rational modal fitting and superconducting rescaling of the scattering parameters for the two axion frequency cases considered in this work. Upper row: comparison between the simulated scattering parameters of the cryogenic copper cavity and the corresponding rational fit. Lower row: predicted scattering parameters for the superconducting niobium cavity obtained by rescaling the intrinsic quality factors while keeping the coupling coefficients, modal residues, resonant frequencies, and non-resonant background fixed.}
\label{fig:Sparameters_fit_sc}
\end{figure*}

In this way, the resonant frequencies, residues, and smooth background obtained from the rational fit provide a compact representation of the simulated response, while the quality factors can be consistently modified to predict the response expected for a SRF cavity. The resulting frequency-domain model preserves the physical interpretation of the poles and zeros of the scattering matrix resulting in a convenient framework for comparing the simulated copper response with the expected superconducting performance.

A low-order polynomial may be insufficient to reproduce a varying baseline, whereas an excessively high order could absorb resonant features that should instead be represented by poles. Therefore, $P$ must be chosen so that the polynomial captures only the smooth part of the response, leaving the narrow resonant structures to the modal term. The number of modes $M$ corresponds to the number of physical poles retained within the fitted band.

Once the modal parameters have been obtained, the loaded quality factor can be related to the intrinsic quality factor and to the external coupling coefficients. We have used the convention
\begin{equation}
\beta_{\ell,m}=\frac{Q_{0,m}}{Q_{e,\ell,m}},
\label{eq:beta_definition}
\end{equation}
where $\ell=1,2$ denotes the port index,  $Q_{e,\ell,m}$ is the external quality factor associated with port $\ell$, and the loaded quality factor of mode $m$ is given by
\begin{equation}
Q_{L,m}=\frac{Q_{0,m}}{1+\beta_{1,m}+\beta_{2,m}}.
\label{eq:loaded_q}
\end{equation}
This relation allows the fitted loaded quality factor to be interpreted in terms of intrinsic losses and port loading. For the superconducting prediction, the coupling coefficients are assumed to remain fixed, since they are mainly determined by the geometry of the ports and by the field distribution of the mode. This assumption serves as a first-order approximation, although in reality changes in the port geometry —in this case, by extracting the coaxial probe to reduce coupling— will alter the frequency response. The intrinsic quality factor is then rescaled according to the ratio between the surface resistance used in the reference simulation $R_s^{\mathrm{sim}}$, and the surface resistance of the superconducting material $R_s^{\mathrm{SRF}}$,
\begin{equation}
Q_{0,m}^{\mathrm{SRF}}
=
Q_{0,m}^{\mathrm{sim}}
\frac{R_s^{\mathrm{sim}}}{R_s^{\mathrm{SRF}}}.
\label{eq:q0_srf_scaling}
\end{equation}
The corresponding superconducting loaded quality factor is then obtained as
\begin{equation}
Q_{L,m}^{\mathrm{SRF}}
=
\frac{Q_{0,m}^{\mathrm{SRF}}}{1+\beta_{1,m}+\beta_{2,m}}.
\label{eq:ql_srf}
\end{equation}
The complete fitting and rescaling procedure is illustrated in Fig.~\ref{fig:Sparameters_fit_sc}, where the upper row shows the scattering parameters obtained for the cryogenic copper cavity together with the corresponding rational fits for the two axion frequency cases considered in this work. The lower row shows the superconducting niobium predictions obtained after rescaling the intrinsic quality factors according to Eq.~\eqref{eq:q0_srf_scaling}, also for the two cases. The rational model reproduces the simulated copper response providing the corresponding high-Q superconducting prediction without modifying the fitted modal residues and the non-resonant background.

The coupling-rescaling procedure was validated using cryogenic copper simulations for $\beta_r=5.42$, 9.23, 15.35, and 24.67. The predicted pump leakage was compared with direct electromagnetic simulations using the corresponding modified probe geometries. The mean absolute deviation was 1.8 dB, with a maximum discrepancy of 5.1 dB and deviations below 1.3 dB for $\beta_r\geq9.23$. These results support the use of the rescaled S-parameters as a first-order approximation of the modified port response.
%

\input{references.bbl}
\end{document}

%% file: references.bbl
%

%% file: main.bbl
\begin{thebibliography}{104}%
\makeatletter
\providecommand \@ifxundefined [1]{%
 \@ifx{#1\undefined}
}%
\providecommand \@ifnum [1]{%
 \ifnum #1\expandafter \@firstoftwo
 \else \expandafter \@secondoftwo
 \fi
}%
\providecommand \@ifx [1]{%
 \ifx #1\expandafter \@firstoftwo
 \else \expandafter \@secondoftwo
 \fi
}%
\providecommand \natexlab [1]{#1}%
\providecommand \enquote  [1]{``#1''}%
\providecommand \bibnamefont  [1]{#1}%
\providecommand \bibfnamefont [1]{#1}%
\providecommand \citenamefont [1]{#1}%
\providecommand \href@noop [0]{\@secondoftwo}%
\providecommand \href [0]{\begingroup \@sanitize@url \@href}%
\providecommand \@href[1]{\@@startlink{#1}\@@href}%
\providecommand \@@href[1]{\endgroup#1\@@endlink}%
\providecommand \@sanitize@url [0]{\catcode `\\12\catcode `\$12\catcode `\&12\catcode `\#12\catcode `\^12\catcode `\_12\catcode `\%12\relax}%
\providecommand \@@startlink[1]{}%
\providecommand \@@endlink[0]{}%
\providecommand \url  [0]{\begingroup\@sanitize@url \@url }%
\providecommand \@url [1]{\endgroup\@href {#1}{\urlprefix }}%
\providecommand \urlprefix  [0]{URL }%
\providecommand \Eprint [0]{\href }%
\providecommand \doibase [0]{https://doi.org/}%
\providecommand \selectlanguage [0]{\@gobble}%
\providecommand \bibinfo  [0]{\@secondoftwo}%
\providecommand \bibfield  [0]{\@secondoftwo}%
\providecommand \translation [1]{[#1]}%
\providecommand \BibitemOpen [0]{}%
\providecommand \bibitemStop [0]{}%
\providecommand \bibitemNoStop [0]{.\EOS\space}%
\providecommand \EOS [0]{\spacefactor3000\relax}%
\providecommand \BibitemShut  [1]{\csname bibitem#1\endcsname}%
\let\auto@bib@innerbib\@empty
\bibitem [{\citenamefont {Peccei}\ and\ \citenamefont {Quinn}(1977)}]{Peccei_Quinn}%
  \BibitemOpen
  \bibfield  {author} {\bibinfo {author} {\bibfnamefont {R.~D.}\ \bibnamefont {Peccei}}\ and\ \bibinfo {author} {\bibfnamefont {H.~R.}\ \bibnamefont {Quinn}},\ }\bibfield  {title} {\bibinfo {title} {$\mathrm{CP}$ {Conservation} in the {Presence} of {Pseudoparticles}},\ }\href {https://doi.org/10.1103/PhysRevLett.38.1440} {\bibfield  {journal} {\bibinfo  {journal} {Phys. Rev. Lett.}\ }\textbf {\bibinfo {volume} {38}},\ \bibinfo {pages} {1440} (\bibinfo {year} {1977})}\BibitemShut {NoStop}%
\bibitem [{\citenamefont {Kim}\ and\ \citenamefont {Carosi}(2010)}]{Kim:2008hd}%
  \BibitemOpen
  \bibfield  {author} {\bibinfo {author} {\bibfnamefont {J.~E.}\ \bibnamefont {Kim}}\ and\ \bibinfo {author} {\bibfnamefont {G.}~\bibnamefont {Carosi}},\ }\bibfield  {title} {\bibinfo {title} {{Axions and the Strong CP Problem}},\ }\href {https://doi.org/10.1103/RevModPhys.82.557} {\bibfield  {journal} {\bibinfo  {journal} {Rev. Mod. Phys.}\ }\textbf {\bibinfo {volume} {82}},\ \bibinfo {pages} {557} (\bibinfo {year} {2010})},\ \bibinfo {note} {[Erratum: Rev.Mod.Phys. 91, 049902 (2019)]},\ \Eprint {https://arxiv.org/abs/0807.3125} {arXiv:0807.3125 [hep-ph]} \BibitemShut {NoStop}%
\bibitem [{\citenamefont {Weinberg}(1978)}]{Weinberg}%
  \BibitemOpen
  \bibfield  {author} {\bibinfo {author} {\bibfnamefont {S.}~\bibnamefont {Weinberg}},\ }\bibfield  {title} {\bibinfo {title} {{A} {N}ew {L}ight {B}oson?},\ }\href {https://doi.org/10.1103/PhysRevLett.40.223} {\bibfield  {journal} {\bibinfo  {journal} {Phys. Rev. Lett.}\ }\textbf {\bibinfo {volume} {40}},\ \bibinfo {pages} {223} (\bibinfo {year} {1978})}\BibitemShut {NoStop}%
\bibitem [{\citenamefont {Wilczek}(1978)}]{Wilczek}%
  \BibitemOpen
  \bibfield  {author} {\bibinfo {author} {\bibfnamefont {F.}~\bibnamefont {Wilczek}},\ }\bibfield  {title} {\bibinfo {title} {{P}roblem of {S}trong {$P$} and {$T$} {I}nvariance in the {P}resence of {I}nstantons},\ }\href {https://doi.org/10.1103/PhysRevLett.40.279} {\bibfield  {journal} {\bibinfo  {journal} {Phys. Rev. Lett.}\ }\textbf {\bibinfo {volume} {40}},\ \bibinfo {pages} {279} (\bibinfo {year} {1978})}\BibitemShut {NoStop}%
\bibitem [{\citenamefont {Preskill}\ \emph {et~al.}(1983)\citenamefont {Preskill}, \citenamefont {Wise},\ and\ \citenamefont {Wilczek}}]{Willy_1}%
  \BibitemOpen
  \bibfield  {author} {\bibinfo {author} {\bibfnamefont {J.}~\bibnamefont {Preskill}}, \bibinfo {author} {\bibfnamefont {M.~B.}\ \bibnamefont {Wise}},\ and\ \bibinfo {author} {\bibfnamefont {F.}~\bibnamefont {Wilczek}},\ }\bibfield  {title} {\bibinfo {title} {{Cosmology of the Invisible Axion}},\ }\href {https://doi.org/10.1016/0370-2693(83)90637-8} {\bibfield  {journal} {\bibinfo  {journal} {Phys. Lett. B}\ }\textbf {\bibinfo {volume} {120}},\ \bibinfo {pages} {127} (\bibinfo {year} {1983})}\BibitemShut {NoStop}%
\bibitem [{\citenamefont {Abbott}\ and\ \citenamefont {Sikivie}(1983)}]{Willy_2}%
  \BibitemOpen
  \bibfield  {author} {\bibinfo {author} {\bibfnamefont {L.~F.}\ \bibnamefont {Abbott}}\ and\ \bibinfo {author} {\bibfnamefont {P.}~\bibnamefont {Sikivie}},\ }\bibfield  {title} {\bibinfo {title} {{A Cosmological Bound on the Invisible Axion}},\ }\href {https://doi.org/10.1016/0370-2693(83)90638-X} {\bibfield  {journal} {\bibinfo  {journal} {Phys. Lett. B}\ }\textbf {\bibinfo {volume} {120}},\ \bibinfo {pages} {133} (\bibinfo {year} {1983})}\BibitemShut {NoStop}%
\bibitem [{\citenamefont {Dine}\ and\ \citenamefont {Fischler}(1983)}]{Willy_3}%
  \BibitemOpen
  \bibfield  {author} {\bibinfo {author} {\bibfnamefont {M.}~\bibnamefont {Dine}}\ and\ \bibinfo {author} {\bibfnamefont {W.}~\bibnamefont {Fischler}},\ }\bibfield  {title} {\bibinfo {title} {{The Not So Harmless Axion}},\ }\href {https://doi.org/10.1016/0370-2693(83)90639-1} {\bibfield  {journal} {\bibinfo  {journal} {Phys. Lett. B}\ }\textbf {\bibinfo {volume} {120}},\ \bibinfo {pages} {137} (\bibinfo {year} {1983})}\BibitemShut {NoStop}%
\bibitem [{\citenamefont {Du}\ \emph {et~al.}(2018)\citenamefont {Du} \emph {et~al.}}]{ADMX}%
  \BibitemOpen
  \bibfield  {author} {\bibinfo {author} {\bibfnamefont {N.}~\bibnamefont {Du}} \emph {et~al.} (\bibinfo {collaboration} {ADMX}),\ }\bibfield  {title} {\bibinfo {title} {{{A} {S}earch for {I}nvisible {A}xion {D}ark {M}atter with the {A}xion {D}ark {M}atter {E}xperiment}},\ }\href {https://doi.org/10.1103/PhysRevLett.120.151301} {\bibfield  {journal} {\bibinfo  {journal} {Phys. Rev. Lett.}\ }\textbf {\bibinfo {volume} {120}},\ \bibinfo {pages} {151301} (\bibinfo {year} {2018})},\ \Eprint {https://arxiv.org/abs/1804.05750} {arXiv:1804.05750 [hep-ex]} \BibitemShut {NoStop}%
\bibitem [{\citenamefont {Kwon}\ \emph {et~al.}(2021)\citenamefont {Kwon} \emph {et~al.}}]{CAPP_2020}%
  \BibitemOpen
  \bibfield  {author} {\bibinfo {author} {\bibfnamefont {O.}~\bibnamefont {Kwon}} \emph {et~al.} (\bibinfo {collaboration} {CAPP}),\ }\bibfield  {title} {\bibinfo {title} {{First Results from an Axion Haloscope at CAPP around 10.7 $\mu$eV}},\ }\href {https://doi.org/10.1103/PhysRevLett.126.191802} {\bibfield  {journal} {\bibinfo  {journal} {Phys. Rev. Lett.}\ }\textbf {\bibinfo {volume} {126}},\ \bibinfo {pages} {191802} (\bibinfo {year} {2021})},\ \Eprint {https://arxiv.org/abs/2012.10764} {arXiv:2012.10764 [hep-ex]} \BibitemShut {NoStop}%
\bibitem [{\citenamefont {Ahn}\ \emph {et~al.}(2024)\citenamefont {Ahn} \emph {et~al.}}]{CAPP_2024}%
  \BibitemOpen
  \bibfield  {author} {\bibinfo {author} {\bibfnamefont {S.}~\bibnamefont {Ahn}} \emph {et~al.} (\bibinfo {collaboration} {CAPP}),\ }\bibfield  {title} {\bibinfo {title} {{Extensive Search for Axion Dark Matter over 1~GHz with CAPP'S Main Axion Experiment}},\ }\href {https://doi.org/10.1103/PhysRevX.14.031023} {\bibfield  {journal} {\bibinfo  {journal} {Phys. Rev. X}\ }\textbf {\bibinfo {volume} {14}},\ \bibinfo {pages} {031023} (\bibinfo {year} {2024})},\ \Eprint {https://arxiv.org/abs/2402.12892} {arXiv:2402.12892 [hep-ex]} \BibitemShut {NoStop}%
\bibitem [{\citenamefont {Backes}\ \emph {et~al.}(2021)\citenamefont {Backes} \emph {et~al.}}]{HAYSTAC}%
  \BibitemOpen
  \bibfield  {author} {\bibinfo {author} {\bibfnamefont {K.~M.}\ \bibnamefont {Backes}} \emph {et~al.} (\bibinfo {collaboration} {HAYSTAC}),\ }\bibfield  {title} {\bibinfo {title} {{{A} quantum-enhanced search for dark matter axions}},\ }\href {https://doi.org/10.1038/s41586-021-03226-7} {\bibfield  {journal} {\bibinfo  {journal} {Nature}\ }\textbf {\bibinfo {volume} {590}},\ \bibinfo {pages} {238} (\bibinfo {year} {2021})},\ \Eprint {https://arxiv.org/abs/2008.01853} {arXiv:2008.01853 [quant-ph]} \BibitemShut {NoStop}%
\bibitem [{\citenamefont {Rettaroli}\ \emph {et~al.}(2024)\citenamefont {Rettaroli} \emph {et~al.}}]{QUAX:2024fut}%
  \BibitemOpen
  \bibfield  {author} {\bibinfo {author} {\bibfnamefont {A.}~\bibnamefont {Rettaroli}} \emph {et~al.} (\bibinfo {collaboration} {QUAX}),\ }\bibfield  {title} {\bibinfo {title} {{Search for axion dark matter with the QUAX--LNF tunable haloscope}},\ }\href {https://doi.org/10.1103/PhysRevD.110.022008} {\bibfield  {journal} {\bibinfo  {journal} {Phys. Rev. D}\ }\textbf {\bibinfo {volume} {110}},\ \bibinfo {pages} {022008} (\bibinfo {year} {2024})},\ \Eprint {https://arxiv.org/abs/2402.19063} {arXiv:2402.19063 [physics.ins-det]} \BibitemShut {NoStop}%
\bibitem [{\citenamefont {McAllister}\ \emph {et~al.}(2017)\citenamefont {McAllister}, \citenamefont {Flower}, \citenamefont {Ivanov}, \citenamefont {Goryachev}, \citenamefont {Bourhill},\ and\ \citenamefont {Tobar}}]{ORGAN}%
  \BibitemOpen
  \bibfield  {author} {\bibinfo {author} {\bibfnamefont {B.~T.}\ \bibnamefont {McAllister}}, \bibinfo {author} {\bibfnamefont {G.}~\bibnamefont {Flower}}, \bibinfo {author} {\bibfnamefont {E.~N.}\ \bibnamefont {Ivanov}}, \bibinfo {author} {\bibfnamefont {M.}~\bibnamefont {Goryachev}}, \bibinfo {author} {\bibfnamefont {J.}~\bibnamefont {Bourhill}},\ and\ \bibinfo {author} {\bibfnamefont {M.~E.}\ \bibnamefont {Tobar}},\ }\bibfield  {title} {\bibinfo {title} {{T}he {ORGAN} experiment: {A}n axion haloscope above 15 {GH}z},\ }\href {https://doi.org/https://doi.org/10.1016/j.dark.2017.09.010} {\bibfield  {journal} {\bibinfo  {journal} {Physics of the Dark Universe}\ }\textbf {\bibinfo {volume} {18}},\ \bibinfo {pages} {67} (\bibinfo {year} {2017})}\BibitemShut {NoStop}%
\bibitem [{\citenamefont {Álvarez Melcón}\ \emph {et~al.}()\citenamefont {Álvarez Melcón}, \citenamefont {Cuendis}, \citenamefont {Cogollos}, \citenamefont {Díaz-Morcillo}, \citenamefont {Döbrich}, \citenamefont {Gallego}, \citenamefont {Gimeno}, \citenamefont {Irastorza}, \citenamefont {Lozano-Guerrero}, \citenamefont {Malbrunot}, \citenamefont {Navarro}, \citenamefont {Garay}, \citenamefont {Redondo}, \citenamefont {Vafeiadis},\ and\ \citenamefont {Wuensch}}]{RADES_1}%
  \BibitemOpen
  \bibfield  {author} {\bibinfo {author} {\bibfnamefont {A.}~\bibnamefont {Álvarez Melcón}}, \bibinfo {author} {\bibfnamefont {S.~A.}\ \bibnamefont {Cuendis}}, \bibinfo {author} {\bibfnamefont {C.}~\bibnamefont {Cogollos}}, \bibinfo {author} {\bibfnamefont {A.}~\bibnamefont {Díaz-Morcillo}}, \bibinfo {author} {\bibfnamefont {B.}~\bibnamefont {Döbrich}}, \bibinfo {author} {\bibfnamefont {J.~D.}\ \bibnamefont {Gallego}}, \bibinfo {author} {\bibfnamefont {B.}~\bibnamefont {Gimeno}}, \bibinfo {author} {\bibfnamefont {I.~G.}\ \bibnamefont {Irastorza}}, \bibinfo {author} {\bibfnamefont {A.~J.}\ \bibnamefont {Lozano-Guerrero}}, \bibinfo {author} {\bibfnamefont {C.}~\bibnamefont {Malbrunot}}, \bibinfo {author} {\bibfnamefont {P.}~\bibnamefont {Navarro}}, \bibinfo {author} {\bibfnamefont {C.~P.}\ \bibnamefont {Garay}}, \bibinfo {author} {\bibfnamefont {J.}~\bibnamefont {Redondo}}, \bibinfo {author} {\bibfnamefont {T.}~\bibnamefont {Vafeiadis}},\ and\ \bibinfo {author} {\bibfnamefont {W.}~\bibnamefont {Wuensch}},\
  }\bibfield  {title} {\bibinfo {title} {{A}xion searches with microwave filters: the {RADES} project},\ }\href {https://doi.org/10.1088/1475-7516/2018/05/040} {\bibfield  {journal} {\bibinfo  {journal} {Journal of Cosmology and Astroparticle Physics}\ }\textbf {\bibinfo {volume} {2018}}\bibinfo  {number} { ((05))},\ \bibinfo {pages} {040}}\BibitemShut {NoStop}%
\bibitem [{\citenamefont {Melc\'on}\ \emph {et~al.}()\citenamefont {Melc\'on} \emph {et~al.}}]{RADES_2}%
  \BibitemOpen
\bibfield  {number} {  }\bibfield  {author} {\bibinfo {author} {\bibfnamefont {A.~A.}\ \bibnamefont {Melc\'on}} \emph {et~al.},\ }\bibfield  {title} {\bibinfo {title} {{Scalable haloscopes for axion dark matter detection in the 30$\mu$eV range with RADES}},\ }\href {https://doi.org/10.1007/JHEP07(2020)084} {\bibfield  {journal} {\bibinfo  {journal} {JHEP}\ }\textbf {\bibinfo {volume} {07}},\ \bibinfo {pages} {084}},\ \Eprint {https://arxiv.org/abs/2002.07639} {arXiv:2002.07639 [hep-ex]} \BibitemShut {NoStop}%
\bibitem [{\citenamefont {Aja}\ \emph {et~al.}()\citenamefont {Aja} \emph {et~al.}}]{CADEx}%
  \BibitemOpen
  \bibfield  {author} {\bibinfo {author} {\bibfnamefont {B.}~\bibnamefont {Aja}} \emph {et~al.},\ }\bibfield  {title} {\bibinfo {title} {{The {C}anfranc {A}xion {D}etection {E}xperiment ({CADE}x): search for axions at 90 {GH}z with {K}inetic {I}nductance {D}etectors}},\ }\href {https://doi.org/10.1088/1475-7516/2022/11/044} {\bibfield  {journal} {\bibinfo  {journal} {JCAP}\ }\textbf {\bibinfo {volume} {11}},\ \bibinfo {pages} {044}},\ \Eprint {https://arxiv.org/abs/2206.02980} {arXiv:2206.02980 [hep-ex]} \BibitemShut {NoStop}%
\bibitem [{\citenamefont {Irastorza}\ and\ \citenamefont {Redondo}(2018)}]{IRASTORZA201889}%
  \BibitemOpen
  \bibfield  {author} {\bibinfo {author} {\bibfnamefont {I.~G.}\ \bibnamefont {Irastorza}}\ and\ \bibinfo {author} {\bibfnamefont {J.}~\bibnamefont {Redondo}},\ }\bibfield  {title} {\bibinfo {title} {New experimental approaches in the search for axion-like particles},\ }\href {https://doi.org/https://doi.org/10.1016/j.ppnp.2018.05.003} {\bibfield  {journal} {\bibinfo  {journal} {Progress in Particle and Nuclear Physics}\ }\textbf {\bibinfo {volume} {102}},\ \bibinfo {pages} {89} (\bibinfo {year} {2018})}\BibitemShut {NoStop}%
\bibitem [{\citenamefont {Bradley}\ \emph {et~al.}(2003)\citenamefont {Bradley}, \citenamefont {Clarke}, \citenamefont {Kinion}, \citenamefont {Rosenberg}, \citenamefont {van Bibber}, \citenamefont {Matsuki}, \citenamefont {M\"uck},\ and\ \citenamefont {Sikivie}}]{RevModPhys.75.777}%
  \BibitemOpen
  \bibfield  {author} {\bibinfo {author} {\bibfnamefont {R.}~\bibnamefont {Bradley}}, \bibinfo {author} {\bibfnamefont {J.}~\bibnamefont {Clarke}}, \bibinfo {author} {\bibfnamefont {D.}~\bibnamefont {Kinion}}, \bibinfo {author} {\bibfnamefont {L.~J.}\ \bibnamefont {Rosenberg}}, \bibinfo {author} {\bibfnamefont {K.}~\bibnamefont {van Bibber}}, \bibinfo {author} {\bibfnamefont {S.}~\bibnamefont {Matsuki}}, \bibinfo {author} {\bibfnamefont {M.}~\bibnamefont {M\"uck}},\ and\ \bibinfo {author} {\bibfnamefont {P.}~\bibnamefont {Sikivie}},\ }\bibfield  {title} {\bibinfo {title} {Microwave cavity searches for dark-matter axions},\ }\href {https://doi.org/10.1103/RevModPhys.75.777} {\bibfield  {journal} {\bibinfo  {journal} {Rev. Mod. Phys.}\ }\textbf {\bibinfo {volume} {75}},\ \bibinfo {pages} {777} (\bibinfo {year} {2003})}\BibitemShut {NoStop}%
\bibitem [{\citenamefont {Sikivie}\ \emph {et~al.}(2014)\citenamefont {Sikivie}, \citenamefont {Sullivan},\ and\ \citenamefont {Tanner}}]{LC_resonator1}%
  \BibitemOpen
  \bibfield  {author} {\bibinfo {author} {\bibfnamefont {P.}~\bibnamefont {Sikivie}}, \bibinfo {author} {\bibfnamefont {N.}~\bibnamefont {Sullivan}},\ and\ \bibinfo {author} {\bibfnamefont {D.~B.}\ \bibnamefont {Tanner}},\ }\bibfield  {title} {\bibinfo {title} {{Proposal for {A}xion {D}ark {M}atter {D}etection {U}sing an {LC} {C}ircuit}},\ }\href {https://doi.org/10.1103/PhysRevLett.112.131301} {\bibfield  {journal} {\bibinfo  {journal} {Phys. Rev. Lett.}\ }\textbf {\bibinfo {volume} {112}},\ \bibinfo {pages} {131301} (\bibinfo {year} {2014})},\ \Eprint {https://arxiv.org/abs/1310.8545} {arXiv:1310.8545 [hep-ph]} \BibitemShut {NoStop}%
\bibitem [{\citenamefont {Kahn}\ \emph {et~al.}(2016)\citenamefont {Kahn}, \citenamefont {Safdi},\ and\ \citenamefont {Thaler}}]{LC_resonator2}%
  \BibitemOpen
  \bibfield  {author} {\bibinfo {author} {\bibfnamefont {Y.}~\bibnamefont {Kahn}}, \bibinfo {author} {\bibfnamefont {B.~R.}\ \bibnamefont {Safdi}},\ and\ \bibinfo {author} {\bibfnamefont {J.}~\bibnamefont {Thaler}},\ }\bibfield  {title} {\bibinfo {title} {{Broadband and {R}esonant {A}pproaches to {A}xion {D}ark {M}atter {D}etection}},\ }\href {https://doi.org/10.1103/PhysRevLett.117.141801} {\bibfield  {journal} {\bibinfo  {journal} {Phys. Rev. Lett.}\ }\textbf {\bibinfo {volume} {117}},\ \bibinfo {pages} {141801} (\bibinfo {year} {2016})},\ \Eprint {https://arxiv.org/abs/1602.01086} {arXiv:1602.01086 [hep-ph]} \BibitemShut {NoStop}%
\bibitem [{\citenamefont {Lasenby}(2020)}]{Lasenby_2020}%
  \BibitemOpen
  \bibfield  {author} {\bibinfo {author} {\bibfnamefont {R.}~\bibnamefont {Lasenby}},\ }\bibfield  {title} {\bibinfo {title} {Microwave cavity searches for low-frequency axion dark matter},\ }\href {https://doi.org/10.1103/PhysRevD.102.015008} {\bibfield  {journal} {\bibinfo  {journal} {Phys. Rev. D}\ }\textbf {\bibinfo {volume} {102}},\ \bibinfo {pages} {015008} (\bibinfo {year} {2020})}\BibitemShut {NoStop}%
\bibitem [{\citenamefont {Lasenby}(2021)}]{Lasenby_2021}%
  \BibitemOpen
  \bibfield  {author} {\bibinfo {author} {\bibfnamefont {R.}~\bibnamefont {Lasenby}},\ }\bibfield  {title} {\bibinfo {title} {Parametrics of electromagnetic searches for axion dark matter},\ }\href {https://doi.org/10.1103/PhysRevD.103.075007} {\bibfield  {journal} {\bibinfo  {journal} {Phys. Rev. D}\ }\textbf {\bibinfo {volume} {103}},\ \bibinfo {pages} {075007} (\bibinfo {year} {2021})}\BibitemShut {NoStop}%
\bibitem [{\citenamefont {Berlin}\ \emph {et~al.}(2021)\citenamefont {Berlin}, \citenamefont {D'Agnolo}, \citenamefont {Ellis},\ and\ \citenamefont {Zhou}}]{Heterodyne_2021}%
  \BibitemOpen
  \bibfield  {author} {\bibinfo {author} {\bibfnamefont {A.}~\bibnamefont {Berlin}}, \bibinfo {author} {\bibfnamefont {R.~T.}\ \bibnamefont {D'Agnolo}}, \bibinfo {author} {\bibfnamefont {S.~A.~R.}\ \bibnamefont {Ellis}},\ and\ \bibinfo {author} {\bibfnamefont {K.}~\bibnamefont {Zhou}},\ }\bibfield  {title} {\bibinfo {title} {Heterodyne broadband detection of axion dark matter},\ }\href {https://doi.org/10.1103/PhysRevD.104.L111701} {\bibfield  {journal} {\bibinfo  {journal} {Phys. Rev. D}\ }\textbf {\bibinfo {volume} {104}},\ \bibinfo {pages} {L111701} (\bibinfo {year} {2021})}\BibitemShut {NoStop}%
\bibitem [{\citenamefont {Goryachev}\ \emph {et~al.}(2019)\citenamefont {Goryachev}, \citenamefont {McAllister},\ and\ \citenamefont {Tobar}}]{Upconversion_original}%
  \BibitemOpen
  \bibfield  {author} {\bibinfo {author} {\bibfnamefont {M.}~\bibnamefont {Goryachev}}, \bibinfo {author} {\bibfnamefont {B.~T.}\ \bibnamefont {McAllister}},\ and\ \bibinfo {author} {\bibfnamefont {M.~E.}\ \bibnamefont {Tobar}},\ }\bibfield  {title} {\bibinfo {title} {Axion detection with precision frequency metrology},\ }\href {https://doi.org/https://doi.org/10.1016/j.dark.2019.100345} {\bibfield  {journal} {\bibinfo  {journal} {Physics of the Dark Universe}\ }\textbf {\bibinfo {volume} {26}},\ \bibinfo {pages} {100345} (\bibinfo {year} {2019})}\BibitemShut {NoStop}%
\bibitem [{\citenamefont {Thomson}\ \emph {et~al.}(2021)\citenamefont {Thomson}, \citenamefont {McAllister}, \citenamefont {Goryachev}, \citenamefont {Ivanov},\ and\ \citenamefont {Tobar}}]{Upconversion_frequency}%
  \BibitemOpen
  \bibfield  {author} {\bibinfo {author} {\bibfnamefont {C.~A.}\ \bibnamefont {Thomson}}, \bibinfo {author} {\bibfnamefont {B.~T.}\ \bibnamefont {McAllister}}, \bibinfo {author} {\bibfnamefont {M.}~\bibnamefont {Goryachev}}, \bibinfo {author} {\bibfnamefont {E.~N.}\ \bibnamefont {Ivanov}},\ and\ \bibinfo {author} {\bibfnamefont {M.~E.}\ \bibnamefont {Tobar}},\ }\bibfield  {title} {\bibinfo {title} {Upconversion {L}oop {O}scillator {A}xion {D}etection {E}xperiment: {A} {P}recision {F}requency {I}nterferometric {A}xion {D}ark {M}atter {S}earch with a {C}ylindrical {M}icrowave {C}avity},\ }\href {https://doi.org/10.1103/PhysRevLett.126.081803} {\bibfield  {journal} {\bibinfo  {journal} {Phys. Rev. Lett.}\ }\textbf {\bibinfo {volume} {126}},\ \bibinfo {pages} {081803} (\bibinfo {year} {2021})}\BibitemShut {NoStop}%
\bibitem [{\citenamefont {Sikivie}(2013)}]{Sikivie_SRF}%
  \BibitemOpen
  \bibfield  {author} {\bibinfo {author} {\bibfnamefont {P.}~\bibnamefont {Sikivie}},\ }\href {https://arxiv.org/abs/1009.0762} {\bibinfo {title} {Superconducting {R}adio {F}requency {C}avities as {A}xion {D}ark {M}atter {D}etectors}} (\bibinfo {year} {2013}),\ \Eprint {https://arxiv.org/abs/1009.0762} {arXiv:1009.0762 [hep-ph]} \BibitemShut {NoStop}%
\bibitem [{\citenamefont {Thomson}\ \emph {et~al.}(2023)\citenamefont {Thomson}, \citenamefont {Goryachev}, \citenamefont {McAllister}, \citenamefont {Ivanov}, \citenamefont {Altin},\ and\ \citenamefont {Tobar}}]{Upconversion_australianos}%
  \BibitemOpen
  \bibfield  {author} {\bibinfo {author} {\bibfnamefont {C.~A.}\ \bibnamefont {Thomson}}, \bibinfo {author} {\bibfnamefont {M.}~\bibnamefont {Goryachev}}, \bibinfo {author} {\bibfnamefont {B.~T.}\ \bibnamefont {McAllister}}, \bibinfo {author} {\bibfnamefont {E.~N.}\ \bibnamefont {Ivanov}}, \bibinfo {author} {\bibfnamefont {P.}~\bibnamefont {Altin}},\ and\ \bibinfo {author} {\bibfnamefont {M.~E.}\ \bibnamefont {Tobar}},\ }\bibfield  {title} {\bibinfo {title} {Searching for low-mass axions using resonant upconversion},\ }\href {https://doi.org/10.1103/PhysRevD.107.112003} {\bibfield  {journal} {\bibinfo  {journal} {Phys. Rev. D}\ }\textbf {\bibinfo {volume} {107}},\ \bibinfo {pages} {112003} (\bibinfo {year} {2023})}\BibitemShut {NoStop}%
\bibitem [{\citenamefont {Berlin}\ \emph {et~al.}(2023)\citenamefont {Berlin}, \citenamefont {Blas}, \citenamefont {D'Agnolo}, \citenamefont {Ellis}, \citenamefont {Harnik}, \citenamefont {Kahn}, \citenamefont {Sch\"utte-Engel},\ and\ \citenamefont {Wentzel}}]{MAGO_2.0}%
  \BibitemOpen
  \bibfield  {author} {\bibinfo {author} {\bibfnamefont {A.}~\bibnamefont {Berlin}}, \bibinfo {author} {\bibfnamefont {D.}~\bibnamefont {Blas}}, \bibinfo {author} {\bibfnamefont {R.~T.}\ \bibnamefont {D'Agnolo}}, \bibinfo {author} {\bibfnamefont {S.~A.~R.}\ \bibnamefont {Ellis}}, \bibinfo {author} {\bibfnamefont {R.}~\bibnamefont {Harnik}}, \bibinfo {author} {\bibfnamefont {Y.}~\bibnamefont {Kahn}}, \bibinfo {author} {\bibfnamefont {J.}~\bibnamefont {Sch\"utte-Engel}},\ and\ \bibinfo {author} {\bibfnamefont {M.}~\bibnamefont {Wentzel}},\ }\bibfield  {title} {\bibinfo {title} {Electromagnetic cavities as mechanical bars for gravitational waves},\ }\href {https://doi.org/10.1103/PhysRevD.108.084058} {\bibfield  {journal} {\bibinfo  {journal} {Phys. Rev. D}\ }\textbf {\bibinfo {volume} {108}},\ \bibinfo {pages} {084058} (\bibinfo {year} {2023})}\BibitemShut {NoStop}%
\bibitem [{\citenamefont {Tobar}\ \emph {et~al.}(2022)\citenamefont {Tobar}, \citenamefont {Thomson}, \citenamefont {Campbell}, \citenamefont {Quiskamp}, \citenamefont {Bourhill}, \citenamefont {McAllister}, \citenamefont {Ivanov},\ and\ \citenamefont {Goryachev}}]{GW_upconversion}%
  \BibitemOpen
  \bibfield  {author} {\bibinfo {author} {\bibfnamefont {M.~E.}\ \bibnamefont {Tobar}}, \bibinfo {author} {\bibfnamefont {C.~A.}\ \bibnamefont {Thomson}}, \bibinfo {author} {\bibfnamefont {W.~M.}\ \bibnamefont {Campbell}}, \bibinfo {author} {\bibfnamefont {A.}~\bibnamefont {Quiskamp}}, \bibinfo {author} {\bibfnamefont {J.~F.}\ \bibnamefont {Bourhill}}, \bibinfo {author} {\bibfnamefont {B.~T.}\ \bibnamefont {McAllister}}, \bibinfo {author} {\bibfnamefont {E.~N.}\ \bibnamefont {Ivanov}},\ and\ \bibinfo {author} {\bibfnamefont {M.}~\bibnamefont {Goryachev}},\ }\bibfield  {title} {\bibinfo {title} {Comparing instrument spectral sensitivity of dissimilar electromagnetic haloscopes to axion dark matter and high frequency gravitational waves},\ }\bibfield  {journal} {\bibinfo  {journal} {Symmetry}\ }\textbf {\bibinfo {volume} {14}},\ \href {https://doi.org/10.3390/sym14102165} {10.3390/sym14102165} (\bibinfo {year} {2022})\BibitemShut {NoStop}%
\bibitem [{\citenamefont {Gu{\'e}}\ \emph {et~al.}(2026)\citenamefont {Gu{\'e}}, \citenamefont {Krokotsch},\ and\ \citenamefont {Moortgat-Pick}}]{Gue:2026kga}%
  \BibitemOpen
  \bibfield  {author} {\bibinfo {author} {\bibfnamefont {J.}~\bibnamefont {Gu{\'e}}}, \bibinfo {author} {\bibfnamefont {T.}~\bibnamefont {Krokotsch}},\ and\ \bibinfo {author} {\bibfnamefont {G.}~\bibnamefont {Moortgat-Pick}},\ }\href {https://arxiv.org/abs/2602.08507} {\bibinfo {title} {{Covariant eigenmode overlap formalism for gravitational wave signals in electromagnetic cavities}}} (\bibinfo {year} {2026}),\ \Eprint {https://arxiv.org/abs/2602.08507} {arXiv:2602.08507 [gr-qc]} \BibitemShut {NoStop}%
\bibitem [{\citenamefont {Kim}\ \emph {et~al.}(2019)\citenamefont {Kim}, \citenamefont {Kim}, \citenamefont {Jeong}, \citenamefont {Kim}, \citenamefont {Shin},\ and\ \citenamefont {Semertzidis}}]{kim_CAPP_2019}%
  \BibitemOpen
  \bibfield  {author} {\bibinfo {author} {\bibfnamefont {Y.}~\bibnamefont {Kim}}, \bibinfo {author} {\bibfnamefont {D.}~\bibnamefont {Kim}}, \bibinfo {author} {\bibfnamefont {J.}~\bibnamefont {Jeong}}, \bibinfo {author} {\bibfnamefont {J.}~\bibnamefont {Kim}}, \bibinfo {author} {\bibfnamefont {Y.~C.}\ \bibnamefont {Shin}},\ and\ \bibinfo {author} {\bibfnamefont {Y.~K.}\ \bibnamefont {Semertzidis}},\ }\bibfield  {title} {\bibinfo {title} {Effective approximation of electromagnetism for axion haloscope searches},\ }\href {https://doi.org/10.1016/j.dark.2019.100362} {\bibfield  {journal} {\bibinfo  {journal} {Physics of the Dark Universe}\ }\textbf {\bibinfo {volume} {26}},\ \bibinfo {pages} {1} (\bibinfo {year} {2019})}\BibitemShut {NoStop}%
\bibitem [{\citenamefont {Primakoff}(1951)}]{Primakoff_1}%
  \BibitemOpen
  \bibfield  {author} {\bibinfo {author} {\bibfnamefont {H.}~\bibnamefont {Primakoff}},\ }\bibfield  {title} {\bibinfo {title} {Photoproduction of neutral mesons in nuclear electric fields and the mean life of the neutral meson},\ }\href {https://doi.org/10.1103/PhysRev.81.899} {\bibfield  {journal} {\bibinfo  {journal} {Phys. Rev.}\ }\textbf {\bibinfo {volume} {81}},\ \bibinfo {pages} {899} (\bibinfo {year} {1951})}\BibitemShut {NoStop}%
\bibitem [{\citenamefont {Sikivie}(1983)}]{Sikivie_1983}%
  \BibitemOpen
  \bibfield  {author} {\bibinfo {author} {\bibfnamefont {P.}~\bibnamefont {Sikivie}},\ }\bibfield  {title} {\bibinfo {title} {{Experimental Tests of the Invisible Axion}},\ }\href {https://doi.org/10.1103/PhysRevLett.51.1415} {\bibfield  {journal} {\bibinfo  {journal} {Phys. Rev. Lett.}\ }\textbf {\bibinfo {volume} {51}},\ \bibinfo {pages} {1415} (\bibinfo {year} {1983})},\ \bibinfo {note} {[Erratum: Phys.Rev.Lett. 52, 695 (1984)]}\BibitemShut {NoStop}%
\bibitem [{\citenamefont {Jackson}(1999)}]{jackson}%
  \BibitemOpen
  \bibfield  {author} {\bibinfo {author} {\bibfnamefont {J.~D.}\ \bibnamefont {Jackson}},\ }\href@noop {} {\emph {\bibinfo {title} {Classical Electrodynamics}}},\ \bibinfo {edition} {3rd}\ ed.\ (\bibinfo  {publisher} {John Wiley and Sons, Inc.},\ \bibinfo {year} {1999})\BibitemShut {NoStop}%
\bibitem [{\citenamefont {Collin}(1992)}]{collin_FMI}%
  \BibitemOpen
  \bibfield  {author} {\bibinfo {author} {\bibfnamefont {R.~E.}\ \bibnamefont {Collin}},\ }\href@noop {} {\emph {\bibinfo {title} {Foundations for Microwave Engineering}}},\ \bibinfo {edition} {2nd}\ ed.\ (\bibinfo  {publisher} {McGraw-Hill, Inc.},\ \bibinfo {year} {1992})\BibitemShut {NoStop}%
\bibitem [{\citenamefont {Collin}(1991)}]{collin_FTGW}%
  \BibitemOpen
  \bibfield  {author} {\bibinfo {author} {\bibfnamefont {R.~E.}\ \bibnamefont {Collin}},\ }\href@noop {} {\emph {\bibinfo {title} {Field Theory of Guided Waves}}},\ \bibinfo {edition} {2nd}\ ed.\ (\bibinfo  {publisher} {IEEE Press},\ \bibinfo {year} {1991})\BibitemShut {NoStop}%
\bibitem [{\citenamefont {Sikivie}(2021)}]{Sikivie_2021}%
  \BibitemOpen
  \bibfield  {author} {\bibinfo {author} {\bibfnamefont {P.}~\bibnamefont {Sikivie}},\ }\bibfield  {title} {\bibinfo {title} {{Invisible axion search methods}},\ }\href {https://doi.org/0034-6861=2021=93(1)=015004(36)} {\bibfield  {journal} {\bibinfo  {journal} {Rev. Modern Physics}\ }\textbf {\bibinfo {volume} {93}},\ \bibinfo {pages} {015004} (\bibinfo {year} {2021})}\BibitemShut {NoStop}%
\bibitem [{\citenamefont {{Berlin}}\ \emph {et~al.}(2020)\citenamefont {{Berlin}}, \citenamefont {{D'Agnolo}}, \citenamefont {{Ellis}}, \citenamefont {{Nantista}}, \citenamefont {{Neilson}}, \citenamefont {{Schuster}}, \citenamefont {{Tantawi}}, \citenamefont {{Toro}},\ and\ \citenamefont {{Zhou}}}]{Superconducting_upconversion}%
  \BibitemOpen
  \bibfield  {author} {\bibinfo {author} {\bibfnamefont {A.}~\bibnamefont {{Berlin}}}, \bibinfo {author} {\bibfnamefont {R.~T.}\ \bibnamefont {{D'Agnolo}}}, \bibinfo {author} {\bibfnamefont {S.~A.~R.}\ \bibnamefont {{Ellis}}}, \bibinfo {author} {\bibfnamefont {C.}~\bibnamefont {{Nantista}}}, \bibinfo {author} {\bibfnamefont {J.}~\bibnamefont {{Neilson}}}, \bibinfo {author} {\bibfnamefont {P.}~\bibnamefont {{Schuster}}}, \bibinfo {author} {\bibfnamefont {S.}~\bibnamefont {{Tantawi}}}, \bibinfo {author} {\bibfnamefont {N.}~\bibnamefont {{Toro}}},\ and\ \bibinfo {author} {\bibfnamefont {K.}~\bibnamefont {{Zhou}}},\ }\bibfield  {title} {\bibinfo {title} {{Axion dark matter detection by superconducting resonant frequency conversion}},\ }\bibfield  {journal} {\bibinfo  {journal} {Journal of High Energy Physics}\ }\textbf {\bibinfo {volume} {2020}},\ \href {https://doi.org/10.1007/JHEP07(2020)088} {10.1007/JHEP07(2020)088} (\bibinfo {year} {2020}),\ \Eprint {https://arxiv.org/abs/1912.11048} {arXiv:1912.11048
  [hep-ph]} \BibitemShut {NoStop}%
\bibitem [{\citenamefont {Tobar}\ \emph {et~al.}(2024)\citenamefont {Tobar}, \citenamefont {Thomson}, \citenamefont {McAllister}, \citenamefont {Goryachev}, \citenamefont {Sokolov},\ and\ \citenamefont {Ringwald}}]{Upconversion_annalen}%
  \BibitemOpen
  \bibfield  {author} {\bibinfo {author} {\bibfnamefont {M.~E.}\ \bibnamefont {Tobar}}, \bibinfo {author} {\bibfnamefont {C.~A.}\ \bibnamefont {Thomson}}, \bibinfo {author} {\bibfnamefont {B.~T.}\ \bibnamefont {McAllister}}, \bibinfo {author} {\bibfnamefont {M.}~\bibnamefont {Goryachev}}, \bibinfo {author} {\bibfnamefont {A.~V.}\ \bibnamefont {Sokolov}},\ and\ \bibinfo {author} {\bibfnamefont {A.}~\bibnamefont {Ringwald}},\ }\bibfield  {title} {\bibinfo {title} {Sensitivity of {R}esonant {A}xion {H}aloscopes to {Q}uantum {E}lectromagnetodynamics},\ }\href {https://doi.org/https://doi.org/10.1002/andp.202200594} {\bibfield  {journal} {\bibinfo  {journal} {Annalen der Physik}\ }\textbf {\bibinfo {volume} {536}},\ \bibinfo {pages} {2200594} (\bibinfo {year} {2024})}\BibitemShut {NoStop}%
\bibitem [{\citenamefont {Hatch}\ and\ \citenamefont {Williams}(1958)}]{Hatch1958}%
  \BibitemOpen
  \bibfield  {author} {\bibinfo {author} {\bibfnamefont {A.~J.}\ \bibnamefont {Hatch}}\ and\ \bibinfo {author} {\bibfnamefont {H.~B.}\ \bibnamefont {Williams}},\ }\bibfield  {title} {\bibinfo {title} {Multipacting modes of high-frequency gaseous breakdown},\ }\href@noop {} {\bibfield  {journal} {\bibinfo  {journal} {Physical Review}\ }\textbf {\bibinfo {volume} {112}},\ \bibinfo {pages} {681} (\bibinfo {year} {1958})}\BibitemShut {NoStop}%
\bibitem [{\citenamefont {Vaughan}(1988)}]{Vaughan1988}%
  \BibitemOpen
  \bibfield  {author} {\bibinfo {author} {\bibfnamefont {J.~R.~M.}\ \bibnamefont {Vaughan}},\ }\bibfield  {title} {\bibinfo {title} {Multipactor},\ }\href@noop {} {\bibfield  {journal} {\bibinfo  {journal} {IEEE Transactions on Electron Devices}\ }\textbf {\bibinfo {volume} {35}},\ \bibinfo {pages} {1172} (\bibinfo {year} {1988})}\BibitemShut {NoStop}%
\bibitem [{\citenamefont {Woode}\ and\ \citenamefont {Petit}(1989)}]{Woode1989}%
  \BibitemOpen
  \bibfield  {author} {\bibinfo {author} {\bibfnamefont {A.}~\bibnamefont {Woode}}\ and\ \bibinfo {author} {\bibfnamefont {J.}~\bibnamefont {Petit}},\ }\href@noop {} {\emph {\bibinfo {title} {Diagnostic Investigations into the Multipactor Effect, Susceptibility Zone Measurements and Parameters Affecting a Discharge}}},\ \bibinfo {type} {Tech. Rep.}\ \bibinfo {number} {Working Paper 1556}\ (\bibinfo  {institution} {European Space Agency (ESA)},\ \bibinfo {address} {Noordwijk, The Netherlands},\ \bibinfo {year} {1989})\BibitemShut {NoStop}%
\bibitem [{\citenamefont {Kishek}\ \emph {et~al.}(1998)\citenamefont {Kishek}, \citenamefont {Lau}, \citenamefont {Ang}, \citenamefont {Valfells},\ and\ \citenamefont {Gilgenbach}}]{Kishek1998}%
  \BibitemOpen
  \bibfield  {author} {\bibinfo {author} {\bibfnamefont {R.}~\bibnamefont {Kishek}}, \bibinfo {author} {\bibfnamefont {Y.~Y.}\ \bibnamefont {Lau}}, \bibinfo {author} {\bibfnamefont {L.~K.}\ \bibnamefont {Ang}}, \bibinfo {author} {\bibfnamefont {A.}~\bibnamefont {Valfells}},\ and\ \bibinfo {author} {\bibfnamefont {R.~M.}\ \bibnamefont {Gilgenbach}},\ }\bibfield  {title} {\bibinfo {title} {Multipactor discharge on metals and dielectrics: Historical review and recent theories},\ }\href@noop {} {\bibfield  {journal} {\bibinfo  {journal} {Physics of Plasmas}\ }\textbf {\bibinfo {volume} {5}},\ \bibinfo {pages} {2120} (\bibinfo {year} {1998})}\BibitemShut {NoStop}%
\bibitem [{ECS(2003)}]{ECSS2003}%
  \BibitemOpen
  \href@noop {} {\emph {\bibinfo {title} {Multipacting Design and Test}}},\ \bibinfo {type} {Tech. Rep.}\ \bibinfo {number} {ECSS-20-01A}\ (\bibinfo  {institution} {European Cooperation for Space Standardization (ECSS)},\ \bibinfo {address} {Noordwijk, The Netherlands},\ \bibinfo {year} {2003})\ \bibinfo {note} {eSA-ESTEC, ESA Publication Division}\BibitemShut {NoStop}%
\bibitem [{\citenamefont {Perez}\ \emph {et~al.}(2009)\citenamefont {Perez} \emph {et~al.}}]{Perez2009}%
  \BibitemOpen
  \bibfield  {author} {\bibinfo {author} {\bibfnamefont {A.~M.}\ \bibnamefont {Perez}} \emph {et~al.},\ }\bibfield  {title} {\bibinfo {title} {Prediction of multipactor breakdown thresholds in coaxial transmission lines for traveling, standing, and mixed waves},\ }\href@noop {} {\bibfield  {journal} {\bibinfo  {journal} {IEEE Transactions on Plasma Science}\ }\textbf {\bibinfo {volume} {37}},\ \bibinfo {pages} {2031} (\bibinfo {year} {2009})}\BibitemShut {NoStop}%
\bibitem [{\citenamefont {Gonz{\'a}lez-Iglesias}\ \emph {et~al.}(2016)\citenamefont {Gonz{\'a}lez-Iglesias}, \citenamefont {Monerris}, \citenamefont {Mart{\'\i}nez}, \citenamefont {D{\'\i}az}, \citenamefont {Boria},\ and\ \citenamefont {Iglesias}}]{GonzalezIglesias2016}%
  \BibitemOpen
  \bibfield  {author} {\bibinfo {author} {\bibfnamefont {D.}~\bibnamefont {Gonz{\'a}lez-Iglesias}}, \bibinfo {author} {\bibfnamefont {{\'O}.}~\bibnamefont {Monerris}}, \bibinfo {author} {\bibfnamefont {B.~G.}\ \bibnamefont {Mart{\'\i}nez}}, \bibinfo {author} {\bibfnamefont {M.~E.}\ \bibnamefont {D{\'\i}az}}, \bibinfo {author} {\bibfnamefont {V.~E.}\ \bibnamefont {Boria}},\ and\ \bibinfo {author} {\bibfnamefont {P.~M.}\ \bibnamefont {Iglesias}},\ }\bibfield  {title} {\bibinfo {title} {Multipactor rf breakdown in coaxial transmission lines with digitally modulated signals},\ }\href@noop {} {\bibfield  {journal} {\bibinfo  {journal} {IEEE Transactions on Electron Devices}\ }\textbf {\bibinfo {volume} {63}},\ \bibinfo {pages} {4096} (\bibinfo {year} {2016})}\BibitemShut {NoStop}%
\bibitem [{\citenamefont {Semenov}\ \emph {et~al.}(2018)\citenamefont {Semenov}, \citenamefont {Rakova}, \citenamefont {Sorolla}, \citenamefont {González-Iglesias}, \citenamefont {Monerris}, \citenamefont {Gimeno}, \citenamefont {Puech},\ and\ \citenamefont {Sombrin}}]{Semenov2018Multipactor}%
  \BibitemOpen
  \bibfield  {author} {\bibinfo {author} {\bibfnamefont {V.~E.}\ \bibnamefont {Semenov}}, \bibinfo {author} {\bibfnamefont {E.~I.}\ \bibnamefont {Rakova}}, \bibinfo {author} {\bibfnamefont {E.}~\bibnamefont {Sorolla}}, \bibinfo {author} {\bibfnamefont {D.}~\bibnamefont {González-Iglesias}}, \bibinfo {author} {\bibfnamefont {{\'O}.}~\bibnamefont {Monerris}}, \bibinfo {author} {\bibfnamefont {B.}~\bibnamefont {Gimeno}}, \bibinfo {author} {\bibfnamefont {J.}~\bibnamefont {Puech}},\ and\ \bibinfo {author} {\bibfnamefont {J.~B.}\ \bibnamefont {Sombrin}},\ }\bibfield  {title} {\bibinfo {title} {Enhancement of the multipactor threshold inside nonrectangular iris},\ }\href@noop {} {\bibfield  {journal} {\bibinfo  {journal} {IEEE Transactions on Electron Devices}\ }\textbf {\bibinfo {volume} {65}} (\bibinfo {year} {2018})}\BibitemShut {NoStop}%
\bibitem [{\citenamefont {Vague}\ \emph {et~al.}(2018)\citenamefont {Vague}, \citenamefont {Melgarejo}, \citenamefont {Guglielmi}, \citenamefont {Boria}, \citenamefont {Anza}, \citenamefont {Vicente}, \citenamefont {Moreno}, \citenamefont {Taroncher}, \citenamefont {Mart{\'\i}nez},\ and\ \citenamefont {Raboso}}]{Vague2018}%
  \BibitemOpen
  \bibfield  {author} {\bibinfo {author} {\bibfnamefont {J.}~\bibnamefont {Vague}}, \bibinfo {author} {\bibfnamefont {J.~C.}\ \bibnamefont {Melgarejo}}, \bibinfo {author} {\bibfnamefont {M.}~\bibnamefont {Guglielmi}}, \bibinfo {author} {\bibfnamefont {V.~E.}\ \bibnamefont {Boria}}, \bibinfo {author} {\bibfnamefont {S.}~\bibnamefont {Anza}}, \bibinfo {author} {\bibfnamefont {C.}~\bibnamefont {Vicente}}, \bibinfo {author} {\bibfnamefont {M.~R.}\ \bibnamefont {Moreno}}, \bibinfo {author} {\bibfnamefont {M.}~\bibnamefont {Taroncher}}, \bibinfo {author} {\bibfnamefont {B.~G.}\ \bibnamefont {Mart{\'\i}nez}},\ and\ \bibinfo {author} {\bibfnamefont {D.}~\bibnamefont {Raboso}},\ }\bibfield  {title} {\bibinfo {title} {Multipactor effect characterization of dielectric materials for space applications},\ }\href@noop {} {\bibfield  {journal} {\bibinfo  {journal} {IEEE Transactions on Microwave Theory and Techniques}\ }\textbf {\bibinfo {volume} {66}} (\bibinfo {year} {2018})}\BibitemShut {NoStop}%
\bibitem [{\citenamefont {Gonz{\'a}lez-Iglesias}\ \emph {et~al.}(2024)\citenamefont {Gonz{\'a}lez-Iglesias}, \citenamefont {Gimeno}, \citenamefont {Esperante}, \citenamefont {Mart{\'\i}nez-Reviriego}, \citenamefont {Mart{\'\i}n-Luna}, \citenamefont {Fuster-Mart{\'\i}nez}, \citenamefont {Blanch}, \citenamefont {Mart{\'\i}nez}, \citenamefont {Menendez}, \citenamefont {Fuster} \emph {et~al.}}]{GonzalezIglesias2024}%
  \BibitemOpen
  \bibfield  {author} {\bibinfo {author} {\bibfnamefont {D.}~\bibnamefont {Gonz{\'a}lez-Iglesias}}, \bibinfo {author} {\bibfnamefont {B.}~\bibnamefont {Gimeno}}, \bibinfo {author} {\bibfnamefont {D.}~\bibnamefont {Esperante}}, \bibinfo {author} {\bibfnamefont {P.}~\bibnamefont {Mart{\'\i}nez-Reviriego}}, \bibinfo {author} {\bibfnamefont {P.}~\bibnamefont {Mart{\'\i}n-Luna}}, \bibinfo {author} {\bibfnamefont {N.}~\bibnamefont {Fuster-Mart{\'\i}nez}}, \bibinfo {author} {\bibfnamefont {C.}~\bibnamefont {Blanch}}, \bibinfo {author} {\bibfnamefont {E.}~\bibnamefont {Mart{\'\i}nez}}, \bibinfo {author} {\bibfnamefont {A.}~\bibnamefont {Menendez}}, \bibinfo {author} {\bibfnamefont {J.}~\bibnamefont {Fuster}}, \emph {et~al.},\ }\bibfield  {title} {\bibinfo {title} {Non-resonant ultra-fast multipactor regime in dielectric-assist accelerating structures},\ }\href@noop {} {\bibfield  {journal} {\bibinfo  {journal} {Results in Physics}\ }\textbf {\bibinfo {volume} {56}},\ \bibinfo {pages} {107245} (\bibinfo {year}
  {2024})}\BibitemShut {NoStop}%
\bibitem [{\citenamefont {Kilpatrick}(1957)}]{Kilpatrick1957}%
  \BibitemOpen
  \bibfield  {author} {\bibinfo {author} {\bibfnamefont {W.~D.}\ \bibnamefont {Kilpatrick}},\ }\bibfield  {title} {\bibinfo {title} {Criterion for vacuum sparking designed to include both rf and dc},\ }\href@noop {} {\bibfield  {journal} {\bibinfo  {journal} {Review of Scientific Instruments}\ }\textbf {\bibinfo {volume} {28}},\ \bibinfo {pages} {824} (\bibinfo {year} {1957})}\BibitemShut {NoStop}%
\bibitem [{\citenamefont {Germain}\ and\ \citenamefont {Rohrbach}(1968)}]{Germain1968}%
  \BibitemOpen
  \bibfield  {author} {\bibinfo {author} {\bibfnamefont {C.}~\bibnamefont {Germain}}\ and\ \bibinfo {author} {\bibfnamefont {F.}~\bibnamefont {Rohrbach}},\ }\bibfield  {title} {\bibinfo {title} {High voltage breakdown in vacuum},\ }\href@noop {} {\bibfield  {journal} {\bibinfo  {journal} {Vacuum}\ }\textbf {\bibinfo {volume} {18}},\ \bibinfo {pages} {371} (\bibinfo {year} {1968})}\BibitemShut {NoStop}%
\bibitem [{\citenamefont {Bane}\ \emph {et~al.}(2005)\citenamefont {Bane}, \citenamefont {Dolgashev}, \citenamefont {Raubenheimer}, \citenamefont {Stupakov},\ and\ \citenamefont {Wu}}]{Bane2005}%
  \BibitemOpen
  \bibfield  {author} {\bibinfo {author} {\bibfnamefont {K.~L.~F.}\ \bibnamefont {Bane}}, \bibinfo {author} {\bibfnamefont {V.~A.}\ \bibnamefont {Dolgashev}}, \bibinfo {author} {\bibfnamefont {T.}~\bibnamefont {Raubenheimer}}, \bibinfo {author} {\bibfnamefont {G.~V.}\ \bibnamefont {Stupakov}},\ and\ \bibinfo {author} {\bibfnamefont {J.}~\bibnamefont {Wu}},\ }\bibfield  {title} {\bibinfo {title} {Dark currents and their effect on the primary beam in an x-band linac},\ }\href@noop {} {\bibfield  {journal} {\bibinfo  {journal} {Physical Review Special Topics - Accelerators and Beams}\ }\textbf {\bibinfo {volume} {8}},\ \bibinfo {pages} {064401} (\bibinfo {year} {2005})}\BibitemShut {NoStop}%
\bibitem [{\citenamefont {Grudiev}\ \emph {et~al.}(2009)\citenamefont {Grudiev}, \citenamefont {Calatroni},\ and\ \citenamefont {Wuensch}}]{Grudiev2009}%
  \BibitemOpen
  \bibfield  {author} {\bibinfo {author} {\bibfnamefont {A.}~\bibnamefont {Grudiev}}, \bibinfo {author} {\bibfnamefont {S.}~\bibnamefont {Calatroni}},\ and\ \bibinfo {author} {\bibfnamefont {W.}~\bibnamefont {Wuensch}},\ }\bibfield  {title} {\bibinfo {title} {New local field quantity describing the high gradient limit of accelerating structures},\ }\href@noop {} {\bibfield  {journal} {\bibinfo  {journal} {Physical Review Special Topics - Accelerators and Beams}\ }\textbf {\bibinfo {volume} {12}},\ \bibinfo {pages} {102001} (\bibinfo {year} {2009})}\BibitemShut {NoStop}%
\bibitem [{\citenamefont {Wuensch}\ \emph {et~al.}(2017)\citenamefont {Wuensch}, \citenamefont {Degiovanni}, \citenamefont {Calatroni}, \citenamefont {Korsb{\"a}ck}, \citenamefont {Djurabekova}, \citenamefont {Rajam{\"a}ki},\ and\ \citenamefont {Giner-Navarro}}]{Wuensch2017}%
  \BibitemOpen
  \bibfield  {author} {\bibinfo {author} {\bibfnamefont {W.}~\bibnamefont {Wuensch}}, \bibinfo {author} {\bibfnamefont {A.}~\bibnamefont {Degiovanni}}, \bibinfo {author} {\bibfnamefont {S.}~\bibnamefont {Calatroni}}, \bibinfo {author} {\bibfnamefont {A.}~\bibnamefont {Korsb{\"a}ck}}, \bibinfo {author} {\bibfnamefont {F.}~\bibnamefont {Djurabekova}}, \bibinfo {author} {\bibfnamefont {R.}~\bibnamefont {Rajam{\"a}ki}},\ and\ \bibinfo {author} {\bibfnamefont {J.}~\bibnamefont {Giner-Navarro}},\ }\bibfield  {title} {\bibinfo {title} {Statistics of vacuum breakdown in the high-gradient and low-rate regime},\ }\href@noop {} {\bibfield  {journal} {\bibinfo  {journal} {Physical Review Accelerators and Beams}\ }\textbf {\bibinfo {volume} {20}},\ \bibinfo {pages} {011007} (\bibinfo {year} {2017})}\BibitemShut {NoStop}%
\bibitem [{\citenamefont {Mart{\'\i}nez-Reviriego}\ \emph {et~al.}(2023)\citenamefont {Mart{\'\i}nez-Reviriego}, \citenamefont {Fuster-Mart{\'\i}nez}, \citenamefont {Esperante}, \citenamefont {Boronat}, \citenamefont {Gimeno}, \citenamefont {Blanch}, \citenamefont {Gonz{\'a}lez-Iglesias}, \citenamefont {Mart{\'\i}n-Luna}, \citenamefont {Mart{\'\i}nez}, \citenamefont {Menendez}, \citenamefont {Pedraza}, \citenamefont {Fern{\'a}ndez}, \citenamefont {Fuster}, \citenamefont {Grudiev}, \citenamefont {Lasheras},\ and\ \citenamefont {Wuensch}}]{MartinezReviriego2023}%
  \BibitemOpen
  \bibfield  {author} {\bibinfo {author} {\bibfnamefont {P.}~\bibnamefont {Mart{\'\i}nez-Reviriego}}, \bibinfo {author} {\bibfnamefont {N.}~\bibnamefont {Fuster-Mart{\'\i}nez}}, \bibinfo {author} {\bibfnamefont {D.}~\bibnamefont {Esperante}}, \bibinfo {author} {\bibfnamefont {M.}~\bibnamefont {Boronat}}, \bibinfo {author} {\bibfnamefont {B.}~\bibnamefont {Gimeno}}, \bibinfo {author} {\bibfnamefont {C.}~\bibnamefont {Blanch}}, \bibinfo {author} {\bibfnamefont {D.}~\bibnamefont {Gonz{\'a}lez-Iglesias}}, \bibinfo {author} {\bibfnamefont {P.}~\bibnamefont {Mart{\'\i}n-Luna}}, \bibinfo {author} {\bibfnamefont {E.}~\bibnamefont {Mart{\'\i}nez}}, \bibinfo {author} {\bibfnamefont {A.}~\bibnamefont {Menendez}}, \bibinfo {author} {\bibfnamefont {L.}~\bibnamefont {Pedraza}}, \bibinfo {author} {\bibfnamefont {J.}~\bibnamefont {Fern{\'a}ndez}}, \bibinfo {author} {\bibfnamefont {J.}~\bibnamefont {Fuster}}, \bibinfo {author} {\bibfnamefont {A.}~\bibnamefont {Grudiev}}, \bibinfo {author} {\bibfnamefont {N.~C.}\ \bibnamefont
  {Lasheras}},\ and\ \bibinfo {author} {\bibfnamefont {W.}~\bibnamefont {Wuensch}},\ }\bibfield  {title} {\bibinfo {title} {High-power performance studies of an s-band high-gradient accelerating cavity for medical applications},\ }\href@noop {} {\bibfield  {journal} {\bibinfo  {journal} {Nuclear Engineering and Technology}\ } (\bibinfo {year} {2023})}\BibitemShut {NoStop}%
\bibitem [{\citenamefont {Bogorad}\ \emph {et~al.}(2019)\citenamefont {Bogorad}, \citenamefont {Hook}, \citenamefont {Kahn},\ and\ \citenamefont {Soreq}}]{SRF_Borogad}%
  \BibitemOpen
  \bibfield  {author} {\bibinfo {author} {\bibfnamefont {Z.}~\bibnamefont {Bogorad}}, \bibinfo {author} {\bibfnamefont {A.}~\bibnamefont {Hook}}, \bibinfo {author} {\bibfnamefont {Y.}~\bibnamefont {Kahn}},\ and\ \bibinfo {author} {\bibfnamefont {Y.}~\bibnamefont {Soreq}},\ }\bibfield  {title} {\bibinfo {title} {Probing {A}xionlike {P}articles and the {A}xiverse with {S}uperconducting {R}adio-{F}requency {C}avities},\ }\href {https://doi.org/10.1103/PhysRevLett.123.021801} {\bibfield  {journal} {\bibinfo  {journal} {Phys. Rev. Lett.}\ }\textbf {\bibinfo {volume} {123}},\ \bibinfo {pages} {021801} (\bibinfo {year} {2019})}\BibitemShut {NoStop}%
\bibitem [{\citenamefont {Janish}\ \emph {et~al.}(2019)\citenamefont {Janish}, \citenamefont {Narayan}, \citenamefont {Rajendran},\ and\ \citenamefont {Riggins}}]{SRF_Janish}%
  \BibitemOpen
  \bibfield  {author} {\bibinfo {author} {\bibfnamefont {R.}~\bibnamefont {Janish}}, \bibinfo {author} {\bibfnamefont {V.}~\bibnamefont {Narayan}}, \bibinfo {author} {\bibfnamefont {S.}~\bibnamefont {Rajendran}},\ and\ \bibinfo {author} {\bibfnamefont {P.}~\bibnamefont {Riggins}},\ }\bibfield  {title} {\bibinfo {title} {Axion production and detection with superconducting rf cavities},\ }\href {https://doi.org/10.1103/PhysRevD.100.015036} {\bibfield  {journal} {\bibinfo  {journal} {Phys. Rev. D}\ }\textbf {\bibinfo {volume} {100}},\ \bibinfo {pages} {015036} (\bibinfo {year} {2019})}\BibitemShut {NoStop}%
\bibitem [{\citenamefont {Posen}\ and\ \citenamefont {Hall}(2017)}]{NbSn_paper}%
  \BibitemOpen
  \bibfield  {author} {\bibinfo {author} {\bibfnamefont {S.}~\bibnamefont {Posen}}\ and\ \bibinfo {author} {\bibfnamefont {D.~L.}\ \bibnamefont {Hall}},\ }\bibfield  {title} {\bibinfo {title} {{Nb$_3$Sn superconducting radiofrequency cavities: fabrication, results, properties, and prospects}},\ }\href {https://doi.org/10.1088/1361-6668/30/3/033004} {\bibfield  {journal} {\bibinfo  {journal} {Supercond. Sci. Technol.}\ }\textbf {\bibinfo {volume} {30}},\ \bibinfo {pages} {033004} (\bibinfo {year} {2017})}\BibitemShut {NoStop}%
\bibitem [{\citenamefont {Romanenko}\ and\ \citenamefont {Schuster}(2017)}]{Romanenko2017}%
  \BibitemOpen
  \bibfield  {author} {\bibinfo {author} {\bibfnamefont {A.}~\bibnamefont {Romanenko}}\ and\ \bibinfo {author} {\bibfnamefont {D.~I.}\ \bibnamefont {Schuster}},\ }\bibfield  {title} {\bibinfo {title} {{Understanding Quality Factor Degradation in Superconducting Niobium Cavities at Low Microwave Field Amplitudes}},\ }\href {https://doi.org/10.1103/PhysRevLett.119.264801} {\bibfield  {journal} {\bibinfo  {journal} {Phys. Rev. Lett.}\ }\textbf {\bibinfo {volume} {119}},\ \bibinfo {pages} {264801} (\bibinfo {year} {2017})}\BibitemShut {NoStop}%
\bibitem [{\citenamefont {Ciovati}\ and\ \citenamefont {Kneisel}(2006)}]{Ciovati2006}%
  \BibitemOpen
  \bibfield  {author} {\bibinfo {author} {\bibfnamefont {G.}~\bibnamefont {Ciovati}}\ and\ \bibinfo {author} {\bibfnamefont {P.}~\bibnamefont {Kneisel}},\ }\bibfield  {title} {\bibinfo {title} {{Measurement of the high-field Q drop in the TM010 and TE011 modes in a niobium cavity}},\ }\href {https://doi.org/10.1103/PhysRevSTAB.9.042001} {\bibfield  {journal} {\bibinfo  {journal} {Phys. Rev. ST Accel. Beams}\ }\textbf {\bibinfo {volume} {9}},\ \bibinfo {pages} {042001} (\bibinfo {year} {2006})}\BibitemShut {NoStop}%
\bibitem [{\citenamefont {Gonnella}\ \emph {et~al.}(2016)\citenamefont {Gonnella}, \citenamefont {Kaufman},\ and\ \citenamefont {Liepe}}]{Gonnella2016}%
  \BibitemOpen
  \bibfield  {author} {\bibinfo {author} {\bibfnamefont {D.}~\bibnamefont {Gonnella}}, \bibinfo {author} {\bibfnamefont {J.}~\bibnamefont {Kaufman}},\ and\ \bibinfo {author} {\bibfnamefont {M.}~\bibnamefont {Liepe}},\ }\bibfield  {title} {\bibinfo {title} {{Sensitivity of Niobium Superconducting Cavities to Trapped Magnetic Flux Dissipation}},\ }\href {https://doi.org/10.1063/1.4941944} {\bibfield  {journal} {\bibinfo  {journal} {J. Appl. Phys.}\ }\textbf {\bibinfo {volume} {119}},\ \bibinfo {pages} {073904} (\bibinfo {year} {2016})}\BibitemShut {NoStop}%
\bibitem [{\citenamefont {Sharping}\ \emph {et~al.}(2020)\citenamefont {Sharping}, \citenamefont {Pate}, \citenamefont {Parker}, \citenamefont {Thompson}, \citenamefont {Martinez}, \citenamefont {Castelli},\ and\ \citenamefont {Chiao}}]{Sharping2020}%
  \BibitemOpen
  \bibfield  {author} {\bibinfo {author} {\bibfnamefont {J.~E.}\ \bibnamefont {Sharping}}, \bibinfo {author} {\bibfnamefont {J.}~\bibnamefont {Pate}}, \bibinfo {author} {\bibfnamefont {J.}~\bibnamefont {Parker}}, \bibinfo {author} {\bibfnamefont {J.~J.}\ \bibnamefont {Thompson}}, \bibinfo {author} {\bibfnamefont {L.~A.}\ \bibnamefont {Martinez}}, \bibinfo {author} {\bibfnamefont {A.~R.}\ \bibnamefont {Castelli}},\ and\ \bibinfo {author} {\bibfnamefont {R.~Y.}\ \bibnamefont {Chiao}},\ }\bibfield  {title} {\bibinfo {title} {{Joints and shape imperfections in high-Q 3D SRF cavities for RF optomechanics}},\ }\href {https://doi.org/10.1063/5.0007451} {\bibfield  {journal} {\bibinfo  {journal} {J. Appl. Phys.}\ }\textbf {\bibinfo {volume} {128}},\ \bibinfo {pages} {073906} (\bibinfo {year} {2020})}\BibitemShut {NoStop}%
\bibitem [{\citenamefont {Bernard}\ \emph {et~al.}(2000)\citenamefont {Bernard}, \citenamefont {Gemme}, \citenamefont {Parodi},\ and\ \citenamefont {Picasso}}]{Bernard2000PACO}%
  \BibitemOpen
  \bibfield  {author} {\bibinfo {author} {\bibfnamefont {P.}~\bibnamefont {Bernard}}, \bibinfo {author} {\bibfnamefont {G.}~\bibnamefont {Gemme}}, \bibinfo {author} {\bibfnamefont {R.}~\bibnamefont {Parodi}},\ and\ \bibinfo {author} {\bibfnamefont {E.}~\bibnamefont {Picasso}},\ }\href@noop {} {\bibinfo {title} {The {RF} control and detection system for {PACO} the parametric converter detector}} (\bibinfo {year} {2000}),\ \Eprint {https://arxiv.org/abs/physics/0004031} {arXiv:physics/0004031 [physics.ins-det]} \BibitemShut {NoStop}%
\bibitem [{\citenamefont {Pozar}(2012)}]{pozar}%
  \BibitemOpen
  \bibfield  {author} {\bibinfo {author} {\bibfnamefont {D.~M.}\ \bibnamefont {Pozar}},\ }\href@noop {} {\emph {\bibinfo {title} {Microwave Engineering}}},\ \bibinfo {edition} {4th}\ ed.\ (\bibinfo  {publisher} {John Wiley and Sons, Inc.},\ \bibinfo {year} {2012})\BibitemShut {NoStop}%
\bibitem [{\citenamefont {Kim}\ \emph {et~al.}()\citenamefont {Kim}, \citenamefont {Jeong}, \citenamefont {Youn}, \citenamefont {Kim},\ and\ \citenamefont {Semertzidis}}]{CAPP_revisiting}%
  \BibitemOpen
  \bibfield  {author} {\bibinfo {author} {\bibfnamefont {D.}~\bibnamefont {Kim}}, \bibinfo {author} {\bibfnamefont {J.}~\bibnamefont {Jeong}}, \bibinfo {author} {\bibfnamefont {S.}~\bibnamefont {Youn}}, \bibinfo {author} {\bibfnamefont {Y.}~\bibnamefont {Kim}},\ and\ \bibinfo {author} {\bibfnamefont {Y.}~\bibnamefont {Semertzidis}},\ }\bibfield  {title} {\bibinfo {title} {Revisiting the detection rate for axion haloscopes},\ }\href {https://doi.org/10.1088/1475-7516/2020/03/066} {\bibfield  {journal} {\bibinfo  {journal} {Journal of Cosmology and Astroparticle Physics}\ }\textbf {\bibinfo {volume} {2020}},\ \bibinfo {pages} {066}}\BibitemShut {NoStop}%
\bibitem [{\citenamefont {Conciauro}\ \emph {et~al.}(2000)\citenamefont {Conciauro}, \citenamefont {Guglielimi},\ and\ \citenamefont {Sorrentino}}]{conciauro}%
  \BibitemOpen
  \bibfield  {author} {\bibinfo {author} {\bibfnamefont {G.}~\bibnamefont {Conciauro}}, \bibinfo {author} {\bibfnamefont {M.}~\bibnamefont {Guglielimi}},\ and\ \bibinfo {author} {\bibfnamefont {R.}~\bibnamefont {Sorrentino}},\ }\href@noop {} {\emph {\bibinfo {title} {Advanced Modal Analysis. CAD Techniques for Waveguide Components and Filters}}},\ \bibinfo {edition} {1st}\ ed.\ (\bibinfo  {publisher} {John Wiley and Sons, Ltd},\ \bibinfo {year} {2000})\BibitemShut {NoStop}%
\bibitem [{\citenamefont {Arcioni}\ \emph {et~al.}(1995)\citenamefont {Arcioni}, \citenamefont {Bressan},\ and\ \citenamefont {Perregrini}}]{birme3d_3Dcavities}%
  \BibitemOpen
  \bibfield  {author} {\bibinfo {author} {\bibfnamefont {P.}~\bibnamefont {Arcioni}}, \bibinfo {author} {\bibfnamefont {M.}~\bibnamefont {Bressan}},\ and\ \bibinfo {author} {\bibfnamefont {L.}~\bibnamefont {Perregrini}},\ }\bibfield  {title} {\bibinfo {title} {A new boundary integral approach to the determination of resonant modes of arbitrarily shaped cavities},\ }\href {https://doi.org/10.1109/22.402270} {\bibfield  {journal} {\bibinfo  {journal} {IEEE Transactions on Microwave Theory and Techniques}\ }\textbf {\bibinfo {volume} {MTT-43}},\ \bibinfo {pages} {1848} (\bibinfo {year} {1995})}\BibitemShut {NoStop}%
\bibitem [{\citenamefont {Arcioni}\ \emph {et~al.}(2002)\citenamefont {Arcioni}, \citenamefont {Bozzi}, \citenamefont {Bressan}, \citenamefont {Conciauro},\ and\ \citenamefont {Perregrini}}]{birme3d_3Dcavities_ports}%
  \BibitemOpen
  \bibfield  {author} {\bibinfo {author} {\bibfnamefont {P.}~\bibnamefont {Arcioni}}, \bibinfo {author} {\bibfnamefont {M.}~\bibnamefont {Bozzi}}, \bibinfo {author} {\bibfnamefont {M.}~\bibnamefont {Bressan}}, \bibinfo {author} {\bibfnamefont {G.}~\bibnamefont {Conciauro}},\ and\ \bibinfo {author} {\bibfnamefont {L.}~\bibnamefont {Perregrini}},\ }\bibfield  {title} {\bibinfo {title} {Frequency/time-domain modelling of 3d waveguide structures by a bi-rme approach},\ }\href {https://doi.org/10.1002/jnm.429} {\bibfield  {journal} {\bibinfo  {journal} {International Journal of Numerical Modeling}\ }\textbf {\bibinfo {volume} {15}},\ \bibinfo {pages} {3} (\bibinfo {year} {2002})}\BibitemShut {NoStop}%
\bibitem [{\citenamefont {San-Blas}\ \emph {et~al.}(2017)\citenamefont {San-Blas}, \citenamefont {Gimeno},\ and\ \citenamefont {Boria}}]{birme3d_angel_multipactor}%
  \BibitemOpen
  \bibfield  {author} {\bibinfo {author} {\bibfnamefont {A.}~\bibnamefont {San-Blas}}, \bibinfo {author} {\bibfnamefont {B.}~\bibnamefont {Gimeno}},\ and\ \bibinfo {author} {\bibfnamefont {V.}~\bibnamefont {Boria}},\ }\bibfield  {title} {\bibinfo {title} {Study of the multipactor phenomenon using a full-wave integral equation technique},\ }\href {https://doi.org/10.1016/j.aeue.2017.06.009} {\bibfield  {journal} {\bibinfo  {journal} {International Journal of Electronics and Communications}\ }\textbf {\bibinfo {volume} {79}},\ \bibinfo {pages} {286} (\bibinfo {year} {2017})}\BibitemShut {NoStop}%
\bibitem [{\citenamefont {{Ángel A.~San~Blas}}\ \emph {et~al.}(2007)\citenamefont {{Ángel A.~San~Blas}}, \citenamefont {Mira}, \citenamefont {Boria}, \citenamefont {Gimeno}, \citenamefont {Bressan},\ and\ \citenamefont {Arcioni}}]{birme3d_angel_posts}%
  \BibitemOpen
  \bibfield  {author} {\bibinfo {author} {\bibnamefont {{Ángel A.~San~Blas}}}, \bibinfo {author} {\bibfnamefont {F.}~\bibnamefont {Mira}}, \bibinfo {author} {\bibfnamefont {V.~E.}\ \bibnamefont {Boria}}, \bibinfo {author} {\bibfnamefont {B.}~\bibnamefont {Gimeno}}, \bibinfo {author} {\bibfnamefont {M.}~\bibnamefont {Bressan}},\ and\ \bibinfo {author} {\bibfnamefont {P.}~\bibnamefont {Arcioni}},\ }\bibfield  {title} {\bibinfo {title} {On the fast and rigorous analysis of compensated waveguide junctions using off-centered partial-height metallic posts},\ }\href {https://doi.org/10.1109/TMTT.2006.886928} {\bibfield  {journal} {\bibinfo  {journal} {IEEE Transactions on Microwave Theory and Techniques}\ }\textbf {\bibinfo {volume} {55}},\ \bibinfo {pages} {168} (\bibinfo {year} {2007})}\BibitemShut {NoStop}%
\bibitem [{\citenamefont {Mira}\ \emph {et~al.}(2005)\citenamefont {Mira}, \citenamefont {Bressan}, \citenamefont {Conciauro}, \citenamefont {Gimeno},\ and\ \citenamefont {Boria}}]{birme3d_fermin}%
  \BibitemOpen
  \bibfield  {author} {\bibinfo {author} {\bibfnamefont {F.}~\bibnamefont {Mira}}, \bibinfo {author} {\bibfnamefont {M.}~\bibnamefont {Bressan}}, \bibinfo {author} {\bibfnamefont {G.}~\bibnamefont {Conciauro}}, \bibinfo {author} {\bibfnamefont {B.}~\bibnamefont {Gimeno}},\ and\ \bibinfo {author} {\bibfnamefont {V.}~\bibnamefont {Boria}},\ }\bibfield  {title} {\bibinfo {title} {Fast s-domain modeling of rectangular waveguides with radially symmetric metal insets},\ }\href {https://doi.org/10.1109/TMTT.2005.845762} {\bibfield  {journal} {\bibinfo  {journal} {IEEE Transactions on Microwave Theory and Techniques}\ }\textbf {\bibinfo {volume} {53}},\ \bibinfo {pages} {1294} (\bibinfo {year} {2005})}\BibitemShut {NoStop}%
\bibitem [{\citenamefont {Gil}\ \emph {et~al.}(2009)\citenamefont {Gil}, \citenamefont {Blas}, \citenamefont {Vicente}, \citenamefont {Gimeno}, \citenamefont {Bressan}, \citenamefont {Boria}, \citenamefont {Conciauro},\ and\ \citenamefont {Maestre}}]{birme3d_jordi_MTT}%
  \BibitemOpen
  \bibfield  {author} {\bibinfo {author} {\bibfnamefont {J.}~\bibnamefont {Gil}}, \bibinfo {author} {\bibfnamefont {A.~S.}\ \bibnamefont {Blas}}, \bibinfo {author} {\bibfnamefont {C.}~\bibnamefont {Vicente}}, \bibinfo {author} {\bibfnamefont {B.}~\bibnamefont {Gimeno}}, \bibinfo {author} {\bibfnamefont {M.}~\bibnamefont {Bressan}}, \bibinfo {author} {\bibfnamefont {V.}~\bibnamefont {Boria}}, \bibinfo {author} {\bibfnamefont {G.}~\bibnamefont {Conciauro}},\ and\ \bibinfo {author} {\bibfnamefont {M.}~\bibnamefont {Maestre}},\ }\bibfield  {title} {\bibinfo {title} {Full-wave analysis and design of dielectric-loaded waveguide filters using a state-space integral-equation method},\ }\href {https://doi.org/10.1109/TMTT.2008.2008974} {\bibfield  {journal} {\bibinfo  {journal} {IEEE Transactions on Microwave Theory and Techniques}\ }\textbf {\bibinfo {volume} {57}},\ \bibinfo {pages} {109} (\bibinfo {year} {2009})}\BibitemShut {NoStop}%
\bibitem [{\citenamefont {Gil}\ \emph {et~al.}(2006)\citenamefont {Gil}, \citenamefont {Pérez}, \citenamefont {Gimeno}, \citenamefont {Bressan}, \citenamefont {Boria},\ and\ \citenamefont {Conciauro}}]{birme3d_jordi_MWCL}%
  \BibitemOpen
  \bibfield  {author} {\bibinfo {author} {\bibfnamefont {J.}~\bibnamefont {Gil}}, \bibinfo {author} {\bibfnamefont {A.~M.}\ \bibnamefont {Pérez}}, \bibinfo {author} {\bibfnamefont {B.}~\bibnamefont {Gimeno}}, \bibinfo {author} {\bibfnamefont {M.}~\bibnamefont {Bressan}}, \bibinfo {author} {\bibfnamefont {V.}~\bibnamefont {Boria}},\ and\ \bibinfo {author} {\bibfnamefont {G.}~\bibnamefont {Conciauro}},\ }\bibfield  {title} {\bibinfo {title} {Analysis of cylindrical dielectric resonators in rectangular cavities using a state-space integral-equation method},\ }\href {https://doi.org/10.1109/LMWC.2006.885584} {\bibfield  {journal} {\bibinfo  {journal} {IEEE Microwave and Wireless Components Letters}\ }\textbf {\bibinfo {volume} {16}},\ \bibinfo {pages} {636} (\bibinfo {year} {2006})}\BibitemShut {NoStop}%
\bibitem [{\citenamefont {Arcioni}\ \emph {et~al.}(2014)\citenamefont {Arcioni}, \citenamefont {Bozzi}, \citenamefont {Bressan}, \citenamefont {Conciauro},\ and\ \citenamefont {Perregrini}}]{birme3d_overview}%
  \BibitemOpen
  \bibfield  {author} {\bibinfo {author} {\bibfnamefont {P.}~\bibnamefont {Arcioni}}, \bibinfo {author} {\bibfnamefont {M.}~\bibnamefont {Bozzi}}, \bibinfo {author} {\bibfnamefont {M.}~\bibnamefont {Bressan}}, \bibinfo {author} {\bibfnamefont {G.}~\bibnamefont {Conciauro}},\ and\ \bibinfo {author} {\bibfnamefont {L.}~\bibnamefont {Perregrini}},\ }\bibfield  {title} {\bibinfo {title} {The bi-rme method: an historical overview},\ }\bibfield  {journal} {\bibinfo  {journal} {2014 International Conference on Numerical Electromagnetic Modeling and Optimization for RF, Microwave, and Terahertz Applications (NEMO)}\ }\href {https://doi.org/10.1109/NEMO.2014.6995653} {10.1109/NEMO.2014.6995653} (\bibinfo {year} {2014})\BibitemShut {NoStop}%
\bibitem [{\citenamefont {Bressan}\ \emph {et~al.}(2018)\citenamefont {Bressan}, \citenamefont {Battistutta}, \citenamefont {Bozzi},\ and\ \citenamefont {Perregrini}}]{birme3d_pavia_MTT}%
  \BibitemOpen
  \bibfield  {author} {\bibinfo {author} {\bibfnamefont {M.}~\bibnamefont {Bressan}}, \bibinfo {author} {\bibfnamefont {S.}~\bibnamefont {Battistutta}}, \bibinfo {author} {\bibfnamefont {M.}~\bibnamefont {Bozzi}},\ and\ \bibinfo {author} {\bibfnamefont {L.}~\bibnamefont {Perregrini}},\ }\bibfield  {title} {\bibinfo {title} {Modeling of inhomogeneous and lossy waveguide components by the segmentation technique combined with the calculation of green’s function by ewald’s method},\ }\href {https://doi.org/10.1109/TMTT.2017.2787587} {\bibfield  {journal} {\bibinfo  {journal} {IEEE Transactions on Microwave Theory and Techniques}\ }\textbf {\bibinfo {volume} {66}},\ \bibinfo {pages} {633} (\bibinfo {year} {2018})}\BibitemShut {NoStop}%
\bibitem [{\citenamefont {Navarro}\ \emph {et~al.}(2022)\citenamefont {Navarro}, \citenamefont {Gimeno}, \citenamefont {Álvarez Melcón}, \citenamefont {Cuendis}, \citenamefont {Cogollos}, \citenamefont {Díaz-Morcillo}, \citenamefont {Gallego}, \citenamefont {Barceló}, \citenamefont {Golm}, \citenamefont {Irastorza}, \citenamefont {Guerrero},\ and\ \citenamefont {Garay}}]{bi-rme3d_axions}%
  \BibitemOpen
  \bibfield  {author} {\bibinfo {author} {\bibfnamefont {P.}~\bibnamefont {Navarro}}, \bibinfo {author} {\bibfnamefont {B.}~\bibnamefont {Gimeno}}, \bibinfo {author} {\bibfnamefont {A.}~\bibnamefont {Álvarez Melcón}}, \bibinfo {author} {\bibfnamefont {S.~A.}\ \bibnamefont {Cuendis}}, \bibinfo {author} {\bibfnamefont {C.}~\bibnamefont {Cogollos}}, \bibinfo {author} {\bibfnamefont {A.}~\bibnamefont {Díaz-Morcillo}}, \bibinfo {author} {\bibfnamefont {J.}~\bibnamefont {Gallego}}, \bibinfo {author} {\bibfnamefont {J.~G.}\ \bibnamefont {Barceló}}, \bibinfo {author} {\bibfnamefont {J.}~\bibnamefont {Golm}}, \bibinfo {author} {\bibfnamefont {I.}~\bibnamefont {Irastorza}}, \bibinfo {author} {\bibfnamefont {A.~L.}\ \bibnamefont {Guerrero}},\ and\ \bibinfo {author} {\bibfnamefont {C.~P.}\ \bibnamefont {Garay}},\ }\bibfield  {title} {\bibinfo {title} {{Wide-band full-wave electromagnetic modal analysis of the coupling between dark-matter axions and photons in microwave resonators}},\ }\bibfield  {journal} {\bibinfo
  {journal} {Physics of the Dark Universe}\ }\textbf {\bibinfo {volume} {36, 101001}},\ \href {https://doi.org/https://doi.org/10.1016/j.dark.2022.101001} {https://doi.org/10.1016/j.dark.2022.101001} (\bibinfo {year} {2022})\BibitemShut {NoStop}%
\bibitem [{\citenamefont {Tai}(1991)}]{chentotai}%
  \BibitemOpen
  \bibfield  {author} {\bibinfo {author} {\bibfnamefont {C.-T.}\ \bibnamefont {Tai}},\ }\href@noop {} {\emph {\bibinfo {title} {Generalized Vector and Dyadic Analysis. Applied Mathematics in Field Theory}}},\ \bibinfo {edition} {1st}\ ed.\ (\bibinfo  {publisher} {IEEE Press},\ \bibinfo {year} {1991})\BibitemShut {NoStop}%
\bibitem [{Note1()}]{Note1}%
  \BibitemOpen
  \bibinfo {note} {The characteristic impedance of the coaxial line is given by $Z_0 = \protect \frac {Z_w}{2\pi }\ln \protect \!\left (\protect \frac {R_{\protect \mathrm {out}}}{R_{\protect \mathrm {in}}}\right )$}\BibitemShut {NoStop}%
\bibitem [{\citenamefont {Duffy}\ \emph {et~al.}(2006)\citenamefont {Duffy}, \citenamefont {Sikivie}, \citenamefont {Tanner}, \citenamefont {Asztalos}, \citenamefont {Hagmann}, \citenamefont {Kinion}, \citenamefont {Rosenberg}, \citenamefont {van Bibber}, \citenamefont {Yu},\ and\ \citenamefont {Bradley}}]{ADMX_LF}%
  \BibitemOpen
  \bibfield  {author} {\bibinfo {author} {\bibfnamefont {L.~D.}\ \bibnamefont {Duffy}}, \bibinfo {author} {\bibfnamefont {P.}~\bibnamefont {Sikivie}}, \bibinfo {author} {\bibfnamefont {D.~B.}\ \bibnamefont {Tanner}}, \bibinfo {author} {\bibfnamefont {S.~J.}\ \bibnamefont {Asztalos}}, \bibinfo {author} {\bibfnamefont {C.}~\bibnamefont {Hagmann}}, \bibinfo {author} {\bibfnamefont {D.}~\bibnamefont {Kinion}}, \bibinfo {author} {\bibfnamefont {L.~J.}\ \bibnamefont {Rosenberg}}, \bibinfo {author} {\bibfnamefont {K.}~\bibnamefont {van Bibber}}, \bibinfo {author} {\bibfnamefont {D.~B.}\ \bibnamefont {Yu}},\ and\ \bibinfo {author} {\bibfnamefont {R.~F.}\ \bibnamefont {Bradley}},\ }\bibfield  {title} {\bibinfo {title} {High resolution search for dark-matter axions},\ }\bibfield  {journal} {\bibinfo  {journal} {Physical Review D}\ }\textbf {\bibinfo {volume} {74}},\ \href {https://doi.org/10.1103/physrevd.74.012006} {10.1103/physrevd.74.012006} (\bibinfo {year} {2006})\BibitemShut {NoStop}%
\bibitem [{\citenamefont {Alesini}\ \emph {et~al.}(2023)\citenamefont {Alesini}, \citenamefont {Babusci}, \citenamefont {Beltrame}, \citenamefont {Bossi}, \citenamefont {Ciambrone}, \citenamefont {D’Elia}, \citenamefont {{Di Gioacchino}}, \citenamefont {{Di Pirro}}, \citenamefont {Döbrich}, \citenamefont {Falferi}, \citenamefont {Gatti}, \citenamefont {Giannotti}, \citenamefont {Gianotti}, \citenamefont {Lamanna}, \citenamefont {Ligi}, \citenamefont {Maccarrone}, \citenamefont {Mazzitelli}, \citenamefont {Mirizzi}, \citenamefont {Mueck}, \citenamefont {Nardi}, \citenamefont {Nguyen}, \citenamefont {Rettaroli}, \citenamefont {Rezvani}, \citenamefont {Teofilo}, \citenamefont {Tocci}, \citenamefont {Tomassini}, \citenamefont {Visinelli},\ and\ \citenamefont {Zantedeschi}}]{FLASH}%
  \BibitemOpen
  \bibfield  {author} {\bibinfo {author} {\bibfnamefont {D.}~\bibnamefont {Alesini}}, \bibinfo {author} {\bibfnamefont {D.}~\bibnamefont {Babusci}}, \bibinfo {author} {\bibfnamefont {P.}~\bibnamefont {Beltrame}}, \bibinfo {author} {\bibfnamefont {F.}~\bibnamefont {Bossi}}, \bibinfo {author} {\bibfnamefont {P.}~\bibnamefont {Ciambrone}}, \bibinfo {author} {\bibfnamefont {A.}~\bibnamefont {D’Elia}}, \bibinfo {author} {\bibfnamefont {D.}~\bibnamefont {{Di Gioacchino}}}, \bibinfo {author} {\bibfnamefont {G.}~\bibnamefont {{Di Pirro}}}, \bibinfo {author} {\bibfnamefont {B.}~\bibnamefont {Döbrich}}, \bibinfo {author} {\bibfnamefont {P.}~\bibnamefont {Falferi}}, \bibinfo {author} {\bibfnamefont {C.}~\bibnamefont {Gatti}}, \bibinfo {author} {\bibfnamefont {M.}~\bibnamefont {Giannotti}}, \bibinfo {author} {\bibfnamefont {P.}~\bibnamefont {Gianotti}}, \bibinfo {author} {\bibfnamefont {G.}~\bibnamefont {Lamanna}}, \bibinfo {author} {\bibfnamefont {C.}~\bibnamefont {Ligi}}, \bibinfo {author} {\bibfnamefont
  {G.}~\bibnamefont {Maccarrone}}, \bibinfo {author} {\bibfnamefont {G.}~\bibnamefont {Mazzitelli}}, \bibinfo {author} {\bibfnamefont {A.}~\bibnamefont {Mirizzi}}, \bibinfo {author} {\bibfnamefont {M.}~\bibnamefont {Mueck}}, \bibinfo {author} {\bibfnamefont {E.}~\bibnamefont {Nardi}}, \bibinfo {author} {\bibfnamefont {F.}~\bibnamefont {Nguyen}}, \bibinfo {author} {\bibfnamefont {A.}~\bibnamefont {Rettaroli}}, \bibinfo {author} {\bibfnamefont {J.}~\bibnamefont {Rezvani}}, \bibinfo {author} {\bibfnamefont {F.~E.}\ \bibnamefont {Teofilo}}, \bibinfo {author} {\bibfnamefont {S.}~\bibnamefont {Tocci}}, \bibinfo {author} {\bibfnamefont {S.}~\bibnamefont {Tomassini}}, \bibinfo {author} {\bibfnamefont {L.}~\bibnamefont {Visinelli}},\ and\ \bibinfo {author} {\bibfnamefont {M.}~\bibnamefont {Zantedeschi}},\ }\bibfield  {title} {\bibinfo {title} {The future search for low-frequency axions and new physics with the {FLASH} resonant cavity experiment at {F}rascati {N}ational {L}aboratories},\ }\href
  {https://doi.org/https://doi.org/10.1016/j.dark.2023.101370} {\bibfield  {journal} {\bibinfo  {journal} {Physics of the Dark Universe}\ }\textbf {\bibinfo {volume} {42}},\ \bibinfo {pages} {101370} (\bibinfo {year} {2023})}\BibitemShut {NoStop}%
\bibitem [{\citenamefont {Ahyoune}\ \emph {et~al.}(2023)\citenamefont {Ahyoune} \emph {et~al.}}]{RADES_BabyIAXO}%
  \BibitemOpen
  \bibfield  {author} {\bibinfo {author} {\bibfnamefont {S.}~\bibnamefont {Ahyoune}} \emph {et~al.},\ }\bibfield  {title} {\bibinfo {title} {{A {P}roposal for a {L}ow-{F}requency {A}xion {S}earch in the 1-2 $\mu$e{V} {R}ange and {B}elow with the {B}aby{IAXO} {M}agnet}},\ }\href {https://doi.org/10.1002/andp.202300326} {\bibfield  {journal} {\bibinfo  {journal} {Annalen Phys.}\ }\textbf {\bibinfo {volume} {535}},\ \bibinfo {pages} {2300326} (\bibinfo {year} {2023})},\ \Eprint {https://arxiv.org/abs/2306.17243} {arXiv:2306.17243 [physics.ins-det]} \BibitemShut {NoStop}%
\bibitem [{\citenamefont {Abeln}\ \emph {et~al.}(2021)\citenamefont {Abeln}, \citenamefont {Altenmüller}, \citenamefont {Cuendis}, \citenamefont {Armengaud}, \citenamefont {Attié}, \citenamefont {Aune}, \citenamefont {Basso}, \citenamefont {Bergé}, \citenamefont {Biasuzzi}, \citenamefont {de~Sousa}, \citenamefont {Brun}, \citenamefont {Bykovskiy}, \citenamefont {Calvet}, \citenamefont {Carmona}, \citenamefont {Castel}, \citenamefont {Cebrián}, \citenamefont {Chernov}, \citenamefont {Christensen}, \citenamefont {Civitani},\ and\ \citenamefont {Yanes-Díaz}}]{BabyIAXO2021}%
  \BibitemOpen
  \bibfield  {author} {\bibinfo {author} {\bibfnamefont {A.}~\bibnamefont {Abeln}}, \bibinfo {author} {\bibfnamefont {K.}~\bibnamefont {Altenmüller}}, \bibinfo {author} {\bibfnamefont {S.}~\bibnamefont {Cuendis}}, \bibinfo {author} {\bibfnamefont {E.}~\bibnamefont {Armengaud}}, \bibinfo {author} {\bibfnamefont {D.}~\bibnamefont {Attié}}, \bibinfo {author} {\bibfnamefont {S.}~\bibnamefont {Aune}}, \bibinfo {author} {\bibfnamefont {S.}~\bibnamefont {Basso}}, \bibinfo {author} {\bibfnamefont {L.}~\bibnamefont {Bergé}}, \bibinfo {author} {\bibfnamefont {B.}~\bibnamefont {Biasuzzi}}, \bibinfo {author} {\bibfnamefont {P.~B.}\ \bibnamefont {de~Sousa}}, \bibinfo {author} {\bibfnamefont {P.}~\bibnamefont {Brun}}, \bibinfo {author} {\bibfnamefont {N.}~\bibnamefont {Bykovskiy}}, \bibinfo {author} {\bibfnamefont {D.}~\bibnamefont {Calvet}}, \bibinfo {author} {\bibfnamefont {J.}~\bibnamefont {Carmona}}, \bibinfo {author} {\bibfnamefont {J.}~\bibnamefont {Castel}}, \bibinfo {author} {\bibfnamefont {S.}~\bibnamefont
  {Cebrián}}, \bibinfo {author} {\bibfnamefont {V.}~\bibnamefont {Chernov}}, \bibinfo {author} {\bibfnamefont {F.}~\bibnamefont {Christensen}}, \bibinfo {author} {\bibfnamefont {M.}~\bibnamefont {Civitani}},\ and\ \bibinfo {author} {\bibfnamefont {A.}~\bibnamefont {Yanes-Díaz}},\ }\bibfield  {title} {\bibinfo {title} {Conceptual design of {B}aby{IAXO}, the intermediate stage towards the {I}nternational {A}xion {O}bservatory},\ }\bibfield  {journal} {\bibinfo  {journal} {Journal of High Energy Physics}\ }\textbf {\bibinfo {volume} {2021}},\ \href {https://doi.org/10.1007/JHEP05(2021)137} {10.1007/JHEP05(2021)137} (\bibinfo {year} {2021})\BibitemShut {NoStop}%
\bibitem [{Note2()}]{Note2}%
  \BibitemOpen
  \bibinfo {note} {Since the cavity geometry is not canonical and incorporates tuning mechanisms, the resonant modes are renamed as quasi-TE and quasi-TM, abbreviated as qTE and qTM, respectively.}\BibitemShut {Stop}%
\bibitem [{\citenamefont {Reina-Valero}\ \emph {et~al.}(2025)\citenamefont {Reina-Valero}, \citenamefont {Navarro-Madrid}, \citenamefont {Blas}, \citenamefont {D\'{\i}az-Morcillo}, \citenamefont {Irastorza}, \citenamefont {Gimeno},\ and\ \citenamefont {Monz\'o-Cabrera}}]{BabyIAXO_HFGW}%
  \BibitemOpen
  \bibfield  {author} {\bibinfo {author} {\bibfnamefont {J.}~\bibnamefont {Reina-Valero}}, \bibinfo {author} {\bibfnamefont {J.~R.}\ \bibnamefont {Navarro-Madrid}}, \bibinfo {author} {\bibfnamefont {D.}~\bibnamefont {Blas}}, \bibinfo {author} {\bibfnamefont {A.}~\bibnamefont {D\'{\i}az-Morcillo}}, \bibinfo {author} {\bibfnamefont {I.~G.}\ \bibnamefont {Irastorza}}, \bibinfo {author} {\bibfnamefont {B.}~\bibnamefont {Gimeno}},\ and\ \bibinfo {author} {\bibfnamefont {J.}~\bibnamefont {Monz\'o-Cabrera}},\ }\bibfield  {title} {\bibinfo {title} {High-frequency gravitational waves detection with the {B}aby{IAXO} haloscopes},\ }\href {https://doi.org/10.1103/PhysRevD.111.043024} {\bibfield  {journal} {\bibinfo  {journal} {Phys. Rev. D}\ }\textbf {\bibinfo {volume} {111}},\ \bibinfo {pages} {043024} (\bibinfo {year} {2025})}\BibitemShut {NoStop}%
\bibitem [{CST()}]{CST}%
  \BibitemOpen
  \href@noop {} {\bibinfo {title} {\uppercase{CST STUDIO SUITE}: Electromagnetic field simulation software, \url{https://www.3ds.com/products-services/simulia/products/cst-studio-suite/}}}\BibitemShut {NoStop}%
\bibitem [{Note3()}]{Note3}%
  \BibitemOpen
  \bibinfo {note} {Therefore, after selecting these modes, the position of the readout port can be optimized to couple this pair of modes and further reduce the leakage of the pump mode.}\BibitemShut {Stop}%
\bibitem [{\citenamefont {{Rohde \& Schwarz}}(2026)}]{RohdeSchwarz_SMA100B_2026}%
  \BibitemOpen
  \bibfield  {author} {\bibinfo {author} {\bibnamefont {{Rohde \& Schwarz}}},\ }\href@noop {} {\bibinfo {title} {{R\&{S} {SMA}100{B} {RF} and {M}icrowave {S}ignal {G}enerator: {S}pecifications}}} (\bibinfo {year} {2026}),\ \bibinfo {note} {datasheet}\BibitemShut {NoStop}%
\bibitem [{Note4()}]{Note4}%
  \BibitemOpen
  \bibinfo {note} {The value $R_s=1\protect \,\protect \mathrm {n}\Omega $ should be understood as an optimistic SRF benchmark; actual values will depend on material treatment, RF field amplitude, temperature, trapped flux, and residual losses.}\BibitemShut {Stop}%
\bibitem [{\citenamefont {Petersan}\ and\ \citenamefont {Anlage}(1998)}]{Petersan1998Q}%
  \BibitemOpen
  \bibfield  {author} {\bibinfo {author} {\bibfnamefont {P.~J.}\ \bibnamefont {Petersan}}\ and\ \bibinfo {author} {\bibfnamefont {S.~M.}\ \bibnamefont {Anlage}},\ }\bibfield  {title} {\bibinfo {title} {Measurement of resonant frequency and quality factor of microwave resonators: Comparison of methods},\ }\href {https://doi.org/10.1063/1.368498} {\bibfield  {journal} {\bibinfo  {journal} {Journal of Applied Physics}\ }\textbf {\bibinfo {volume} {84}},\ \bibinfo {pages} {3392} (\bibinfo {year} {1998})}\BibitemShut {NoStop}%
\bibitem [{\citenamefont {Gustavsen}\ and\ \citenamefont {Semlyen}(1999)}]{Gustavsen1999VF}%
  \BibitemOpen
  \bibfield  {author} {\bibinfo {author} {\bibfnamefont {B.}~\bibnamefont {Gustavsen}}\ and\ \bibinfo {author} {\bibfnamefont {A.}~\bibnamefont {Semlyen}},\ }\bibfield  {title} {\bibinfo {title} {Rational approximation of frequency domain responses by vector fitting},\ }\href@noop {} {\bibfield  {journal} {\bibinfo  {journal} {IEEE Transactions on Power Delivery}\ }\textbf {\bibinfo {volume} {14}},\ \bibinfo {pages} {1052} (\bibinfo {year} {1999})}\BibitemShut {NoStop}%
\bibitem [{\citenamefont {Ramella}\ \emph {et~al.}(2021)\citenamefont {Ramella}, \citenamefont {Pirola},\ and\ \citenamefont {Corbellini}}]{Ramella2021HighQ}%
  \BibitemOpen
  \bibfield  {author} {\bibinfo {author} {\bibfnamefont {C.}~\bibnamefont {Ramella}}, \bibinfo {author} {\bibfnamefont {M.}~\bibnamefont {Pirola}},\ and\ \bibinfo {author} {\bibfnamefont {S.}~\bibnamefont {Corbellini}},\ }\bibfield  {title} {\bibinfo {title} {Accurate characterization of high-q microwave resonances for metrology applications},\ }\href {https://doi.org/10.1109/JMW.2021.3063247} {\bibfield  {journal} {\bibinfo  {journal} {IEEE Journal of Microwaves}\ }\textbf {\bibinfo {volume} {1}},\ \bibinfo {pages} {610} (\bibinfo {year} {2021})}\BibitemShut {NoStop}%
\bibitem [{\citenamefont {{Micro Lambda Wireless, Inc.}}(2024)}]{BandReject}%
  \BibitemOpen
  \bibfield  {author} {\bibinfo {author} {\bibnamefont {{Micro Lambda Wireless, Inc.}}},\ }\href {https://www.microlambdawireless.com/uploads/pdfs/MLUN%20Ultra%20Notch%20BR.pdf} {\bibinfo {title} {{MLUN Series Ultra Notch YIG Band Reject Filters}}} (\bibinfo {year} {2024}),\ \bibinfo {note} {model MLUN-0503; accessed 29 May 2026}\BibitemShut {NoStop}%
\bibitem [{\citenamefont {{Wainwright Instruments GmbH}}({\natexlab{a}})}]{Wainwright_Tunable}%
  \BibitemOpen
  \bibfield  {author} {\bibinfo {author} {\bibnamefont {{Wainwright Instruments GmbH}}},\ }\href {https://www.wainwright-filters.com/standard-filter/wtrctv10-400-700-110-02-50-25-07-quote} {\bibinfo {title} {{WTRCTV10-400-700-110-02-50-25-07-QUOTE: Manually Tunable Band-Reject/Notch Filter}}} ({\natexlab{a}}),\ \bibinfo {note} {10-section cavity design; accessed June 4, 2026}\BibitemShut {NoStop}%
\bibitem [{\citenamefont {{Wainwright Instruments GmbH}}({\natexlab{b}})}]{Wainwright_Fixed}%
  \BibitemOpen
  \bibfield  {author} {\bibinfo {author} {\bibnamefont {{Wainwright Instruments GmbH}}},\ }\href {https://www.wainwright-filters.com/standard-filter/wrctv10-400-700-10-7-80-28-x-fo-quote} {\bibinfo {title} {{WRCTV10-400-700-10-7-80-28-X-FO-QUOTE Band-Reject/Notch Filter}}} ({\natexlab{b}}),\ \bibinfo {note} {ten-section cavity filter with a customer-defined center frequency between 400 and 700 MHz, a 10 MHz stopband bandwidth, and a minimum stopband attenuation of 80 dB; accessed June 4, 2026}\BibitemShut {NoStop}%
\bibitem [{Cry()}]{Cryostat}%
  \BibitemOpen
  \href@noop {} {\bibinfo {title} {\uppercase{BLUEFORS LD350/LD450}, \url{https://bluefors.com/products/dilution-refrigerator-measurement-systems/ld-dilution-refrigerator-measurement-system/}}}\BibitemShut {NoStop}%
\bibitem [{\citenamefont {Salemi}\ \emph {et~al.}(2021)\citenamefont {Salemi}, \citenamefont {Foster}, \citenamefont {Ouellet}, \citenamefont {Gavin}, \citenamefont {Pappas}, \citenamefont {Cheng}, \citenamefont {Richardson}, \citenamefont {Henning}, \citenamefont {Kahn}, \citenamefont {Nguyen}, \citenamefont {Rodd}, \citenamefont {Safdi},\ and\ \citenamefont {Winslow}}]{Salemi2021ABRACADABRA}%
  \BibitemOpen
  \bibfield  {author} {\bibinfo {author} {\bibfnamefont {C.~P.}\ \bibnamefont {Salemi}}, \bibinfo {author} {\bibfnamefont {J.~W.}\ \bibnamefont {Foster}}, \bibinfo {author} {\bibfnamefont {J.~L.}\ \bibnamefont {Ouellet}}, \bibinfo {author} {\bibfnamefont {A.}~\bibnamefont {Gavin}}, \bibinfo {author} {\bibfnamefont {K.~M.~W.}\ \bibnamefont {Pappas}}, \bibinfo {author} {\bibfnamefont {S.}~\bibnamefont {Cheng}}, \bibinfo {author} {\bibfnamefont {K.~A.}\ \bibnamefont {Richardson}}, \bibinfo {author} {\bibfnamefont {R.}~\bibnamefont {Henning}}, \bibinfo {author} {\bibfnamefont {Y.}~\bibnamefont {Kahn}}, \bibinfo {author} {\bibfnamefont {R.}~\bibnamefont {Nguyen}}, \bibinfo {author} {\bibfnamefont {N.~L.}\ \bibnamefont {Rodd}}, \bibinfo {author} {\bibfnamefont {B.~R.}\ \bibnamefont {Safdi}},\ and\ \bibinfo {author} {\bibfnamefont {L.}~\bibnamefont {Winslow}},\ }\bibfield  {title} {\bibinfo {title} {Search for low-mass axion dark matter with {ABRACADABRA}-10 cm},\ }\href
  {https://doi.org/10.1103/PhysRevLett.127.081801} {\bibfield  {journal} {\bibinfo  {journal} {Physical Review Letters}\ }\textbf {\bibinfo {volume} {127}},\ \bibinfo {pages} {081801} (\bibinfo {year} {2021})},\ \Eprint {https://arxiv.org/abs/2102.06722} {arXiv:2102.06722 [hep-ex]} \BibitemShut {NoStop}%
\bibitem [{\citenamefont {Rapidis}(2023)}]{Rapidis2023DMRadio}%
  \BibitemOpen
  \bibfield  {author} {\bibinfo {author} {\bibfnamefont {N.~M.}\ \bibnamefont {Rapidis}},\ }\bibfield  {title} {\bibinfo {title} {Status of {DMRadio-50L} and {DMRadio-m}$^3$},\ }\href {https://doi.org/10.21468/SciPostPhysProc.12.036} {\bibfield  {journal} {\bibinfo  {journal} {SciPost Physics Proceedings}\ }\textbf {\bibinfo {volume} {12}},\ \bibinfo {pages} {036} (\bibinfo {year} {2023})},\ \Eprint {https://arxiv.org/abs/2210.07215} {arXiv:2210.07215 [hep-ex]} \BibitemShut {NoStop}%
\bibitem [{\citenamefont {Brouwer}\ \emph {et~al.}(2022{\natexlab{a}})\citenamefont {Brouwer} \emph {et~al.}}]{Brouwer2022DMRadiom3}%
  \BibitemOpen
  \bibfield  {author} {\bibinfo {author} {\bibfnamefont {L.}~\bibnamefont {Brouwer}} \emph {et~al.} (\bibinfo {collaboration} {DMRadio Collaboration}),\ }\bibfield  {title} {\bibinfo {title} {Projected sensitivity of {DMRadio-m}$^3$: A search for the {QCD} axion below 1~{$\mu$eV}},\ }\href {https://doi.org/10.1103/PhysRevD.106.103008} {\bibfield  {journal} {\bibinfo  {journal} {Physical Review D}\ }\textbf {\bibinfo {volume} {106}},\ \bibinfo {pages} {103008} (\bibinfo {year} {2022}{\natexlab{a}})},\ \Eprint {https://arxiv.org/abs/2204.13781} {arXiv:2204.13781 [hep-ex]} \BibitemShut {NoStop}%
\bibitem [{\citenamefont {Brouwer}\ \emph {et~al.}(2022{\natexlab{b}})\citenamefont {Brouwer} \emph {et~al.}}]{Brouwer2022DMRadioGUT}%
  \BibitemOpen
  \bibfield  {author} {\bibinfo {author} {\bibfnamefont {L.}~\bibnamefont {Brouwer}} \emph {et~al.} (\bibinfo {collaboration} {DMRadio Collaboration}),\ }\bibfield  {title} {\bibinfo {title} {Proposal for a definitive search for {GUT}-scale {QCD} axions},\ }\href {https://doi.org/10.1103/PhysRevD.106.112003} {\bibfield  {journal} {\bibinfo  {journal} {Physical Review D}\ }\textbf {\bibinfo {volume} {106}},\ \bibinfo {pages} {112003} (\bibinfo {year} {2022}{\natexlab{b}})},\ \Eprint {https://arxiv.org/abs/2203.11246} {arXiv:2203.11246 [hep-ex]} \BibitemShut {NoStop}%
\bibitem [{\citenamefont {Navarro-Madrid}\ \emph {et~al.}(2025)\citenamefont {Navarro-Madrid}, \citenamefont {Reina-Valero}, \citenamefont {Díaz-Morcillo},\ and\ \citenamefont {Gimeno}}]{DarkPhoton_JR}%
  \BibitemOpen
  \bibfield  {author} {\bibinfo {author} {\bibfnamefont {J.~R.}\ \bibnamefont {Navarro-Madrid}}, \bibinfo {author} {\bibfnamefont {J.}~\bibnamefont {Reina-Valero}}, \bibinfo {author} {\bibfnamefont {A.}~\bibnamefont {Díaz-Morcillo}},\ and\ \bibinfo {author} {\bibfnamefont {B.}~\bibnamefont {Gimeno}},\ }\bibfield  {title} {\bibinfo {title} {Enhancement of dark photon haloscope sensitivity with degenerate modes: Toward axion-level form factor and polarization determination},\ }\bibfield  {journal} {\bibinfo  {journal} {Physical Review D}\ }\textbf {\bibinfo {volume} {112}},\ \href {https://doi.org/10.1103/L9CQ-NJ8C} {10.1103/L9CQ-NJ8C} (\bibinfo {year} {2025})\BibitemShut {NoStop}%
\bibitem [{\citenamefont {Hanson}\ and\ \citenamefont {Yakovlev}(2002)}]{hanson_yakovlev}%
  \BibitemOpen
  \bibfield  {author} {\bibinfo {author} {\bibfnamefont {G.~W.}\ \bibnamefont {Hanson}}\ and\ \bibinfo {author} {\bibfnamefont {A.~B.}\ \bibnamefont {Yakovlev}},\ }\href@noop {} {\emph {\bibinfo {title} {Operator Theory for Electromagnetics. An introduction}}},\ \bibinfo {edition} {1st}\ ed.\ (\bibinfo  {publisher} {Springer-Verlag},\ \bibinfo {year} {2002})\BibitemShut {NoStop}%
\bibitem [{\citenamefont {Bachbauer}\ and\ \citenamefont {Gold}(2025)}]{Bachbauer2025Fano}%
  \BibitemOpen
  \bibfield  {author} {\bibinfo {author} {\bibfnamefont {F.}~\bibnamefont {Bachbauer}}\ and\ \bibinfo {author} {\bibfnamefont {G.}~\bibnamefont {Gold}},\ }\bibfield  {title} {\bibinfo {title} {Compensation of fano resonances in microwave resonators},\ }in\ \href {https://doi.org/10.23919/EuMC65286.2025.11234994} {\emph {\bibinfo {booktitle} {Proceedings of the 55th European Microwave Conference (EuMC)}}}\ (\bibinfo  {publisher} {European Microwave Association (EuMA)},\ \bibinfo {address} {Utrecht, The Netherlands},\ \bibinfo {year} {2025})\ pp.\ \bibinfo {pages} {961--964}\BibitemShut {NoStop}%
\bibitem [{\citenamefont {Kancleris}\ \emph {et~al.}(2020)\citenamefont {Kancleris}, \citenamefont {Slekas}, \citenamefont {Seliuta},\ and\ \citenamefont {Kamarauskas}}]{Kancleris2020Fano}%
  \BibitemOpen
  \bibfield  {author} {\bibinfo {author} {\bibfnamefont {Z.}~\bibnamefont {Kancleris}}, \bibinfo {author} {\bibfnamefont {G.}~\bibnamefont {Slekas}}, \bibinfo {author} {\bibfnamefont {D.}~\bibnamefont {Seliuta}},\ and\ \bibinfo {author} {\bibfnamefont {A.}~\bibnamefont {Kamarauskas}},\ }\bibfield  {title} {\bibinfo {title} {Fano resonance in metasurfaces and its application},\ }in\ \href@noop {} {\emph {\bibinfo {booktitle} {Proceedings of the International Radar Symposium (IRS 2020)}}}\ (\bibinfo {year} {2020})\ pp.\ \bibinfo {pages} {328--333}\BibitemShut {NoStop}%
\bibitem [{\citenamefont {Steshenko}\ \emph {et~al.}(2025)\citenamefont {Steshenko}, \citenamefont {Khutoryan}, \citenamefont {Kulik}, \citenamefont {Kuleshov}, \citenamefont {Kirilenko},\ and\ \citenamefont {Ponomarenko}}]{steshenko2025inclinedFano}%
  \BibitemOpen
  \bibfield  {author} {\bibinfo {author} {\bibfnamefont {S.}~\bibnamefont {Steshenko}}, \bibinfo {author} {\bibfnamefont {E.}~\bibnamefont {Khutoryan}}, \bibinfo {author} {\bibfnamefont {D.}~\bibnamefont {Kulik}}, \bibinfo {author} {\bibfnamefont {O.}~\bibnamefont {Kuleshov}}, \bibinfo {author} {\bibfnamefont {A.}~\bibnamefont {Kirilenko}},\ and\ \bibinfo {author} {\bibfnamefont {S.}~\bibnamefont {Ponomarenko}},\ }\href {https://doi.org/10.1109/DIPED66951.2025.11194392} {\bibinfo {title} {Inclined bar in a rectangular waveguide as a position-controllable fano-resonance stop-band component}} (\bibinfo {year} {2025})\BibitemShut {NoStop}%
\end{thebibliography}
